%Paper: gr-qc/9302024
%From: LOUKO@SUHEP.PHY.SYR.EDU
%Date: Fri, 19 Feb 1993 10:46 EST

%%%%%%%%%%%%%%%%%%%%%%%%%%%%%%%%%%%%%%%%%%%%%%%%%%%%%%%%%%%%%%%%%%%%%%%
      \parskip=0pt plus 1pt
     \def\section#1#2{\vskip1truecm\noindent{\bf#1\ #2}\vskip0.5truecm}\indent
     \def\subsection#1#2{\vskip1truecm\noindent{\bf#1\
          #2}\vskip0.5truecm}\indent

      \def\bneq{$$}
      \def\eneq#1{\eqno#1 $$}
      \def\beginalign{$$\eqalign}
      \def\endalign#1{\eqno#1 $$}

%%%%%%%%%%%%%%%%%%%%%%%%%%%%%%%%%%%%%%%%%%%%%%%%%%%%%%%%%%%%%%%%%%%%%%%%%%

      \def\ct{canonical transformation}

      \def\ps{phase space}

% Fields in math mode or text
      \def\A{\hbox{$A_{aA}{}^B$}}

      \def\E{\hbox{$E^a\!{}_A{}^B$}}

      \def\fA{\hbox{${}^4\!A_{aA}{}^B$}}

 %Other symbols
      \def\ab{{\alpha\beta}}
      
      \def\comp{{\rm C}\llap{\vrule height7.1pt width1pt depth-.4pt\phantom t}}
      \def\conj{{\displaystyle\ast}}
      
      \def\d{\partial}
      \def\D{{\cal D}}

      \def\implies{\Rightarrow}

      \def\Lapop{\displaystyle{{\hbox to 0pt{$\sqcup$\hss}}\sqcap}}
      \def\Lie#1{{\cal L}_{\tenrm #1}}
      \def\lint{\int\nolimits}
      \def\loint{\oint\nolimits}
      \def\nab{\nabla\!}
      \def\ovr{\overline}
      \def\real{{\rm I\!R}}
      \def\to{\rightarrow}
      \def\tr{\mathop{\rm tr}}
      \def\tw{\widetilde}
      \def\und{\underline}
      \def\ut#1{\rlap{\lower1ex\hbox{$\sim$}}#1{}}

%\singlespace
%\hsize=6truein\overfullrule=0pt
\magnification=1200

\centerline{\bf Mathematical Problems of}
\centerline{\bf  Non-perturbative Quantum General Relativity}
\bigskip
\centerline{Abhay Ashtekar}
\smallskip
\centerline{Physics Department, Syracuse University, Syracuse, NY 13244-1130}
\bigskip

\centerline{\bf Abstract}

The purpose of these lectures is to discuss in some detail a
new, {\it non-perturbative} approach to quantum gravity. I would like to
present the basic ideas, outline the key results that have
been obtained so far and indicate where we are headed and what the hopes
are. The audience at this summer school had a diverse background; many
came from high energy physics, some from mathematical physics and the rest
from general relativity. Therefore, I have tried to keep the technicalities
--particularly proofs and even the number of equations-- to a minimum.
My hope is that a research student in any of these three fields should be
able to get a bird's eye view of the entire program. In particular, I have
kept all three perspectives in mind while discussing the difficulties one
encounters and strategies one adopts and in evaluating the successes
and limitations of the program.
\footnote{}{Lectures delivered at the 1992 Les Houches summer school on
Gravitation and Quan\-ti\-za\-tion.}
\vfill\break

%
%\input [ijtpd.lh]qgmac
%\singlespace
\section{1.}{${}\quad$ Introduction}%\par
\medskip
{\narrower\narrower\smallskip\noindent
{\sl ... a really new field of experience will always lead to
crystallization of a new system of scientific concepts and laws.  ...
When faced with essentially new intellectual challenges we continually
follow the example of Columbus who possessed the courage to leave the known
world in the almost insane hope of finding land again beyond the sea.}
\smallskip
W. Heisenberg (Changes in the Foundation of Exact Science)\smallskip}
\medskip

\subsection{1.1}{The problem}%\par

Quantum mechanics and general relativity are among the greatest scientific
discoveries ever made. Born early in this century, the two theories
have dictated the course of physics in all the subsequent decades. The
reason is that they have altered the very conceptual fabric that
underlies fundamental physics. Furthermore,
each has been extraordinarily successful in describing the physical phenomena
in its domain. The experimental checks of QED, for example, are legendary. And
we have learned through Thibault Damour's lectures that experimental tests of
general relativity are now reaching even higher levels of accuracy. And yet
the two theories offer us  {\it strikingly} different pictures of
physical reality. The world of general relativity is precise
and geometric. The world of quantum mechanics is governed by intrinsic
uncertainties and probabilities. It is only because the Planck length happens
to be so small that we can maintain a schizophrenic attitude, using general
relativity to discuss cosmological and astrophysical phenomena and then
blissfully switching over to quantum mechanics to examine the sub-atomic
world. The strategy is of course quite appropriate as a practical stand.
However, form a conceptual viewpoint, it is quite unsatisfactory. Everything
in our past experience of physics tells us that the two descriptions of Nature
we currently use must be approximations, special cases which arise as
suitable limits of a single, universal theory. That theory must be based
on a synthesis of the basic principles of general relativity and quantum
mechanics. This would be the quantum theory of gravity that we are
seeking.

The burden on this theory is huge. First, it should adequately describe
{\it all} the known physical phenomena, currently encompassed in the
separate regimes of general relativity and quantum mechanics. Second,
it should adequately handle Planck regime, the common turf on which both
theories lay separate claims.

The general nature of the claims laid by quantum mechanics are quite
familiar to most working physicists. Take Maxwell's theory as an example.
In classical physics, central objects are the solutions to Maxwell's
equations: at each point of space-time, there is a well-defined electric
and magnetic field; these fields propagate via hyperbolic equations;
the fluxes of energy and momentum across any surface are computed
as integrals of smooth fields; and so on. When quantum mechanics is brought
in, it demands a radical shift of the framework. Now, the electric and
magnetic fields are operators, subject to the Heisenberg uncertainty relations,
whence their values can no longer be specified simultaneously; the field
excitations represent particle states --photons; the propagation is via the
Schr\"odinger evolution in the Fock space; and, the energy is expressed as a
sum of contributions from photons, each carrying a quantum $h\nu$. In the
case of gravity, one expects a similar quantum description, far removed from
the geometric world of general relativity.

General relativity, on the other
hand, has its own claims which are perhaps not so familiar. It asks that
the theory be diffeomorphism invariant, that there be no background
geometric structures, that there be no fiducial space-time to anchor the
machinery of quantum mechanics in. It says that, one cannot just postulate
micro-causality because there is no background causal structure in terms
of which to formulate the notion; that one cannot simply evolve states via a
Schr\"odinger equation because a priori there is no time to evolve anything
in; that the familar particles of specific mass and spin cannot be the
{\it fundamental} excitations because there is no Poincar\'e group to
construct these Casimir operators from. These are surprisingly strong
demands. In particular, they cast a grave doubt on direct applicability
of the ideas and techniques from quantum field theory which are rooted
in a background (Minkowskian) geometry. While these methods would be
obviously useful in suitable approximations, suddenly, they seem to be
inappropriate to tackle the problem at a fundamental level.

What questions is such a framework to address? Here are some examples.
I have purposely tried to focus on issues that are rarely on the fore
front in field theoretic discussions of quantum gravity, to emphasize
that the theory has much more than scattering amplitudes to come to grips
with.

Through general relativity, Einstein taught us that the gravitational
field is inseparably intertwined with the geometry of space-time, whence
geometry is a dynamical, physical entity with degrees of freedom of its
own. Any genuine quantum theory of gravity would therefore have to be a
quantum theory of space-time geometry. What does quantum geometry mean?
If it is a physical entity, like this sheet of paper you are looking
at, does it have constituents? What are they? Is it, like the sheet
of paper, mostly empty? How do these constituents manage to group together
to give us the continuum picture under ``normal'' conditions? What happens
to it near what we call, in classical relativity, a singularity? What
replaces the continuum picture then? Is the smooth geometry that we perceive
on macroscopic scales really a ``mean field''? What physical concepts and
mathematical techniques should we use to go beyond the mean field
approximation?

There has been much excitement about the interplay between quantum field
theory in curved space-times and statistical mechanics in black hole physics.
This is a most fascinating development indeed and there is a general belief
that this interplay is something very basic, very fundamental. Is this really
so from the perspective of full quantum gravity? In classical general
relativity, the notion of a black hole is {\it genuinely} global. To know if
given initial conditions lead to the formation of a black hole, for example,
we have to first obtain the {\it maximal} evolution of the data and then
ask if, in the resulting space-time, the past of future null infinity is
the entire space-time. If it is not, the data evolved to form a black hole
and the boundary of the past of future null infinity is the horizon. It is
{\it very} striking how very global the notion is. Now, in quantum mechanics,
due to the basic uncertainty principle, classical trajectories do not play
a fundamental role. Furthermore, in non-trivial, interacting quantum theory,
it is rather rare that there are solutions to the quantum evolution equations
which approximate a classical solution {\it globally in time}. If this is so,
why should anything like a black hole appear in exact quantum gravity? Why
should black hole thermodynamics have significance beyond the semi-classical
approximation? Let me give an analogy. The Bohr model of atom was considered
fundamental in the 1910s. We soon learned, however, that its
ad-hoc mixture of quantization rules and classical (electron)
trajectories has no fundamental place in the framework of full quantum
mechanics. The ideas are useful only in semi-classical approximations. Is
the situation perhaps similar with black hole thermodynamics? Would
black hole thermodynamics turn out to be the ``Bohr atom of quantum
gravity''? After all, these results are also obtained by a somewhat ad-hoc
combination of classical ideas --black hole geometries with their event
horizons and the resulting classical, causal structure-- and quantum
mechanics --second quantization and Bogoliubov transformations. Or, is
there something much deeper going on here? Are there perhaps exact quantum
states of gravity which do resemble black holes globally? Are the laws
of black hole thermodynamics then firmly rooted in the exact framework
of quantum gravity? This is not out of the question, for black holes are
very special entities in the classical theory. One would hope that quantum
gravity can be developed sufficiently in the coming years to decide which of
the two scenarios is likely to occur. Of course, it would be no mean feat
to be ``just'' the Bohr atom of quantum gravity for, although it does not have
much of a role to play in the finished picture, historically, the Bohr atom
did play a key role in the development of quantum mechanics.

Then, there are a number of questions regarding the nature
of time and dynamics. At a fundamental level, since there is no background
metric, there is no a priori notion of time either. What does dynamics and
evolution even mean if there is no background space-time? How is time born
in the framework? Is it only an approximate concept or is it exact? If
it is approximate, is the notion of unitarity also approximate? Why then do
we not perceive any violations? These questions naturally lead one to the
issue of measurement theory. Constructing a mathematical framework
is only half of the story. We need a suitable framework to discuss the
issues of interpretation as well. In absence of a background space-time
geometry, the Copenhagen interpretation is of very little use. What should
replace it? The probabilities for an exhaustive set of mutually exclusive
alternatives should add up to one. In quantum mechanics, this is generally
ensured by using an instant of time to specify such alternatives. What is one
to do when there is no time and no instants? These are fascinating issues.
However, since the entire course of lectures by Jim Hartle is devoted to this
problem of interpretation, I will not elaborate on this point any further.

\goodbreak
\section{1.2}{${}\quad$ Difficulties}

The importance of the problem of unification of general relativity
and quantum theory was recognized as early as 1930's and a great deal
of effort has been devoted to it in the last three decades.
Yet, we seem to be far from seeing the light at the end of the tunnel.
Why is the problem so hard? What are the principal difficulties? Why
can we not apply the quantization techniques that have been successful in
theories of other interactions? In this subsection, I will address these
questions from various perspectives.

The main difficulty, of course, is the lack of experimental data with a
direct bearing on quantum gravity. One can argue that this need not be
an unsurmountable obstacle. After all, one hardly had any experimental
data with a direct bearing on general relativity when the theory was invented.
Furthermore, the main motivation came from the incompatibility of Newtonian
gravity with special relativity. We face a similar situation; we too are
driven by a basic incompatibility. However, it is also clear that the
situation with general relativity is an anomaly rather than a rule. Most
new physical theories --including quantum mechanics -- arose and were
continually guided and shaped by experimental input. In quantum gravity,
we are trying to make a jump by some twenty orders of magnitude, from a
fermi to a Planck length. The hope that there is no dramatically new physics
in the intermediate range is probably just that --hope.

This experimental situation, however, makes the situation even more
puzzling. If there is hardly any experimental data, theorists should have
a ball; without these ``external, bothersome constraints,'' they should be
able to churn
out a theory a week. Why then do we not have a {\it single} theory in spite
of all this work? The brief answer, I think, is that it is {\it very}
difficult to do quantum physics in absence of a background space-time. We have
{\it very} little experience in constructing physically realistic,
diffeomorphism invariant field theories. Indeed, until recently, there
were just a handful of examples, obtained by truncating general relativity
in various ways. It is only in the last two or three years that a significant
number of diffeomorphism invariant models with an infinite number of degrees
of freedom has become available, albeit in low space-time dimensions.

This response may not satisfy the reader. After all, particle theorists,
mathematical physicists and general relativists all have their favorite
strategies to the problem of quantization and it is important to know
why these have not been successful. Therefore, in the rest of this
sub-section, I will discuss the difficulties from the three perspectives.
\medskip

{\sl Particle theorist's strategies}

In the world of particle physics, perturbative quantum field theory has
proved to be enormously useful. The standard model has met with success
every experimental test
to date. It is therefore tempting to apply these methods also to gravity.
This involves splitting the space-time metric $g_{ab}$ in to two parts, a
kinematic (say, flat) background $\eta_{ab}$ and the dynamical field
$h_{ab}$ measuring the deviation of the physical metric from the background.
Thus, one sets $g_{ab} = \eta_{ab} + Gh_{ab}$, where $G$ is Newton's
constant, regards the deviation $h_{ab}$ as representing the ``physical
gravitational field'' and quantizes it on the background space-time
$\eta_{ab}$. The quanta represent zero rest mass, spin-2 particles
--the gravitons. These are subject to interactions whose nature is dictated
by the Einstein-Hilbert action. One can now bring in the arsenal of the
perturbative quantum field theory to calculate the amplitudes for
graviton-graviton scattering. In the process, however, one has split the
dual role played by $g_{ab}$ in general relativity, assigning to $\eta_{ab}$
the geometric, kinematic role and to $h_{ab}$ the dynamical role of the
gravitational potential. The general relativist may object violently to this
split on the ground that it destroys the whole spirit of Einstein's theory.
But the particle theorist can shrug his shoulders and say that such aesthetic
considerations are subjective and that, in any case, this is a small price to
pay to usher in the powerful and highly successful machinery of perturbative
quantum field theory.

This scenario was introduced by Feynman in the early sixties and carried
to completion by a number of workers, notably DeWitt, in the subsequent
decade. Unfortunately, as is now well known, the strategy fails by its
own criteria. The theory is not renormalizable; the perturbative quantum theory
has infinitely many undetermined parameters, rendering it pretty much useless
as far as physical predictions are concerned. In terms of the modern view,
suggested by the theory of critical phenomena, there is {\it new physics}
at short distances which is not captured by perturbative techniques and to
which the theory is highly sensitive. In the Planck regime, there are
new, non-local effects which make the standard perturbative techniques
inappropriate in this regime. It is possible that the theory does not even
exist, i.e., that it is mathematically inconsistent. However, this is by no
means clear. There are examples of interesting theories --Gross-Neveau model
in 3 space-time dimensions for instance (Wightman, 1992)-- which are not
renormalizable but are exactly soluble. Renormalizability is a criterion of
simplicity; such theories are ``short-distance insensitive'' so that, even in
absence of the detailed picture of the microscopic physics, one can make
predictions using just a finite number of {\it effective} parameters which are
relevant to the scale of observation. It is {\it not} a criterion to decide
if a theory is consistent quantum mechanically.

Simplicity is, nonetheless, very desirable. Therefore, it seemed natural
to try to replace general relativity by another theory of gravity
which {\it is} renormalizable. Impetus to this idea came from the
Weinberg-Salam model. The Fermi theory of weak interactions is not
renormalizable although, at low energies, it agrees with the experimental data
very well. The Weinberg-Salam model agrees with these low energy predictions
and is in addition renormalizable. It marshals new processes at high
energies, ignored by the Fermi model, and these cure the ultra-violet problems.
The moral of the story was clear. Just because a theory is in excellent
agreement with experiments at large distances does not mean that it correctly
captures the physics at small scale. The parallel between the Fermi theory and
general relativity seemed obvious and pursuit began to find new theories of
gravity which would agree with general relativity in the large but usher in
new processes in the small making it renormalizable. And such theories were
in fact found. Their Lagrangians contain higher derivative terms (quadratic,
rather than just linear in curvature) which become important only at high
energies but which suffice to make the theories renormalizable. It turned out,
however, that these theories are not unitary. Their Hamiltonians unbounded
below, causing them to be dramatically unstable.

Another modification of general relativity is suggested by supersymmetry.
Here, there are no higher derivative terms in the gravitational action. Rather,
the {\it number} of fields is increased using supersymmetry. In particular,
new spin 3/2 gravitino fields arise as supersymmetric partners of the spin 2
graviton. The resulting theory --supergravity-- has a Hamiltonian which is
manifestly positive at least formally and instabilities encountered in the
higher derivative theories are avoided. Moreover, it appeared at first
that the infinities of bosonic fields would cancel those of
fermionic fields, making not just gravity but gravity coupled with matter
renormalizable. The hopes ran high. In his Lucasian Chair inaugural
address entitled {\it Is the end in sight for theoretical physics?} Hawking
argued that the $N=8$ supergravity theory would be the ``theory of
everything''. The confidence in this scenario was so strong that in the
Einstein Centennial conference in Princeton, both plenary lectures on quantum
gravity were devoted to supergravity. Unfortunately, the theory turned out
to be non-renormalizable at the two loop level. Again, as I emphasized above,
this does not mean that the theory is necessarily inconsistent; it may
very well exist non-perturbatively. However, the whole reason for
abandoning general relativity in the first place is now considerably weakened.

As is well known, in the eighties, another, qualitatively different idea
dominated ---String theory as the theory of everything. It is obvious even
to a casual observer that this theory is vastly richer than quantum field
theories encountered so far in particle physics and, at least potentially,
extremely powerful. It is a very ``tight'' theory with relatively little fed
in at the outset. And it has had {\it very} significant mathematical successes.
But it is not at all obvious that string theory will indeed provide us with a
consistent, unambiguous quantum theory of gravity in four dimensions. Most
effort that has gone into it is perturbative and there is a growing
realization that non-perturbative effects hold a key to further progress. As
is clear from the scientific program of this school, for example, there has
been a major shift in the direction of research over the last two or three
years. The effort is now concentrated on lower dimensional models which
{\it can} be tackled non-perturbatively. I believe that this work is very
valuable but still in a beginning stage.

One's first reaction to the effort which has gone into the problem over
the last three decades may be that of disappointment. I believe that this
is unwarranted. It is important that one tries out the ``obvious''
avenues before advocating radical revisions. Furthermore, the perturbative
methods that were developed in the context of quantum general relativity
{\it have} turned out to be extremely valuable to the treatment of gauge
theories. Within quantum gravity itself, we {\it have} learnt important
lessons. Perhaps the
most significant of these is that we should not assume right at the outset
what the micro-structure of space-time ought to be. At shorter and shorter
length scales, we find ultra-violet problems, virtual processes involving
higher and higher energies. Why should a smooth continuum be then a good
approximation to represent the space-time geometry? We should not pre-suppose
what the micro-structure should be. Rather, we should let the theory itself
tell us what this structure is. This in turn means that when one comes to
gravity, the basic assumptions of perturbation theory are flawed. We need
non-perturbative approaches. Although we arrived at this lesson starting
from mathematical considerations in quantum field theory, we are led to
the same conclusion by the conceptual questions raised in the last sub-section.
There is much more to quantum gravity than graviton-graviton scattering.
\medskip
\goodbreak

{\sl Mathematical physicist's methods}

The failure of perturbation theory would, presumably, not upset the
mathematical physicist a great deal. After all, she knows that in spite of
perturbative non-renorma\-liza\-bility, a quantum field theory can exist
non-perturbatively. Indeed, as I will indicate below, there is some evidence
from numerical simulations that quantum general relativity itself may have
this feature. One might therefore wonder: why have the standard methods
developed by mathematical physicists not been applied to the problem of
quantum gravity? What are the obstacles? Let me therefore consider some
natural strategies that one may be tempted to try and indicate the type of
difficulties one encounters.

First, one might imagine {\it defining} the goal properly by writing down
a set of axioms. In Minkowskian field theories the Wightman and the
Haag-Kastler axioms serve this purpose. Can we write down an analogous
system for quantum gravity, thereby spelling out the goals in a clean
fashion? Problems arise right away because both the system of axioms are
rooted in the geometry of Minkowski space and in the associated Poincar\'e
group. Let me consider the Wightman system (Streater \& Wightman 1964) for
concreteness. The zeroth axiom asks that the Hilbert space of states carry an
unitary representation of the Poincar\'e group and that the 4-momentum
operator have a spectrum in the future cone; the second axiom states how the
field operators should transform; the third axiom introduces micro-causality,
i.e., the condition that field operators should commute at space-like
separations; and, the fourth and the last axiom requires asymptotic
completeness, i.e., that the Hilbert spaces ${\cal H}^\pm$ of asymptotic
states be isomorphic with the total Hilbert space. Thus, four of the five
axioms derive their meaning directly from Minkowskian structure. It seems
extremely difficult to extend the zeroth and the fourth axioms already to
quantum field theory in topologically non-trivial space-times, leave alone to
the context in which there is no background metric what so ever. And if we
just drop these axioms, we are left with a framework that is too loose to
be useful.

The situation is similar with the Osterwalder-Schrader system. One might
imagine foregoing the use of specific axioms and just using techniques from
Euclidean quantum field theory to construct a suitable mathematical framework.
This is the view recently adapted by some groups using computer simulations.
These methods
have had a great deal of success in certain exactly soluble 2-dimensional
models. The techniques involve dynamical triangulations and have been
extended to the Einstein theory in 4 dimensions (see, e.g., Agishtein \&
Migdal (1991)). Furthermore, there is some numerical evidence that there is
a critical point in the 2-dimensional parameter space spanned by Newton's
constant and the cosmological constant, suggesting that the continuum limit of
the theory may well exist. This is an exciting development and interesting
results have now been obtained by several groups. Let us be optimistic and
suppose that a well-defined Euclidean quantum theory of gravity can actually
be constructed. This would be a major achievement. Unfortunately, it wouldn't
quite solve the problem at hand. The main obstacle is that, as of now, there
is no obvious way to pass from the Euclidean to the Lorentzian regime! The
standard strategy of performing a Wick-rotation simply does not work. First,
we don't know which time coordinate to Wick-rotate. Second, even if we just
choose one and perform the rotation, generically, the resulting metric will
not be Lorentzian but {\it complex}. The overall situation is the following.
Given an analytic Lorentzian metric, one can complexify the manifold and
extend the
metric analytically. However, the resulting complex manifold need not admit
{\it any} Euclidean section. (Conversely, we may analytically continue an
Euclidean metric and the resulting complex space-time need not have any
Lorentzian section.) This is not just an esoteric, technical problem. Even
the Lorentzian Kerr metric, which is {\it stationary}, does not admit an
Euclidean section. Thus, even if one did manage to solve the highly
non-trivial problem of actually constructing an Euclidean theory using the
hints provided by the computer work, without a brand new idea we would still
not be able to answer physical questions that refer to the Lorentzian world.

Why not then try canonical quantization? This method lacks manifest
covariance. Nonetheless, as we will see, one can construct a Hamiltonian
framework without having to introduce any background fields, thereby
respecting the diffeomorphism invariance of the theory. However, from the
perspective of a mathematical physicist, the structure of the resulting
framework has an unusual feature which one has never seen in
any of the familiar field theories: it is a dynamically constrained system.
That is, most of the non-trivial content of the theory is in its constraints.
On physical states, the Hamiltonian vanishes identically if the spatial
topology is compact and equals a surface term --analogous to the charge
integral in QED-- if space-time is asymptotically flat
\footnote{1}{This arises as follows. Constraints normally generate gauge.
However, now the effective gauge group is the group diffeomorphisms whence
``motions within space-time'' are also generated by constraints. Thus, as
might be expected, the origin of this unusual feature lies in diffeomorphism
invariance of the theory.}. Suddenly, then, one finds oneself
in an unfamiliar territory and all the experience one has gained from
canonical quantization of field theories in 2 or 3 dimensions begins to
look not so relevant. Normally, the key problem is that of finding a suitable
representation of the CCRs --or, an appropriate measure on the space of
states-- which lets the Hamiltonian be self-adjoint. Now, the Hamiltonian
seems trivial but all the difficulty seems concentrated in the quantum
constraints. Furthermore, while the representation of the CCRs is to be
chosen {\it prior} to the imposition of constraints, the inner product on
the space of states has physical meaning only {\it after} the constraints
are solved. One thus needs to develop new strategies and modify the familiar
quantization program appropriately.

One may imagine using techniques from geometrical quantization. However,
since the key difficulty lies in the constraints of the theory, a ``correct''
polarization would be the one which is preserved, in an appropriate sense,
by the Hamiltonian flows generated by the constraint functions on the phase
space (Ashtekar \& Stillermann, 1986). The problem of finding such a
polarization is closely related to that of obtaining a general solution to
Einstein's equation and therefore seems hopelessly difficult in 4 dimensions.
(In 3 dimensions, the strategy does work but only because every solution to
the field equations is flat.) One may imagine using instead the Dirac (1964)
approach to quantization of constrained systems. We will see however that the
representation best suited for solving the quantum constraints in this
framework does not arise from any polarization what so ever on the phase
space. Thus, unfortunately, geometric quantization techniques do not seem
to be well suited to this problem. Finally, one might imagine group theoretic
method of quantization. An appropriate canonical group was in fact found and
it does ``interact'' well with the constraints of the theory which generate
spatial diffeomorphisms and triad rotations (Isham \& Kakas, 1984a,b). However,
the problem of incorporating the so called Hamiltonian constraint which
generates time translations again seems hopelessly difficult.
\medbreak

{\sl General relativist's approaches}

To general relativists, a central lesson of Einstein's theory is that
the gravitational field is inseparably intertwined with the geometry of
space-time. Indeed, the most spectacular predictions of the theory, from big
bang to black holes, stem from this synthesis. It therefore seems natural
not to tamper with it in the process of quantization. One is thus extremely
reluctant to split the metric into a kinematical and a dynamical part, and
more generally, to introduce any background fields in the theory at a
fundamental level. Consequently, most work by relativists has occurred in
the canonical or the path integral framework.

The canonical framework was introduced by Dirac and Bergmann in the late 50's
and then developed more fully by several others, most notably Arnowitt,
Deser and Misner in the 60's. (See, e.g., their review (1962)). The 3-metric
(or, the first fundamental form) on
a space-like surface plays the role of the configuration variable while the
extrinsic curvature (or, the second fundamental form), of the momentum. As
indicated above, the theory is subject to constraints which generate
diffeomorphisms. The framework is geometric and fully non-perturbative right
from the start. It provides a complete description of how 3-metrics evolve in
Einstein's theory and was therefore baptized by Wheeler as {\it
geometrodynamics}. The idea is to go to quantum theory by considering wave
functions $\Psi (q_{ab})$ of 3-metrics $q_{ab}$ and to select as physical
states those wave functions which are annihilated by the quantum constraint
operators. The first equation is similar in form to the Gauss constraint of
Yang-Mills theory and requires that the states $\Psi(q_{ab})$ be invariant
under spatial diffeomorphisms. The second constraint --called the
Wheeler-DeWitt equation-- is more complicated. Its classical analog generates
dynamics. Therefore, it is believed to contain the time-evolution equation of
the theory. (For details, see, e.g. Wheeler (1964) and Kucha\v r (1981).)

As emphasized earlier, however, there is no background space-time
and therefore, in particular, no a priori notion of time. How can one speak of
dynamics and time-evolution then? The idea is that a suitable component of the
argument $q_{ab}$ of the wave function is to play the role of time and the
Wheeler-DeWitt equation is to tell us how the wave function evolves with
respect to that time. The counting goes as follows. $q_{ab}$ has six
components. Roughly speaking the condition of (3-dimensional) diffeomorphism
invariance tells us that a physical state can depend on only 3 of the 6
components of $q_{ab}$. Two of these are the true, dynamical degrees while the
third represents ``time''. The Wheeler-DeWitt equation is thus to tell us
how the dependence of the wave function on the true degrees of freedom changes
as the variable representing time increases, and these changes are to be
interpreted as time evolution. Thus, time, in spite of its name, is to be
an ``internal'' variable, not an external clock. Until we isolate time, the
Wheeler-DeWitt equation is just a constraint on the allowable wave functions.
Once time is isolated, the same equation can be interpreted as providing
evolution; dynamics is then born.

These ideas seem extremely appealing. However, in this geometrodynamical
framework, to date, no one has found a single solution to the Wheeler-DeWitt
equation in the full theory. Furthermore, even at a formal level, the
component of the metric which is to represent time has not been isolated and
the idea that the Wheeler-DeWitt equation can be re-interpreted as the
Schr\"odinger equation still remains a hope.

However, in the seventies several models, obtained by freezing all but a
finite number of degrees of freedom of general relativity via imposition of
symmetries, were analysed in detail and the Wheeler quantization program
was successfully completed. The success provided --and continues to provide--
confidence in the general setup. It should be emphasized, however, that these
models --called minisuperspaces-- are exceedingly simple compared to the full
theory. In particular, they are free of divergences simply by virtue of the
fact that one has thrown out, by hand, most of the degrees of freedom,
thereby reducing the quantum field theory to quantum mechanics. (There {\it is}
an interesting model with an infinite number of degrees of freedom which has
been fully analysed: the ``midisuperspace'' of cylindrical waves in the
spatially open case, or, of Gowdy space-times in the spatially compact case.
However, they could be treated exactly because, after appropriate gauge
fixing, the field equations can be made linear and the genuine complications
of quantum field theory can be again avoided.) A rather common viewpoint,
however,
was that this was just a technical convenience. Once one could satisfactorily
treat a sufficiently complicated minisuperspace (e.g., the Bianchi type IX
model) one would be ``almost home''. That is, the key difficulties were
thought to lie in the conceptual challenges such as those posed by absence of
a grandfather time, rather than in quantum field theoretic problems. (For a
critical discussion of this attitude, see, e.g., Isham \& Kakas (1984a,b).)
Thus, the relative weights assigned to various difficulties were almost
completely reversed from those assigned by mathematical and particle
physicists. This is a key reason why
a genuine communication gap persisted between the two communities, a gap which
has begun to close only in recent years. In retrospect, it seems clear that
these extreme positions are flawed and both sets of problems must be faced
squarely. We must address satisfactorily the conceptual problems mentioned
above {\it and}, at the same time, learn to come to grips with the field
theoretic divergences.

Let us now turn to path integral quantization%
\footnote{2}{As was emphasized by Gregg Zuckermann in his lectures, the
canonical and the path integral approaches are not as much in competition
as they are complementary to one another. Path integrals let us compute
transition amplitudes but one has to supplement them with the appropriate
sets of (in and out) states which, in a non-perturbative treatment,
typically come from canonical quantization. Conversely, path integrals let
us compute probability amplitudes for such processes as topology change
which are hard to fit in the canonical framework.}.
Until the recent computer work mentioned above, developments in the path
integral approach had also remained formal,
and perhaps to an even greater degree. In Minkowskian field theories,
path integrals are defined by wick-rotating to the Euclidean space and,
in most ``realistic'' theories, by further making use of perturbation theory.
Now, as we saw above, the notion of Wick rotation faces a difficulty. One
might tentatively adopt a view --advocated by Hawking in the early eighties--
that it is the Euclidean domain that is fundamental-- and see if the path
integrals can be given meaning without worrying about what the results might
imply in the Lorentzian regime. But now the theory is not renormalizable
and one cannot take recourse to  perturbation theory to give meaning to
the path integral. Consequently, as far as the full theory is concerned, most
of the work in this area has remained formal. Again, as in geometrodynamics,
considerable effort has been devoted to minisuperspaces where, due to the
presence of a finite number of degrees of freedom the most pressing of the
worries disappear. However, even with this truncation, the functional
integrals could not be calculated exactly and one often had to take recourse
to (WKB type) approximation methods. Nonetheless, with all these caveats,
calculations can be completed and one can hope to draw some qualitative
lessons by interpreting the results appropriately.
\medskip

This completes the discussion of the difficulties in constructing a quantum
theory of gravity. I do not wish to imply that these are unsurmountable.
Indeed, we will see in these notes and in other contributions to this volume
that some of them have been overcome already. Rather, my goal has been to
point out why ``obvious'' strategies don't quite work, where one gets stuck
and the sort of obstacles one faces.

\goodbreak
\subsection{1.3}{Overview}

Over the past six years, the canonical approach has been revived with
three key inputs: i) an extension of the Dirac program for quantization
of constrained systems to address the problems faced by diffeomorphism
invariant theories such as general relativity; ii) a reformulation of
Hamiltonian general relativity as a dynamical theory of connections
rather than of metrics; and iii) the introduction of loop space methods
to analyse the structure of quantum theories of connections
non-perturbatively. In this new version, the program is fully
non-perturbative ---there are no background fields at the basic level.
At the same time, careful attention is paid to quantum field theoretic
issues. The program has led to several interesting mathematical results.
Some of the more striking among these are: i) techniques for regulating
operators in a way that respects the diffeomorphism invariance of the theory;
ii) quantum states which approximate classical geometries on large scales
but exhibit a definite discrete structure at the Planck scale; iii) an
infinite dimensional space of solutions to all quantum constraints; and,
iv) a deep relation between knot theory and physical states of quantum
gravity. In full general relativity, although significant progress has been
made, the program is still far from being complete. However, it {\it has
been} carried to completion in a number of truncated models. In particular,
the problem of time has been successfully addressed in several cases.

Before outlining the plan of these lectures notes, let me just say a few
words about my own viewpoint towards this approach. Although I will deal
exclusively with general relativity, I do not necessarily believe that
quantum general relativity (coupled to matter) would be the correct theory
to describe physics in the Planck regime. Rather, the attitude is the
following. Since nothing that we know to date implies that quantum general
relativity can {\it not} exist non-perturbatively, it is well worth
finding out if it does. If the theory does exist non-perturbatively, then
its viability would be an experimental question. If it does not, we will learn
why not and the exercise will presumably suggest modifications. (This
attitude is in fact very similar to the one adopted in perturbative
treatments in the seventies and eighties.) In any case, the qualitative
insights one would gain --and, indeed, has already gained--
about the nature of diffeomorphism invariant quantum theories will be
valuable no matter what the correct theory turns out to be. Thus, in the
very least, this analysis can be regarded as a non-perturbative investigation
of a physically interesting model in {\it four} dimensions which is {\it much
more} realistic than analogous diffeomorphism invariant models in two or
less dimensions. Finally, I find the relative richness of the program very
attractive: It has led to new insights in a number of topics quite apart
from issues in quantum gravity. Examples are: the mathematical structure of
the space of self dual Einstein manifolds (see, e.g., Samuel 1992);
$C^\star$-algebras of holonomies and their representation theory (Ashtekar
\& Isham 1992); loop space description of the Fock space of
photons (Ashtekar \& Rovelli, 1992); differential geometric techniques to
analyse knot invariants (Di Bortolo et al, 1992); etc.

In section 2, I have collected several mathematical techniques that are
used through out the notes. These include a short introduction to the
symplectic description of constrained systems and to the Dirac program for
quantization of constrained systems, a summary of the algebraic quantization
program which extends the Dirac program in an appropriate fashion, and an
outline of the powerful ``non-linear duality'' between connections and loops.
Section 3 is devoted to 3-dimensional gravity. In this case the algebraic
quantization program can be completed and in particular, the general goals
of Wheeler's program referred to above can be met. This is interesting because,
although mathematically much simpler, the theory has the same overall
structure as 4-dimensional general relativity and hence, a priori, faces the
same conceptual problems. My aim is to use 3-dimensional gravity as a
toy model to introduce the reader both to the mathematical techniques
and to the general directions of thinking that underlie the program for
4-dimensional gravity. In section 4, I present the new Hamiltonian formulation
of 4-dimensional gravity. Now, the configuration variable is a connection
and all of Einstein's equations become polynomial in the basic canonical
variables. The emphasis on connections suggests that we describe the theory
in terms of holonomies and this is how the loops begin to play  significant
role. Thus, the suggestion is that while metrics, distances and null cones
govern the world of classical relativity, connections, holonomies of spinors
around closed loops, knots and links are the basic objects in the Planck
regime. In section 5, I discuss the quantum theory. In the initial
treatments, as is perhaps inevitable when one embarks on a new exploration,
the issue of regularization of operators was often not treated
with due care. As a result, there is now a widespread belief that the
results obtained within quantum theory are formal. {\it This is no longer
true.} By now, several careful calculations have been performed and a general
framework has emerged which does have the precision that one is normally
used in theoretical (as opposed to, say, mathematical) physics. I have
therefore taken some care to make the current level of precision explicit
and also pointed out what needs to be done to make it acceptable by
standards of mathematical physics. I conclude in section 6 with a brief
discussion.

By now there are about 300 papers on the subject and I can not do justice
to all the interesting results that have emerged. Therefore, I apologize in
advance to various authors who might find that their contributions are
not mentioned at all or discussed very briefly. Fortunately, there do exist
several detailed reviews. First, there is a monograph (Ashtekar 1991)
addressed to research students that covers in detail most results that were
obtained by December 1990; there is a more compact review addressed to
standard physics audiences (Rovelli, 1990); a summer school report geared to
particle physicists (Smolin, 1992); a review where one can find a discussion
of the problems of quantum cosmology (Kodama, 1992); and, a summary of the
developments pertaining to supergravity (Nicoli\& Matschull, 1992). While
there is an inevitable overlap between the present notes and these reviews,
by and large, I have tried to focus on recent results and have presented
the material from a slightly different perspective. My goal is to get the
general picture and the overall direction of thinking across rather than
details of any specific results. For references to more detailed treatments,
the reader should consult the reviews mentioned above {\it and} to the papers
they refer to. (See, in particular, the extensive bibliography in Ashtekar
(1991).) By and large, here I will provide references only to recent papers
which may not be included in the reviews.

My conventions will be as follows. The space-time manifold, M, is assumed
to be topologically $\Sigma\times\real$ of some 3-manifold $\Sigma$
(except in section 3 on 3-dimensional gravity where $\Sigma$ is a 2-manifold).
The space-time metric $g_{ab}$ is assumed to have signature -+++ (and
-++ in section 3). Throughout, I have employed Penrose's abstract index
notation (Penrose \& Rindler (1986), see also Wald (1984)). Since density
weights of various fields often play an important role (except when the
notation becomes too cumbersome) I have used an over-tilde to denote an object
with density weight +1 and an under-tilde, an object with density weight -1.

\vfill\break
%\input [ijtpd.lh]qgmac.tex
%\singlespace\overfullrule=0pt
\section{2.}{${}\quad$ Mathematical Preliminaries}
\medskip
{\narrower\narrower\smallskip\noindent
{\sl The beginner ... should not be discouraged if ... he finds
 that he does not have the prerequisite for reading the  prerequisites.}
\smallskip
P. Halmos\smallskip}
\medskip

The purpose of this section is to collect several mathematical techniques
that are used in the rest of these lecture notes. Sections 2.1 and 2.2
recall the symplectic description of constrained classical systems;
2.3 discusses the Dirac quantization program for such systems and points
out its limitations; 2.4 introduces an algebraic quantization program; and
2.5 provides the reader with some powerful tools for analyzing quantum
theories of connections particularly in the context of diffeomorphism
invariance. Familiarity with these ideas is assumed particularly in sections
3 and 5. However, it is {\it not} essential that the readers have a full
grasp of these methods ---particularly those contained in section 2.5--
before embarking on the main text.

Advanced researchers, on the other hand, should skip this section and
proceed directly to section 3.

\subsection{2.1}{Symplectic framework}
The purpose of this section is to recall some notions from symplectic
geometry. In broad terms, the symplectic framework geometrizes the
Hamiltonian description of classical systems, thereby making it
coordinate-independent and suggesting interesting generalizations. What we
will present here is a self-contained but brief and ``practically oriented''
introduction to the subject. (More extensive accounts can be found, e.g., in
monographs by R. Abraham and J.E. Marsden, V.I. Arnold, and, V.W. Guillemin.)

The arena for classical mechanics is a {\it symplectic manifold}, $(\Gamma,
\Omega_\ab)$, where $\Gamma$ is an even-dimensional manifold, and $\Omega_\ab$
a {\it symplectic form}, i.e., a 2-form which is closed and non-degenerate.
Given any torsion-free derivative operator $\nab $, one can express the closure
requirement as:  $\nab_{[\alpha}\Omega_{\beta\gamma]}=0$. The non-degeneracy
condition reads: $\Omega_\ab v^\alpha=0\>\Leftrightarrow\> v^\alpha=0$. If
$\Gamma$ is finite-dimensional, non-degeneracy guarantees that $\Omega_\ab$ has
a unique inverse, $\Omega^\ab$, with $\Omega^\ab\Omega_{\beta\gamma}=
\delta_\gamma{}^\alpha$, or that the mapping $\Omega: T\Gamma\to T^\conj
\Gamma$ from tangent vectors to cotangent vectors, with $\Omega_\ab v^\beta
=v_\alpha$, is an isomorphism.
   \footnote{3}{If $\Gamma$ is infinite-dimensional, one has to be careful
   with functional analysis. The form $\Omega$ is said to be {\it weakly}
   non-degenerate if its kernel consists only of the zero vector and
   {\it strongly} non-degenerate if the mapping it defines from the tangent
   space to the cotangent space is an isomorphism. In what follows, in the
   infinite-dimensional cases, we will assume only that $\Omega$ is weakly
   non-degenerate. Although weak non-degeneracy does {\it not} ensure that
   $\Omega$ admits an inverse, the main ideas to be discussed here
   go through in the weaker case. Roughly, equations which do not
   involve the inverse of the symplectic form continue to hold in the weakly
   non-degenerate case. Therefore, in the equations which hold in the
   finite-dimensional case, one first multiplies both sides by $\Omega$ with
   an index structure so chosen as to eliminate its inverse and {\it then}
   takes over the resulting equation to the infinite-dimensional case.
   However, here, we will not worry about functional analytic rigor.}

Each point of $\Gamma$ represents a possible state of the given classical
system. Dynamics can be therefore specified by introducing a vector field on
$\Gamma$: integral curves of the vector field represent dynamical trajectories
and the affine parameter keeps track of the passage of time. The availability
of
the symplectic form simplifies the task of specifying the dynamical vector
field. For, as we will see, the symplectic form enables one to construct
these vector fields from {\it functions}---the Hamiltonians---on the phase
space. Thus, to specify dynamics on a symplectic manifold, it suffices to
specify a function thereon.

Given a vector field $v^\alpha$ on $\Gamma$, we say that $v^\alpha$ is an {\it
infinitesimal canonical transformation} iff it leaves the symplectic form
invariant, i.e., iff
\bneq
\Lie{v}\Omega_\ab=0.
\eneq{(2.1.1)}
The diffeomorphisms generated by these $v^\alpha$ are called {\it canonical
transformations}. Since they preserve the geometrical structure of $(\Gamma,
\Omega_{\alpha\beta})$, these canonical  transformations are the symmetries
of classical mechanics. Now, it is easy to verify that $v^\alpha$ satisfies
(2.1.1) iff there exists, locally, (and, if the first homology group of
$\Gamma$ is trivial, globally) a function $f$ such that:
\bneq
v^\alpha = X_f^\alpha:=\Omega^\ab\nab_\beta f.
\eneq{(2.1.2)}
The vector field $X_f^\alpha$ so constructed from $f$ is called the {\it
Hamiltonian vector field} of $f$. Thus, all Hamiltonian vector fields generate
infinitesimal \ct s, and all one-parameter families of \ct s are locally
generated by a function, called the {\it Hamiltonian} of the corresponding
transformation%
\footnote{4}{Note that, in symplectic geometry, the term Hamiltonian
has a more general meaning than in physics. {\it Any} function on the
\ps, when used to generate a canonical transformation, is referred to as
a Hamiltonian; the canonical transformation need not correspond to time
evolution. In what follows, the intended sense in which the term
Hamiltonian is used will be clear from the context.}.
In particular, therefore, we have established that, in striking contrast
to, say, metric manifolds, every symplectic manifold admits infinitely many
independent symmetries.

Given two functions $f,g:\Gamma\to\real$, their {\it Poisson bracket} is
defined by
\beginalign{
\{f,g\}:&=\Omega^\ab\nab_\alpha f\nab_\beta g\cr
 &\equiv-\Lie{X_f}g\equiv\Lie{X_g}f.\cr}
\endalign{(2.1.3)}
It is easy to verify that the Poisson bracket operation turns the vector
space of functions on $\Gamma$ into a Lie algebra. Using this Lie-bracket,
we can now state an important property of the map $f\mapsto X_f^\alpha$
that associates to $f$ its Hamiltonian vector field: It takes Poisson
brackets of functions into commutators of vector fields:
\bneq
X_{\{f,g\}}^\alpha=-[X_f,X_g]^\alpha.
\eneq{(2.1.4)}
(Note also that the map is linear and its kernel consists precisely of the
constant functions on $\Gamma$.)

Let us now return to the issue of dynamics. For a large class of physically
interesting systems, the dynamical vector fields are globally Hamiltonian.
That is, time-evolution of physically interesting systems can be generally
specified simply by fixing a function $H$ on $\Gamma$; its Hamiltonian vector
field $X_H^\alpha$ then provides the dynamical vector field {\it everywhere}
on $\Gamma$. Thus, given a point in the phase space representing the initial
state of the system, the dynamical trajectory is simply the integral curve of
the Hamiltonian vector field $X_H^\alpha$ through that point. Using this fact,
it is straightforward to check that the time evolution (in the Heisenberg
picture) of any observable $f$ is given by:
\bneq
\dot f:=\Lie{X_H}f\equiv\{f,H\}.
\eneq{(2.1.5)}

Finally, we note that, since the symplectic form is closed, it can be obtained
locally (and, if the second homology group of $\Gamma$ is trivial, globally)
from a 1-form $\omega_\alpha$, called the {\it symplectic potential}:
$\Omega_\ab=2\nab_{[\alpha}\omega_{\beta]}$. For a given $\Omega_{\ab}$
the symplectic potential is thus determined up to the addition of a gradient.
This potential plays an important role in geometric quantization.
\goodbreak
\subsection{2.2}{First class constraints}
A dynamical system is said to be constrained if its physical states are
restricted to lie in a submanifold $\hat\Gamma$ of the phase space $\Gamma$,
called the {\it constraint surface}. One can specify $\hat\Gamma$ by the
vanishing of a set of functions $C_{\bf i}:\Gamma\to\real$ called the
{\it constraints}
\footnote{5}{The boldface, lower case latin letters $({\bf i}, {\bf
j}...)$, used here are {\it numerical} indices, running over the number of
constraints.}:
  \bneq
  \hat\Gamma:= \{p\in\Gamma\,\vert\,C_{\bf i}(p)=0,\,{\rm for}\
  {\bf i}=1,\ldots,m\}.
  \eneq{(2.2.1)}
Note that $\hat\Gamma$ does not provide a unique choice of constraint
functions $C_{\bf i}$; there is considerable ``coordinate freedom'' in
the selection of constraint functions. In this section we will restrict
ourselves to the so-called first class constraints since these are the
ones that play an important role in gauge theories and general relativity.

A constrained system is said to be of {\it first class} if for all covectors
$n_\alpha$ normal to $\hat\Gamma$, $\Omega^\ab n_\alpha$ is tangent to
$\hat\Gamma$. This characterization is coordinate independent. Given a set
of constraint functions, one can reformulate this definition as follows:
the system is of first class if the constraint functions ``weakly ''commute,
i.e.,
\bneq
 \forall {\bf i},{\bf j}\qquad\{C_{\bf i},C_{\bf j}\}\approx0,
\eneq{(2.2.2)}
where $\approx$ means ``equals when restricted to the constraint surface''.
This implies that there exist functions $f_{{\bf ij}}{}^{\bf k}$ on
$\Gamma$ such that:
\bneq
\forall {\bf i},{\bf j}\qquad\{C_{\bf i},C_{\bf j}\}=
 -f_{{\bf ij}}{}^{\bf k}C_{\bf k}.
\eneq{(2.2.3)}
These functions are called {\it structure functions}. If they happen to be
constants, the constraint functions $C_{\bf i}$ are generators of a sub-Lie
algebra of the set of functions on $\Gamma$. The second definition of first
class constraints is often more useful in practice, while the first one is
more ``covariant'': it shows explicitly that the notion is independent of
the choice of constraint functions.

Let us now consider a first class constrained system $(\Gamma,\hat\Gamma,
\Omega_\ab,H)$, and examine the consequences of the existence of the
constraints for its dynamics. In particular, since the point representing the
state of the system is required to remain on $\hat\Gamma$, we are interested
in knowing to what extent we can consider just $\hat\Gamma$ as our phase
space, instead of the whole of $\Gamma$.

Suppose $\Gamma$ is $2n$-dimensional. By definition of first class
constraints, the $m$ vector fields $X_{\bf i}^\alpha:=\Omega^{\ab}
\nab_\beta C_{\bf i}$, are tangential to $\hat\Gamma$ and, because of the
non-degeneracy of $\Omega_\ab$, they are linearly independent. Thus, at each
$p\in\hat\Gamma$, they span an $m$-dimensional subspace ${\cal G}\subset
T_p\hat\Gamma$. The vector fields in ${\cal G}$ are called {\it constraint
vector fields}. ${\cal G}$ is called a {\it gauge flat}, since, as Dirac
observed, motion along any of the directions contained in ${\cal G}$
corresponds to a gauge transformation of the system. This interpretation is
suggested by the following considerations. Since the physical states of the
system are restricted to $\hat\Gamma$, measurements can reveal the values of
observables, such as the {\it physical Hamiltonian}, $H$, only on $\hat\Gamma$.
There is thus an ambiguity in extending the function ``off'' $\hat\Gamma$: if
$H$ is an extension, so is $H'= H+ f^{\bf j}C_{\bf j}$ where $f^{\bf j}$ are
{\it any} smooth functions on $\Gamma$. This in turn introduces an ambiguity
in dynamics even for the physical states, i.e., points of $\hat\Gamma$: if $H$
defines the Hamiltonian vector field $X_H^\alpha$, $H'$ leads to the vector
field $X_H^\alpha + f^{\bf j}X^\alpha_{\bf j}$ on $\hat\Gamma$. Hence, we are
led to conclude that motions along constraint vector fields,
$X_{\bf j}^\alpha$, represent gauge.

Consider the restriction of $\Omega_\ab$ to $\hat\Gamma$ (i.e., its pullback
$\hat\Omega_\ab=i^\conj\Omega_\ab$ by the inclusion map $i:\hat\Gamma\to
\Gamma$). Then, for any $n_\alpha$ normal to $\hat\Gamma$ and any $\hat
{v}^\alpha$ tangent to $\hat\Gamma$,
 \bneq
   0=\hat v^\alpha n_\alpha
   =\hat v^\alpha\Omega_{\alpha\gamma}\Omega^{\gamma\beta}n_\beta
   =\hat v^\alpha\Omega_{\alpha\gamma}\hat n^\gamma
\eneq{(2.2.4)}
where $\hat n^\alpha:=\Omega^\ab n_\beta$. However, the $n^\alpha$'s are
in 1-1 correspondence with constraint vector fields, so all constraint
vector fields are degenerate directions for $\hat\Omega_\ab$. Conversely,
all degenerate directions of this tensor are of this form. Thus,
$\hat\Omega_\ab$ is $m$-fold degenerate, and it just defines a {\it
presymplectic structure} on $\hat\Gamma$. The practical significance of this
fact is that $\hat\Omega_\ab$ does not have a unique inverse. One may imagine
defining $\hat\Omega^{\alpha\beta}$ to be the inverse if it satisfies
$\hat\Omega_{\alpha\gamma}\hat\Omega^{\gamma\delta}\hat\Omega_{\delta\beta}
=\hat\Omega_{\ab}$. However, for any constraint vector field $X^\alpha$ and
arbitrary $\ovr{T}{}^\alpha$, $\hat\Omega^\ab+X^{[\alpha}\ovr{T}{}^{\beta]}$
can then equally be considered as an inverse. In particular, using just
$\hat\Omega_\ab$, we cannot associate a unique Hamiltonian vector field to
a function on $\hat\Gamma$. We have to replace (2.1.2) by
\bneq
\hat\Omega_\ab X_H^\beta=\hat\nab_\alpha H,
\eneq{(2.2.5)}
which determines $X_H^\alpha$ up to the addition of a constraint vector field.
In particular, if we want to work just on $\hat\Gamma$, the time evolution of a
system is {\it not} determined uniquely by the Hamiltonian. The ambiguity
corresponds precisely to motions along the constraint vector fields. This
provides an alternate version of the motivation for interpreting ${\cal G}$
as gauge flats.

There are however two ways to recover a well-defined evolution. The first
is to eliminate in $\hat\Gamma$ the variables representing the gauge
degrees of freedom, and introduce the so-called {\it reduced \ps},
$\bar\Gamma:=\hat\Gamma/{\cal G}$, the space of orbits of the gauge
diffeomorphisms.
This is possible first because the gauge flats are integrable, which
follows from Frobenius' lemma and
\bneq
   [X_{{\bf i}},X_{{\bf j}}]=-X_{\{C_{{\bf i}},C_{{\bf j}}\}}
   =X_{f_{{\bf ij}}{}^{\bf k}C_{\bf k}}
   \approx f_{{\bf ij}}{}^{\bf k}X_{\bf k},
\eneq{(2.2.6)}
and second because $\Lie{X_{\bf i}}\Omega_\ab = 0\implies \Lie{X_{\bf
i}}\hat\Omega_\ab = 0$. We now have a projection mapping
$\pi:\hat\Gamma\to\bar\Gamma$, and the (non-degenerate) symplectic form
$\bar\Omega_\ab$ on $\bar\Gamma$ is naturally defined by $\bar\Omega_\ab\bar
u^\alpha\bar v^\beta:= \hat\Omega_\ab\hat u^\alpha\hat v^\beta$, where $\hat
u^\alpha$ and $\hat v^\alpha$ are any two vectors projected to $\bar u^\alpha$
and $\bar v^\alpha$, respectively, by the mapping $\pi$. The second way to
obtain a non-degenerate symplectic form from $\hat\Omega_\ab$ is to fix a
gauge, i.e., a global cross-section of $\hat\Gamma$ each point of which
intersects the integral manifold of constraint vector fields once and only
once, and restrict oneself to states which lie on this cross-section. The
pull-back of $\hat\Omega_\ab$ to the gauge-fixed surface is non-degenerate.
The second method is less elegant and the required global cross-section
need not always exist. However, if it does, the method is often simpler to
use.

We conclude with an example. Consider Maxwell fields
in Minkowski space-time. Let us consider a 3+1 formulation of the theory.
The configuration space ${\cal C}$ can then be taken to be the space of
1-forms $A_a$ --the vector potentials-- on a space-like 3-plane $\Sigma$.
The canonically conjugate momenta are then represented by the electric
fields $E^a$ and the symplectic structure $\Omega$ is then given by:
\bneq
\Omega = 2\lint_\Sigma d^3x\-\,  d\! I E^a(x)\wedge d\!I A_a(x)
\eneq{(2.2.7)}
where $d\!I$ denotes the exterior derivative operator and $\wedge$ the
exterior product, on the infinite dimensional phase space. The system is
subject to the Gauss constraint:
\bneq
D_aE^a(x) = 0.
\eneq{(2.2.8)}
Since the left side of (2.2.8) is a field on $\Sigma$, it is not a
real-valued function on $\Gamma$. To use the Hamiltonian framework
directly, it is convenient to smear the left side by smooth functions
$\Lambda(x)$, say of compact support on $\Sigma$
\bneq
 C_\Lambda(A ,E):=\lint_\Sigma\!d^3\!x\,\Lambda (x) D_aE^a(x)
\eneq{(2.2.9)}
and replace (2.2.8) by: $C_\Lambda = 0$ for all $\Lambda(x)$.
To verify that these constraints form a first class system, one must
compute the Poisson brackets between the constraint functions. These are
straightforward to evaluate: Since the functions depend only on $E^a(x)$
and are independent of $A_a$, the Poisson brackets vanish,
\bneq
 \{C_\Lambda(A ,E) , C_\Phi(A , E)\} = 0,
\eneq{(2.2.10)}
whence we do have a first class set. Next, one can show that the
Hamiltonian vector field of the constraint $C_\Lambda$ generates precisely
the gauge transformation of electrodynamics. Since
\bneq
 X_{C_\Lambda} =\lint_\Sigma\!d^3\!x\>\> \left[-(D_a\Lambda (x)) \cdot
  ({\delta\over\delta A_a (x)}) \right],
\eneq{(2.2.11)}
the infinitesimal canonical transformation induced by this vector field is
\bneq
 A_a\mapsto A_a-\epsilon D_a\Lambda, \quad {\rm and} \quad E^a\mapsto E^a.
\eneq{(2.2.12)}
These are precisely the gauge transformations induced by the gauge function
$\Lambda$.
\goodbreak

\subsection{2.3}{Dirac quantization program and its limitations}

In the sixties, Dirac developed a program for quantization of first
class systems which has since been used widely
\footnote{6}{Second class constraints, if any, arise in pairs and the
pull-back of the symplectic structure to the sub-manifold where these
constraints hold is non-degenerate. Therefore, this sub-manifold can be
taken to be the phase space of the system right in the beginning of the
analysis. That is, the second class constraints are eliminated classically
before embarking on the quantization program.}. (See, e.g., Dirac (1964).)
In particular, the attempts at canonical quantization of general relativity
have traditionally followed this route. In this section, I will first outline
the program and then point out some of its limitations. The algebraic
quantization program introduced in the next subsection is an extension of
the Dirac strategy, aimed at overcoming these limitations.

Let us suppose that the phase space $\Gamma$ of the classical system
has the structure of a cotangent bundle. (I am making this restriction for
simplicity of presentation. The program can be implemented in a more general
context using ideas from geometrical quantization. See, e.g., Ashtekar and
Stillermann (1986).) We shall denote the configuration variables
by $q$ and momentum variables by $p$. Let the first class constraints be
$C(q,p)$. The $q, p$ and $C$ are all collective labels. In particular,
$\Gamma$ may be infinite dimensional and there may be infinitely many
constraints as in the example of a Maxwell field, considered above.

To quantize the system, one proceeds in the following steps:
\item{$\bullet$} Ignore the constraint to begin with. Then, it is natural to
consider as states wave functions $\Psi(q)$ on the configuration space,
represent the operators $\hat{q}$ by multiplication, $\hat{q}\cdot
\Psi (q) = q \Psi(q)$, and $\hat{p}$ by derivation, $\hat{p}\cdot\Psi(q)
= -i{\hbar}\delta\Psi/\delta{q}$, so that the classical Poisson brackets
are taken over to $i\hbar$ times the commutators. Denote by $V$ the vector
space spanned by $\Psi(q)$.
\item{$\bullet$} Since constraints are ignored, the elements of $V$ do not
represent physical states. To obtain these, one first promotes the
classical constraints $C(q,p)$ to operators $\hat{C} := C(\hat{q},
\hat{p})$ on the $V$. In general, this step needs a choice of factor
ordering (and, in the case of a system with infinite number of degrees of
freedom, also regularization). Then, one selects physical states $\Psi_P$
as those elements of $V$ which are annihilated by the operators $\hat{C}$.
The space $V_P$ spanned by $\Psi_P$ is the space of states of relevance
to physics. The idea is to do quantum mechanics on this $V_P$.

Let us illustrate the program by applying it to the case of a Maxwell
field considered above. In the first step, one represents quantum states
as functionals of vector potentials $\Psi (A)$ on the spatial 3-plane
$\Sigma$ and defines the operators (or, rather, operator-valued
distributions) $\hat{A}_a(x)$ and $\hat{E}^a(x)$ as: $\hat{A}_a(x)\cdot
\Psi(A)= A_a(x)\Psi(A)$ and $\hat{E}^a(x)\cdot\Psi(A) = -i\hbar \delta\Psi(A)/
\delta A_a(x)$. In the second step, one wants to select the physical states.
To do so, one first promotes the constraint functionals $C_\Lambda (E)$
to concrete operators: $\hat{C}_\Lambda = -i\hbar\lint d^3x (D_a\Lambda(x))
(\delta/\delta A_a(x))$. The physical states are annihilated by all these
operators. A straightforward calculation shows that this condition is
equivalent to demanding that the states $\Psi_P(A)$ be gauge invariant.
Thus, the imposition of the quantum constraint has precisely the desired
effect: in the classical theory, the constraints generate gauge on the
Maxwell phase space while in quantum theory they require that physical
states be gauge invariant.

As a second, and somewhat different example, consider a free particle
of mass $m$ in Minkowski space. The classical phase space is the cotangent
bundle over Minkowski space and there is one constraint, $p^\mu p_\mu +
m^2 =0$, which ensures, in a relativistically invariant way, that the
intuitive expectation that the system should have only 3 degrees of freedom
is correct. In the Dirac program, states can be represented by functions
$\Psi(q)$ on Minkowski space. The physical states $\Psi_P(q)$ must then
satisfy the operator version $\eta^{\mu\nu}\d_\mu \d_\nu \Psi_P(q)
+m^2 \Psi_P(q) = 0$ of the classical constraint. That is, the physical states
are precisely the solutions to the Klein-Gordon equation. From representation
theory of the Poincar\'e group, we know that the space of suitable solutions
to this equation can be interpreted as representing quantum states of a
relativistic particle with mass $m$ and spin zero. Thus, again, the
Dirac imposition of the quantum constraint provides the answer that we expect
on physical grounds.

In spite of these nice features, the program is incomplete in certain
ways and this turns out to be an important limitation for application to
diffeomorphism invariant theories such as general relativity. First,
it provides {\it no} guidelines for introducing the appropriate inner
product on the space of physical states. In examples such as the ones
considered above --fields or particles in Minkowski space-- one does not
encounter a serious difficulty in practice because one can use the
available symmetries to select a preferred inner-product. For examples,
in the case of the Maxwell field, one selects the vacuum state by invoking
Poincar\'e invariance and then uses the vacuum expectation values of
physical operators to obtain the required Hilbert space structure. Similarly,
in the case of a free relativistic particle, the time translation group
enables one to perform a natural positive and negative frequency
decomposition of the solutions to the Klein-Gordon equation and arrive at
the standard Hilbert space structure. In canonical quantum gravity, on the
other hand, such a space-time group of symmetries is simply not available.
(One might imagine using the spatial diffeomorphism group to select the
vacuum. However, the strategy fails because imposition of quantum
constraints implies that {\it every} physical state is invariant under this
group). Therefore, Dirac's program has to be supplemented with a new guiding
principle.

A second problem one often encounters is the following. In interesting
examples, although the phase space $\Gamma$ may indeed have the cotangent
bundle structure, the configuration space ${\cal C}$ is often a non-trivial
manifold (i.e., not diffeomorphic to $\real^n$). When this occurs, there is
no global chart whence a complete set of configuration and momentum
observables is necessarily {\it overcomplete}. That is, there are algebraic
relations between the basic variables which one wants to promote to operators
in the quantum theory. Again, a guiding principle is needed, already in the
first step of the program, to specify how these are to be incorporated in
the quantum theory. Are they to be imposed as ``additional constraints''?
Or, is one to first construct an algebra which already incorporates
these constraints and {\it then} seek its representations on a vector
space $V$? Or, should one do something entirely different? In concrete
examples, this problem may remain hidden if the quantum constraint
operators $\hat{C}$ can be constructed directly without having to go through
the ``elementary'' operators such as $\hat{q}$ and $\hat{p}$. Unfortunately,
as we will see, this simplification does not always arise whence new input
is needed.

\goodbreak
\subsection{2.4}{Algebraic quantization program}

We will now introduce a slight extension of the Dirac program aimed at
overcoming the limitations discussed above. The new ingredients can be
summarized as follows. (For details, see chapter 10 of Ashtekar (1991),
Ashtekar and Tate (1992) and Rendall (1992a).)

First, the program is algebraic in spirit. Thus, one first constructs
an (abstract) algebra of quantum operators and then looks for its
representations. In non-relativistic quantum mechanics (on $\real^n$)
Von-Neumann's uniqueness theorem tells us that there is only one
representation of the canonical commutation relations (CCRs) that is of
direct physical interest. In the context of field theories, on the other
hand, the CCRs admit infinitely many inequivalent representations and one
does not know a priori which of them would encompass the physical situation
under consideration. The algebraic strategy is therefore better suited.

Second, in the very construction of the algebra, one begins with a class
of ``elementary functions'' on the phase space which can be more general
than the $q$ and the $p$ considered above. Not only does this enable one to
consider systems in which the phase space does not necessarily have the
structure of a cotangent bundle, but it also introduces considerable
flexibility. For the Maxwell field considered above, for example, one may
take, as one's elementary functions, self dual electric and magnetic fields,
or even a ``hybrid'' pair consisting of the self dual (and hence, complex)
electric field but a real vector potential. This is important because,
as we will see in section 4, equations of general relativity simplify
considerably if one uses such hybrid variables. Another new feature
is that the set of elementary variables can be overcomplete. The algebraic
relations between them are incorporated right in the beginning, in the very
construction of the algebra of quantum operators. We shall see in section 5
that this flexibility is essential if one wants to carry out quantization in
the loop representation.

Third, the ``reality conditions'' in the classical theory are incorporated
in the $\star$-relations in the quantum algebra. That is, if two elementary
functions on the classical phase space are complex-conjugates of one another,
the corresponding operators are to be $\star$-adjoints of one another.
This enables one to introduce a new guiding principle to select the physical
inner product {\it without} reference to a symmetry group. Given an
irreducible representation of the quantum algebra on a vector space
($V$ or $V_P$), one seeks an inner product such that the abstractly defined
$\star$-relations become concrete Hermitian-adjointness relations on the
resulting Hilbert space of states. {\it Such an inner-product, if it exists,
is unique} even if the system has an infinite number of degrees of freedom
(Rendall (1992a)). If it does not exist, one has to start all over and begin
with a new vector space representation. In a variety of physical examples
considered so far -- 2+1 gravity, photons and gravitons in 3+1-dimensional
Minkowski space, non-trivially constrained systems with a finite number of
degrees of freedom, quantization based on self dual or ``hybrid'' elementary
variables, etc.-- this procedure has successfully picked out the correct
inner-product.

The example of 2+1 dimensional gravity --discussed in section 3--
is particularly interesting because, as in the 3+1 case, the
theory is diffeomorphism invariant. In spite of this feature, contrary to
what was the conventional wisdom, one can arrive at the correct inner-product
{\it without} having to first isolate time and ``deparametrize'' the theory.
Such a deparametrization is likely to be essential to {\it interpret} the
theory properly. However, for such purposes, it may well suffice to have
only an approximate notion of time. Had it been essential to single out time
to get an inner-product, on the other hand, an approximate notion would
not suffice; one cannot do mathematical physics with approximate inner
products! Finally, note that, in elementary quantum mechanics, it is
always the case that the classical complex-conjugation relations are
promoted to hermiticity conditions on operators. Therefore, the above
strategy involving classical ``reality conditions'' may seem trivial at
first. What is new here is that, unlike in elementary quantum mechanics,
we do not have an a priori inner product and the strategy is used to {\it
select} one.

\def\cS{${\cal S}$}
We can now spell out the main steps involved:\hfill\break
\item{1.} As in the Dirac program, ignore the constraints to begin with.
Select a subspace \cS\ of complex-valued functions on the classical
phase-space, closed under the operation of taking Poisson-brackets. Each
element of \cS\ is to be promoted to a quantum operator unambiguously; it
represents {\it an elementary classical variable}. \cS\ has to be ``small
enough'' so that this quantization procedure can be carried out unambiguously
(i.e. without eventual factor ordering problems) and yet ``large enough'' so
that they can serve as (complex) coordinates on the phase-space.
\item{2.} Associate with each element $F$ of \cS\ an abstract operator $\hat
F$.
These are the {\it elementary} quantum operators. Construct the (free) algebra
generated by these elements and impose the canonical commutation relations:
$[\hat F ,\hat G ]-i\hbar\widehat{\{F,G\}}=0$. If there are algebraic relations
between the elementary classical variables, these are to be incorporated in
the very construction of the quantum algebra. For example, if the functions
$F, G$ and $FG$ on the phase space are all elementary classical variables,
then one requires, in the quantum algebra, $\hat{F}\hat{G} +\hat{G}\hat{F}
-2\widehat{FG} =0$. (For further details, see references quoted in the
beginning of this section.) Denote the resulting algebra by ${\cal A}$.
\item{3.} Next, {\it define} a $\star$-relation (i.e. an involution) on this
algebra by requiring that, if two elementary classical variables $F$ and $G$
are complex conjugates of one another, i.e., if $\bar{F}= G$, then $\hat G
= (\hat F)^\star$, and that the $\star$ operation satisfies the properties
of an involution. (Recall that these are: $(\hat{F} +\lambda \hat{G})^\star
= \hat{F}^\star + (\bar{\lambda}) \hat{G}^\star$, $(\hat{F} \hat{G})^\star =
\hat{G}^\star \hat{F}^\star$, and, $(\hat{F}^\star)^\star = F$.) Denote the
resulting $\star$-algebra by ${\cal A}^{(\star)}$.
\item{4.} Find a representation of the algebra ${\cal A}$ by operators on a
complex vector space $V$. Note that in general $V$ may not be equipped with
any inner-product and that the $\star$-relations can be ignored at this stage.
This representation may be obtained by any convenient means. Possible
candidates are: geometric quantization techniques, group theoretical methods
and Gel'fand spectral theory discussed in the next section.
\item{5.} Obtain the quantum analogs of the classical constraints. In general,
this requires the choice of a factor-ordering {\it and} regularization. Find
the linear subspace $V_{phy}$ of $V$ which is annihilated by all quantum
constraints. This is the space of physical quantum states.
\item{6.} Introduce an inner-product on $V_{phy}$ such that the
$\star$-relations --ignored so far-- become Hermitian adjoint relations on
the resulting Hilbert space. Note that the full $\star$-algebra
${\cal A}^{(\star)}$ itself does {\it not} have a well-defined action on the
physical subspace $V_{phy}$; a general element of ${\cal A}$ would not weakly
commute with the constraints and would therefore throw physical states out
of $V_{phy}$. One must therefore find a ``sufficiently large'' set of
operators which weakly commute with the constraints whose $\star$-adjoints
also weakly commute with the constraints. It is the $\star$-relations between
{\it these} ``physical'' operators that are to be taken over to Hermitian
adjointness relations by the inner product. This is a rather involved
procedure.
\footnote{7}{In certain cases, it turns out to be easier to first isolate
the physical operators that weakly commute with constraints and then
introduce $\star$-relations directly on them. In such situations, it is
not necessary to equip the initial operator algebra ${\cal A}$ with a
$\star$-operation.}
\item{7.} Interpret a sufficiently large class of self-adjoint operators;
devise methods to compute their spectra and eigenvectors; analyze if there is
a precise sense in which the 1-parameter family of transformations generated
by the Hamiltonian can be interpreted as ``time evolution''; etc.

If this program can be completed, one would have available a coherent
mathematical framework. Conceptual issues from measurement theory can
be then faced. It is important to emphasize, however, that these steps
are meant to be guidelines rather than a rigid set of rules. The strength
of the program lies in the fact that: i) it provides a broad framework
encompassing many physically interesting systems including gauge theories and
general relativity; and, ii) it isolates the inputs that are required:
Selection of the space ${\cal S}$ of elementary variables in the first step
and the choice of the vector space representation in the fourth step. Once
these choices are made, the quantum theory, if it exists, is unique. This
helps to focus one's efforts. However, one must use judgement and physical
intuition to make this selection. This is as it must be for we are trying
to arrive at a more complete quantum theory starting only from its classical
limit.

\goodbreak
\subsection{2.5}{Connections and loops: a non-linear duality}

The purpose of this subsection is to present a strategy to quantize
theories of connections by exploiting a certain ``non-linear duality'' that
exists between connections and loops. (Throughout this subsection, the
term ``duality '' is used in the sense of ``interplay'' rather than as
a specific mathematical operation.) This strategy is
particularly well suited for diffeomorphism invariant theories of connections
such as general relativity in 3 and 4 dimensions and certain topological
field theories. The technical level of this material is, however,
substantially more sophisticated than that of the material covered so
far in this section. Therefore, the treatment will not be as detailed.
My aim is to provide a bird's eye view and make the reader comfortable
with the multifaceted use of loops in quantum theories of connections.

Let $A_a^i$ be a connection on a manifold $\Sigma$ which takes values in
the Lie algebra of a group $G$. In concrete applications, we will take
$\Sigma$ to be a spatial slice in space-time and choose $G$ to be either
$U(1)$ or $SU(2)$. Let $\gamma$ be a closed curve in $\Sigma$. Then, the
the trace $T[\gamma, A]$ of the holonomy of $A_a^i$ around a closed loop
$\gamma$ is defined to be:
\bneq
T[\gamma, A ] := \tr\- {\cal P} \big(\exp (\loint_\gamma dS^a\- A_a )\big),
\eneq{(2.5.1)}
where, ${\cal P}$ stands for ``path-ordered,'' the trace is taken in some
representation of $G$ and $A_a$ is the matrix valued 1-form representing
$A_a^i$ in the corresponding representation of the Lie algebra. For
a fixed $\gamma$, $T[\gamma , A]$ can be regarded as a function on the
space ${\cal C}$ of connections. However these functions are {\it gauge
invariant} and therefore naturally project down to the quotient ${\cal C}/
{\cal G}$ of ${\cal C}$ by the group ${\cal G}$ of {\it local} gauge
transformations. Similarly, for a fixed $A$, $T[\gamma, A]$ can be regarded
as a function on the space of loops. However, in general, these functions
do not separate loops. Let $\gamma$ be (trace-)equivalent to $\gamma'$
if for all connections $A_a^i$, in the Lie algebra of $G$, $T[\gamma, A] =
T[\gamma', A]$. (This notion of equivalence is sensitive to the choice $G$.
For example, $\gamma$ is equivalent to $\gamma^{-1}$ if $G$ is $SU(2)$ but
not if it is $U(1)$.) Each of these (trace) equivalence classes of loops will
be called a {\it G-troop}. However, for simplicity, in what follows we will
generally not pass to the quotients; we will regard $T[\gamma, A]$ as a
function of loops and connections and just bear in mind that it can be
projected to the space of troops and gauge-equivalence classes of connections.

Note that there is a striking similarity between the functional form
(2.5.1) of $T[\gamma, A]$ and the form of the integral kernel $\exp
(i\vec{k}\cdot \vec{x})$ which enables one to transform functions of
$\vec{x}$ to functions of $\vec{k}$. Indeed, it is possible to use the
trace as the kernel in an integral transform which sends
functions of connections to functions of loops: we can set
\bneq
\psi(\gamma) := \lint_{{\cal C}/{\cal G}}\- d\mu\- T[\gamma ,A] \Psi[A] ,
\eneq{(2.5.2)}
where $d\mu$ is a measure on ${\cal C}/{\cal G}$. (This transform was first
proposed as a powerful but heuristic device by Rovelli and Smolin (1990)
and was put on a precise mathematical footing by Ashtekar and Isham (1992).)
Thus, in analogy to the duality between $\vec k$ and $\vec x$,
we can say that troops are dual to the gauge equivalence class of connections.
In the non-Abelian case, however, the space of troops as well as the space
of gauge equivalence class of connections are {\it non-linear}, whence the
terminology ``non-linear duality''. We will see that this duality is quite
powerful.

Let me begin by illustrating these ideas in the context of the Abelian
(Maxwell) theory in Minkowski space. We will be interested in a canonical
approach. Therefore, all loops and fields will be defined on a spatial
plane, $t={\rm const}$. Let us begin with observables, the electric and
the magnetic fields. These are normally measured in the lab through their
fluxes across 2-surfaces. Let $S$ and $S'$ be two such surfaces bounded by
closed loops $\gamma$ and $\gamma'$. Then a simple calculation shows that
the commutator between the fluxes of the two operators, $[\hat{B}(S),
\hat{E}(S')]$ is given simply by $i\hbar {\cal L}(\gamma, \gamma')$, where
${\cal L}(\gamma, \gamma')$ is the {\it Gauss linking number} between
the closed loops $\gamma$ and $\gamma'$. Hence, the two fluxes are
are subject to the fundamental Heisenberg uncertainty relations:
\bneq
\Delta[\hat{E}(S)]\cdot \Delta[\hat{B}(S')] = \hbar {\cal L}(\gamma ,
\gamma').
\eneq{(2.5.3)}
On the left side, we have physical fields while on the right side,
the simplest topological invariant associated with two loops. The two
are related precisely because the fluxes in question are just the
holonomies of (magnetic and electric) connections around the loops
that feature on the right side. This interplay between connections and
loops opens up new ways of describing the quantum Maxwell field. For
example, the photon states in the Fock space can be represented by suitable
functions on the loop space and the entire physics of the Maxwell field can
be captured in the properties of these functions (Ashtekar \& Rovelli,
1992).

Let us now turn our attention to non-Abelian theories. To be specific, let
the gauge group be $SU(2)$ and let $\Sigma$ be a spatial 3-manifold. For
the present discussion, we will focus just on the Gauss constraint; other
constraints, if any, are to be considered at the end. We can therefore
take the traces of holonomies as gauge invariant configuration variables.
Since we are now thinking of these as functions of connections, labelled by
loops, let us denote them as $T[\gamma](A)$. It is easy to verify that they
form an overcomplete set of functions on the effective configuration space of
the theory, ${\cal C}/{\cal G}$. Following Rovelli and Smolin (1990), we can
also introduce momentum variables which are again gauge invariant and labelled
by loops. (We will not need their explicit expressions here. They are
discussed in detail in sections 4 and 5.) Together, these two sets can serve
as the elementary classical variables in the algebraic quantization program:
the complex vector space they generate is closed under the Poisson bracket and
the variables are overcomplete on the effective phase space $T^\star({\cal C}
/{\cal G})$. It is easy to follow the steps outlined in section 2.4 to
construct the quantum $\star$-algebra corresponding to these variables.

The next task is that of finding the representations of this quantum algebra.
The following program addresses this task systematically. First, one focuses
on the {\it Abelian} part of the algebra, makes it into a $C^\star$ algebra
and uses a powerful framework due to Gel'fand to develop a proper
representation theory of this $C^\star$-algebra. In the next stage, one
brings in the rest of the algebra. Not all representations of the Abelian
algebra carry representations of the full algebra. One focuses only on
those which do. Finally, one takes up the issue of dynamics (or, of
imposition of the remaining constraints, if any) to further weed the remaining
representations. More precisely, one requires that the representation should
be such that the Hamiltonian (or the constraint operators) be well defined.
While this general program is applicable to any field theory, in practice,
the last step is {\it extremely} difficult to carry out. For diffeomorphism
invariant theories, however, one can often succeed basically because dynamics
is either trivial or governed by constraints. An example is given by
3-dimensional general relativity discussed in the next section. Most of the
discussion in the remainder of this section, however, will be devoted to the
first step of this program.

To carry out this step, we have to select an appropriate Abelian subalgebra
of the quantum algebra.
The obvious choice is to use the configuration variables $T[\gamma](A)$;
since they depend only on the connections on the spatial slice and not on
the conjugate momentum, their mutual Poisson brackets vanish. With each
closed loop $\gamma$, let us associate an abstractly defined quantum
operator $\hat{T}[\gamma ]$. Denote by ${\cal A}_0$ the free associative
algebra they generate. Following the algebraic quantization program, we
need to incorporate in ${\cal A}_0$ the algebraic relations that exist
among the classical functions $T[\gamma ]$ on ${\cal C}/{\cal G}$. This
can be easily achieved. (In this step, the troop equivalence is automatically
incorporated.) Finally the sup norm on the space of classical configuration
variables naturally endows the quantum $\star$-algebra with an appropriate
norm. Denote the resulting $C^\star$-algebra by ${\cal A}$. Note that
because we started by considering functions of connections, the generators
of the algebra $\hat{T}[\gamma]$ are labelled by loops.

To find representations, one can use Gel'fand's theory of representations
of {\it Abelian} $C^\star$ algebras. The first key result is that {\it every
such algebra is isomorphic to the $C^\star$ algebra of continuous complex
valued functions on a} compact {\it topological space}. This space is called
the Gel'fand spectrum of the algebra and generally denoted by $\Delta$.
What is even more interesting is that $\Delta$ can be constructed from the
$C^\star$-algebra itself. As a point set, $\Delta$ is the space of maximal
ideals of the algebra, or, equivalently, of (multiplicative) homomorphisms
from the algebra to the complex numbers. The Gel'fand topology on $\Delta$
is the the weakest one which makes the naturally defined functions on $\Delta$,
$f_{\hat a}(h):= h(\hat a)$, continuous. (Here $h\in \Delta$ is regarded
as a homomorphism from the algebra to complexes and $\hat{a}$ is any fixed
element of the algebra.)

Our holonomy algebra ${\cal A}$ is, in particular, isomorphic to the
algebra of continuous functions on its specturm $\Delta$. Furthermore, one
can show (Rendall, 1992) that the classical configuration space ${\cal C}/
{\cal G}$ is {\it densely} embedded in $\Delta$. Now, one intuitively expects
the quantum states to be gauge invariant functions of connections, i.e.,
functions on ${\cal C}/{\cal G}$. This expectation is essentially correct.
The domain space turns out to be $\Delta$: quantum states $\Psi(h)$ are
functions on $\Delta$ and the algebra operates on them via the natural
map: $\hat{a}\cdot\Psi(h) = h(\hat{a})\Psi(h)$, for all $h\in \Delta$ and
$\hat{a}\in {\cal A}$. There is of course a key difference between ${\cal C}
/{\cal G}$ and $\Delta$; while the former is not even locally-compact
in any one of the standard topologies one uses on spaces of connections, the
latter is {\it compact} in it's natural, Gel'fand topology. This in particular
means that $\Delta$ admits regular measures. Let $d\mu$ be one such measure.
Then a $C^\star$-representation of the algebra is given by choosing as the
Hilbert space of states, the Cauchy completion of all continuous functions
with respect to the Hermitian inner product:
\bneq
\langle \Psi_1|\Psi_2 \rangle := \lint_\Delta d\mu \- \ovr{\Psi_1(h)}\-
\Psi_2(h).
\eneq{(2.5.4)}
Note that, since traces of holonomies around any fixed troop are continuous
functions on $\Delta$, it follows that they belong to the Hilbert space.
Hence, we see that the Rovelli-Smolin transform (2.5.2) is well-defined
from ($L^2$)-functions on $\Delta$ to functions on the space of troops.

In quantum mechanics of systems with only a finite number of degrees of
freedom, the classical configuration space turns out to be the domain
space of quantum states. In field theory, one generally has to enlarge the
classical configuration space to obtain the domain space because quantum
states tend to have support on {\it distributional} classical configurations.
This is the essential reason behind the enlargement from ${\cal C}/{\cal G}$
to $\Delta$. Let us put these important technicalities aside for a moment.
Then, the rough picture that emerges is the following. The operator
algebra is generated by $\hat{T}[\gamma]$ labelled by closed loops while
states $\Psi(h)$ are functions of (generalized) connections.

This interplay between loops and connections continues. To see this, let
us first note that there is a precise sense in which the representations
considered in the previous paragraph are the generic ones. The sense is
the following. Given a $C^\star$-algebra, of particular interest are its
{\it cyclic} representations. These are the ones in which one can obtain
(a dense subspace of) the Hilbert space of states by acting successively by
the elements of the algebra on a fiducial vector, called the cyclic vector
or the vacuum. Roughly speaking, cyclic representations are the ``building
blocks'' from which all other representations can be constructed. The key
fact is that {\it every} cyclic representation is of the type discussed above.
Furthermore, the measure $d\mu$ is completely determined by its ``Fourier
transform,'' $\Gamma[\alpha]$, given by
\bneq
\Gamma[\alpha] := \lint_\Delta d\mu(h) \, \- \- h(\hat{T}[\alpha]),
\eneq{(2.5.5)}
where, once again, we have used the fact that every element $h$ of $\Delta$
is a homomorphism on the algebra. Note that $\Gamma[\alpha]$ --called
the generating function of the representation-- is a function on the space
of loops. Furthermore, one can write down directly --i.e., without knowing
the measure $d\mu$-- necessary and sufficient conditions on a loop function
so that it can qualify to be a generating function. That is, one can write
down suitable loop functions $\Gamma[\alpha]$ directly and {\it use} them
to determine the measure $d\mu$ and hence the representation. Often,
physics of the problem naturally leads to candidates for the generating
functions $\Gamma [\alpha]$. For example, in the $SU(2)$ case under
consideration, if we wish to find a diffeomorphism invariant representation,
we could use for $\Gamma[\alpha]$ the following function:
\bneq
\Gamma[\alpha] := \lint_{\cal M} dV(A) \- \, \-  T[\alpha](A),
\eneq{(2.5.6)}
where ${\cal M}$ is the moduli space of {\it flat} $SU(2)$ connections
on the 3-manifold $\Sigma$ and $dV$ a volume element thereon.
This representation plays a key role in the
quantization of the Chern-Simons theory of  $ISO(3)$ (i.e., the Euclidean
group in 3 dimensions) connections. The diffeomorphism group of $\Sigma$
is unitarily represented on the Hilbert space of states. One expects
that other diffeomorphism invariant measures on $\Delta$ can be
obtained by choosing suitable {\it knot invariants} for one's generating
function, $\Gamma[\alpha ]$.

To summarize, in theories of connections, the generators of the Abelian
$C^\star$ algebra ${\cal A}$ are labelled by loops. The states are
functions of (generalized) connections. Suitable functions of loops lead
to measures on the space of (generalized) connections, thereby fixing the
inner product. More precisely, for each choice of the generating function
$\Gamma[\alpha]$, we obtain a cyclic representation of the algebra ${\cal A}$.
The choice of $\Gamma[\alpha]$ is to be dictated by the physics of the
problem. In diffeomorphism invariant theories, there are often natural
candidates available and it appears that there is an interesting interplay
between knots and diffeomorphism invariant measures on the space of
(generalized) connections. We will see in the next section, quantization of
2+1 dimensional general relativity can be carried out by choosing the
appropriate analog of (2.5.6) for the required generating function.

\goodbreak
\vfill\break

%\input [ijtpd.lh]qgmac
%\singlespace
\def\A{A_a^I} \def\E{\tw{E}{}^a_I}

\section{3.}{${}\quad$ Three faces of 2+1 quantum gravity}%\par
\medskip
{\narrower\narrower\smallskip\noindent
{\sl Feynman thought about it himself. Once --uninterested though he was in
fiction or poetry-- he carefully copied out a verse fragment by Vladimir
Nabokov: ``space is a swarming in the eyes; and time, a singing in the ears''.}
\smallskip
James Gleick (Genius).\smallskip}

\subsection{3.1}{Introduction}%\par

Researchers in non-perturbative quantum gravity have, by and large, adopted
three viewpoints. Very roughly, they can be summarized as follows.

In the first and the most abstract of the three, one begins with some
discrete structures at a fundamental level and attempts to
capture physics through combinatorial principles. Examples of such attempts
are those dealing with spin-networks, twistors and causal sets. The underlying
idea is clearly very exciting and thought provoking. However, these approaches
have suffered from the extreme open-endedness of the problem. The second
viewpoint emerged from the canonical approach as practiced by the Bergmann
school in the sixties and the seventies. The viewpoint here is that all
physics is contained in the solutions to the quantum constraint equations;
there is no time and no dynamics. Physical questions are to be phrased and
answered in terms of the Dirac observables --i.e., the operators which (weakly)
commute with the constraints. This ``timeless description'' is very much in
the spirit of classical general relativity where solutions to field equations
represent entire space-times. At a fundamental, covariant level, nothing
``happens''; evolution arises only when we introduce an extra structure such
as a foliation and compare fields on one leaf of the foliation with those
on another. The main problem with this method is the construction of a sensible
measurement theory to handle issues discussed in Jim Hartle's lectures.
The third viewpoint emerged in the seventies from Wheeler's school of thought
also on canonical quantization. As explained in section 1, the idea here
is to interpret the quantum scalar constraint as a Schr\"odinger equation for
evolution of quantum states by isolating one of the components of the
configuration variables as time. The emphasis is on ``depametrization'' of
general relativity: Not only is this step considered essential for
interpretational issues but also
for selecting the correct mathematical ingredients, particularly the
inner-product on physical states. These ideas have been implemented in a
number of truncated models. However, in full, 3+1 dimensional general
relativity, the program faces serious obstacles (see, e.g., Kucha\v r
(1981)).

In this chapter, we will discuss 3-dimensional general relativity. We will
find that the algebraic quantization program discussed in section 2.4 can
be completed and, furthermore, leads to three descriptions which can be
regarded as implementations of the three sets of ideas discussed above.
These are the three faces of 2+1 quantum gravity. The first is
a description without space and without time; the second is a timeless
description on a given spatial manifold; time arises in the third description,
but ``internally,'' as one of the mathematical variables of the theory
and with it are born the notion of dynamics and unitarity. Thus, we have a
concrete example in which the three sets of ideas co-exist in a complementary
fashion. This framework serves to illustrate what we are aiming for in the
4-dimensional theory. We will see in the next two sections that notable
progress has been made on each of the three directions if one regards
general relativity as a dynamical theory of connections rather than of
metrics.

3-dimensional gravity provides us with an interesting example because,
although it is very similar to the 4-dimensional theory in its structure,
it is technically much simpler. As in the 4-dimensional case, the theory
is diffeomorphism invariant; the gravitational field is coded in the very
geometry of space-time; there is no background time to evolve in; and, in
the Hamiltonian formulation, dynamics is governed by first class constraints.
A simple minded dimensional counting suggests that, as in the 3+1 case, the
theory would be perturbatively non-renormalizable. Indeed, there are some
detailed papers in the literature arguing that in the perturbative
(``covariant'') quantization scheme, the theory is as pathological as
3+1-dimensional general relativity. The theory was also analysed canonically
using geometrodynamical variables ---the 2-metric and the extrinsic curvature
of a spatial slice. It was found that the functional form of the constraints
as well as their Poisson algebra is completely analogous to that in the
3+1 dimensional case, whence it was difficult to find solutions to the
quantum scalar constraint, i.e., the Wheeler-DeWitt equation. Thus, up until
mid-eighties, the general consensus was that quantum general relativity is
almost as difficult in 2+1 dimensions as it is in 3+1. On the other hand, it
was known that, since the vanishing of the Einstein tensor implies the
vanishing of the Riemann tensor in 3-dimensions, every solution to the vacuum
equation is {\it flat}. Thus, in 2+1 dimensions, there are no gravity waves
classically, and hence no gravitons quantum mechanically. This suggests that,
technically, the theory should be {\it much} easier to handle. We now know that
this expectation is correct: the theory can in fact be solved exactly. There
are no divergences, and, the general solution to all quantum constraints can be
constructed provided one chooses judicious canonical variables (Achucarro
\& Townsend (1986), Witten (1988), Ashtekar et al (1989).) Furthermore, these
are precisely the analogs of the canonical variables which have simplified
the equations in 3+1 dimensions (Ashtekar (1987)) and which we will use in the
next two chapters.

In this chapter, we will present a treatment which admits an extension to
3+1 dimensions. In section 2, we present the classical Hamiltonian framework
assuming that the spatial sections are compact. In section 3, we construct the
``timeless'' quantum description along the lines of the Bergmann school. In
section 4, for simplicity we restrict ourselves to the 2-torus spatial
topology and recast this description in a form in which only discrete
structures and combinatorial techniques come into play. In section 5, we
reformulate the description by extracting time and reducing the scalar
constraint to a time evolution equation. (The problem of extracting
time in the case of more general topologies is still open.)

There is a very considerable body of research on 2+1 dimensional gravity
particularly by Carlip, Deser, Jackiw and 't Hooft, Hoyosa, Martin and
Moncrief. Much of this work, however, is in either geometrodynamical
framework or in the context of Witten's (1988) reformulation of 2+1
dimensional general relativity as a ISO(2,1) theory. While both
approaches are closely related to the one presented here, neither
generalizes in a direct manner to 3+1 dimensional general relativity.
Since my primary motivation for treating 2+1 dimensional gravity is to
make the reader familiar with certain ideas, techniques and viewpoints
that are useful in 3+1 dimensions, I will refrain from discussing these
other approaches.

\goodbreak
\subsection{3.2}{Hamiltonian formulation}%\par

Let us assume that the space-time manifold has topology $\Sigma\times\real$
where $\Sigma$ is an orientable, compact 2-manifold of an arbitrarily chosen
but fixed genus. We will begin with the Palatini action and then perform a
Legendre transform to pass to the Hamiltonian description. Thus, the basic
variables are co-triads, $e_a^I$, and $SO(2,1)$ connections,${}^3\A$, where
the uppercase latin letters $I, J...$ denote internal $SO(2,1)$ indices (and
label the co-triads, while the lower case latin indices $a,b,...$ denote
space-time (or spatial) indices. To recover general relativity, one must
assume that the co-triads are linearly independent. However, our framework
itself is slightly more general; it can accommodate degenerate metrics. In
that follows, we shall consider this more general theory (except when we
comment on the relation of this framework to geometrodynamics). The
action is given by:
 \bneq
S(e, {}^3\!A):= \lint_M  d^3x\>\- \tw{\eta}{}^{abc}e_a^I \, \-
{}^3\!F_{bcI},
\eneq{(3.2.1)}
where, $\tw{\eta}{}^{abc}$ is the metric independent Levi-Civita density on
$M$ and, ${}^3\!F_{ab I} := 2\d_{[a} {}^3\!A_{b]I} + \epsilon_{IJK} \-
\>{}^3\!A_a^J\>\- {}^3\!A_b^K$ is the field strength of the connection
${}^3\!\A$. The classical equations of motion are:
\bneq
{}^3\!\D_{[a} e_{b]I} = 0  \qquad {\rm and} \qquad {}^3\!F_{ab}^I = 0,
\eneq{(3.2.2)}
where ${}^3\!\D$ is the gauge covariant derivative operator determined by
${}^3\!\A$. The first equation ensures that the connection is compatible with
the triad while the second tells us that the connection is flat. Together,
they tell us that (when the triad is non-degenerate) the 3-metric $g_{ab} :=
e_a^I e_{bI}$ constructed from the triad is flat, i.e., satisfies the
3-dimensional Einstein's equation. Thus, the 2+1 theory has no local degrees
of freedom. As we will see later in this section, it does, however, have
(a finite number of) global, topological degrees of freedom.

As far as classical equations of motion are concerned, the Palatini action
(3.2.1) is of course completely equivalent to the Einstein-Hilbert action.
However, while the Einstein-Hilbert action leads to the geometrodynamical
Hamiltonian framework --based on 2-metrics and extrinsic curvatures--
as we will now see, in 2+1 dimensions, the Palatini action naturally
leads to quite a different Hamiltonian description.

Let us then perform the Legendre transform of this action. The
resulting phase space description is as follows. The configuration space
${\cal C}$ is the space of pull-backs, $\A$, to the 2-dimensional spatial
slice $\Sigma$ of the connection ${}^3\!\A$. The momentum $\E$ conjugate
to $\A$ is the dual of the pull-back of the co-triad $e_b^I$, $\E :=
\tw\eta{}^{ab}e_{bI}$, where $\tw\eta{}^{ab}$ is the (metric independent)
Levi-Civita density of weight $1$ on $\Sigma$. Thus, the phase space
$\Gamma$ consists of pairs $(\A,\E)$ of fields on $\Sigma$; the basic
(non-vanishing) Poisson brackets are simply
\bneq
\{\A (x),\>\tw{E}{}^b_J (y)\} = \delta_a^b\> \delta_J^I \>\delta^2(x,y).
\eneq{(3.2.3)}
Note that although $\E$ does determine the induced metric $q_{ab}$ on the
spatial $2$-manifold $\Sigma$ via $({\rm det}\- \> q)\>q^{ab} = \E\tw{E}^{bI}$,
the momenta $\E$ are {\it not} dyads; the internal index still runs from $1$
to $3$. {\it Thus, we are not assuming that there is a global frame field on}
$\Sigma$. Such an assumption would have restricted the topology of $\Sigma$
severely; our construction, on the other hand, places no such restriction.

The system has first class constraints. These can be read-off from the
Legendre transform of the action, or, more directly, by simply
pulling back the equations of motion (3.2.1) to the spatial slice:
\bneq
\D_a\E = 0 \qquad {\rm and} \qquad F_{ab}^I  = 0,
\eneq{(3.2.4)}
where $F_{ab}^I$ is the curvature of $\A$. The first equation is the Gauss
constraint which ensures that the internal $SO(2,1)$-rotations are gauge
transformations. The ``time'' component, on the internal index, of the second
equation is equivalent to the scalar (or, Hamiltonian) constraint of 2+1
dimensional geometrodynamics, while the ``space'' component gives us the
vector (or, diffeomorphism) constraint. As could have been expected, the
Hamiltonian is a linear combination of constraints. The striking feature of
the constraints is their
simplicity: the first is linear in momentum, the second depends only on
the configuration variable. We will see that this makes the task of
quantization rather straightforward.

We now introduce two sets of functions associated with closed loops $\gamma$
on $\Sigma$. Elements of the first set will be denoted by $T^0[\gamma ]$
and of the second by $T^1[\gamma]$. The $T^0[\gamma ]$ represent configuration
variables in that they depend only on $A_a^I$ while the $T^1[\gamma ]$
represent the momentum variables in that they depend linearly in $\tw E$.
(The superscripts $0$ and $1$ thus refer to the momentum dependence. One can
also construct observables $T^n[\gamma ]$ which are higher order in momenta.
However, it turns out to be redundant to do so.) These configuration and the
momentum variables are defined as follows:
\bneq
T^0[\gamma ](A) := {1\over 2}\tr U_\gamma \qquad {\rm and} \qquad
T^1[\gamma ](A, \tw E) := {1\over 2} \loint_\gamma dS^a\> \tr E_a U_\gamma ,
\eneq{(3.2.5)}
where, $U_\gamma (s) := {\cal P}\exp\lint_\gamma A$ is the holonomy of $\A$
around $\gamma$, evaluated at the point $\gamma (s)$, $E_a := \ut\eta_{ab}
\tw E^b$, and where we have used the 2-dimensional representation of $SO(2,1)$
to take the trace
\footnote{8}{Since I want to present a bird's eye view, I have skipped
detailed definitions of what we mean by loops, holonomies etc. These can be
found in section 4.3.}.
In the case when $\Sigma$ is non-compact, with the
topology of a punctured 2-plane, $T^0$ is essentially the mass and $T^1$, the
angular momentum, of the ``particle'' at the puncture enclosed by the loop
$\gamma$. In the compact case now under consideration, one does not,
on the other hand, have such a {\it direct} physical interpretation.

Nonetheless, even in the compact case these functions have certain properties
which make them central objects in the classical as well as quantum
descriptions within the canonical framework:
\item{$\bullet$}{They are {\it Dirac observables}; their Poisson brackets
with constraints vanishes weakly. Thus, in the language of section 2.2, they
can be projected down unambiguously to the reduced phase space of the system.
This property is straightforward to verify.}
\item{$\bullet$}{They provide us with a {\it complete} set of observables on
the reduced phase space. This point will be discussed at the end of this
section.}
\item{$\bullet$}{They are closed under the Poisson bracket. Furthermore, the
Poisson bracket can be coded in simple operations involving breaking,
re-routing and joining the loops in the arguments of these functions. This
in turn means that the symplectic structure on true degrees of freedom of
general relativity (and also gauge theories) in 2+1 dimensions can be coded
in these simple operations.}

\medskip
The Poisson brackets are given by:
\beginalign{
\{T^0[\gamma ] , T^0[\delta ]\} &= 0; \qquad
\{T^1[\gamma ], T^0[\delta ]\} = \sum_i \Delta_i (\gamma ,\delta )
(T^0[\gamma\circ_i \delta ] - T^0[\gamma\circ_i \delta^{-1} ])\cr
& \{T^1[\gamma ], T^1[\delta ]\} = \sum_i \Delta_i (\gamma , \delta )
(T^1[\gamma\circ_i \delta ] - T^1[\gamma\circ_i \delta^{-1} ]),\cr}
\endalign{(3.2.6)}
where, $i$ labels the points where the two curves $\gamma$ and $\delta$
intersect, $\circ_i$ stands for composition of the two curves at the $i$th
intersection and $\Delta_i$ equals $\pm {1\over 2}$ depending on whether the
dyad formed by the tangent vectors to the two curves at $i$ is right or left
handed. The general structure of these Poisson brackets is characteristic of
the Poisson algebra of configuration and momentum observables on {\it any}
cotangent bundle. However, the important point is that now we are considering
only a subset of configuration and momentum observables, namely the ones
associated with closed loops. It is somewhat remarkable that this subset is
also closed under Poisson brackets.

The completeness as well as a number of other properties of these 2+1
observables are easier to establish if
one notices a certain relation between them. To see this, note first that,
since we are in 2+1 dimensions, the {\it configuration space} ${\cal C}$
itself is equipped with a natural symplectic structure: For any two vectors
$(\delta A)$ and $(\delta A)'$ tangent to ${\cal C}$ with components
$(\delta A)_a^I$ and $(\delta A)'{}_a^I$, set $\Omega ((\delta A),
(\delta A)' ) := \lint_\Sigma d^2x\> \tw\eta^{ab}(\delta A)_a^I\> (\delta
A)'_{bI}$. Since $T^0[\gamma ](A)$ is a function on ${\cal C}$, we can
construct the Hamiltonian vector field $X_{[\gamma ]}$ it generates. We can
now return to the phase space $\Gamma$. With every vector field on ${\cal C}$,
we can associate a function on $\Gamma$ which is linear in momenta. It turns
out that this is precisely our momentum observable $T^1[\gamma ]$ of
(3.2.5). Thus, the symplectic structure $\Omega$ on ${\cal C}$ enables one
to ``construct'' the observables $T^1[\gamma ]$ starting from $T^0[\gamma ]$.
Since $T^0[\gamma ]$ are simply traces of holonomies, it is generally easy to
analyse their properties. The above construction then enables one to extend
these properties to the momentum observables $T^1[\gamma ]$. Thus, the
following properties:
\beginalign{
T^0[0]& = 1,  \qquad T^0[\alpha ] = T^0[\alpha^{-1}],\qquad
\quad T^0[\alpha\circ\beta ] = T^0[\beta\circ\alpha ],\cr
&{\rm and} \qquad 2T^0[\alpha]\cdot T^0[\beta] = T^0[\alpha\circ\beta ] +
T^0[\alpha\circ\beta^{-1}], \cr}
\endalign{(3.2.7a)}
of traces of holonomies (which follow trivially from $SO(2,1)$ trace
identities), immediately imply:
\beginalign{
&T^1[0] = 0,  \qquad T^1[\alpha ] = T^1[\alpha^{-1}],\qquad
 \quad T^1[\alpha\circ\beta ] = T^1[\beta\circ\alpha ],\cr
 &{\rm and} \quad 2T^0[\alpha]\cdot T^1[\beta]+ 2 T^0[\beta ]\cdot
 T^1[\alpha ] = T^1[\alpha\circ\beta ] + T^1[\alpha\circ\beta^{-1}]. \cr}
\endalign{(3.2.7b)}
These are the ``universal'' algebraic identities shared by {\it all}
$T^0$ and $T^1$ observables because they are {\it over}complete. As required
in the algebraic quantization program, they will go over to the quantum
theory.

A second application of the above construction of $T^1[\gamma ]$ is to the
discussion of their completeness as momentum variables. Let us, to begin
with, ignore the constraint $F_{ab}^I= 0$ and concentrate just on the
Gauss constraint. This part of the discussion will be therefore applicable
also to gauge theories. Now, the effective configuration space is the
gauge equivalence classes of connections on $\Sigma$ and the $T^0[\gamma ]$
are well defined functions on this space. Indeed, one generally thinks
of traces of holonomies as containing ``all the gauge invariant information''
that a connection has. Had the gauge group been $SO(3)$ rather than
$SO(2,1)$, we could have labelled {\it any} gauge equivalence class
of connections $\A$ completely by the values that observables $T^0[\gamma ]$
take on that equivalence class. In the present case, the situation is a little
more complicated: while the holonomies themselves do carry all the gauge
invariant information, some of this is lost while taking traces. (In
particular, connections whose holonomy group is the null-rotation subgroup
of $SO(2,1)$ can not be distinguished from the zero connections if one looks
only at traces.) For our purposes, it will suffice to note just that
the $T^0[\gamma ]$ {\it can} be used to separate points almost everywhere
on the effective configuration space; completeness fails only on a set of
measure zero. (For details, see, e.g., Goldberg et al (1992).) While this
failure {\it does} have some interesting consequences in quantum theory, we
will ignore this issue in the present discussion. The construction of
$T^1[\gamma ]$ from $T^0[\gamma ]$ now implies that these two sets of
functions form an (over)complete set of (Gauss-)gauge invariant observables
on the phase space $\Gamma$. A direct proof of completeness of $T^1[\gamma]$
as (Gauss-)gauge invariant momentum variables would have been quite difficult.

Let us now bring in the constraint $F_{ab}^I = 0$ of general relativity
(which is absent in gauge theories). Since the entire set of
constraints is of first class, we can quotient the constraint surface by the
Hamiltonian vector fields generated by the constraint functions and, as in
section 2.3, pass to the reduced phase space $\bar\Gamma$. Points of
$\bar\Gamma$ represent the ``true degrees of freedom'' of 2+1 dimensional
general relativity.  Since the constraints are either linear or independent
of momenta, the quotient construction is a straightforward procedure and
the resulting $\bar\Gamma$ is again a cotangent bundle. However, while
$\Gamma$ is infinite dimensional, $\bar\Gamma$ is only finite dimensional.
This comes about because, while the configuration variable $\A$ has 6
degrees of freedom per point of $\Sigma$, we also have 6 first class
constraints per space point, given by (3.2.4). Thus, 2+1 quantum gravity
resembles quantum mechanics rather than quantum field theory; it has no
local degrees of freedom.

What structure does $\bar\Gamma$ have? Let us begin with the reduced
configuration space $\bar{\cal C}$ over which $\bar\Gamma$ is the cotangent
bundle. $\bar{\cal C}$ is the moduli space of flat $SO(2,1)$ connections:
each point of $\bar{\cal C}$ is an equivalence class $\bar A$ of flat
$SO(2,1)$ connections, where two are regarded as equivalent if they are
related by a local $SO(2,1)$ transformation. Had the gauge group been
$SO(3)$, $\bar{\cal C}$ would have had just one component. In the $SO(2,1)$
case now under consideration, however, the moduli space has several
{\it disconnected} components because, unlike $SO(3)$, the group $SO(2,1)$
admits distinct continuous subgroups and connections with different
sub-groups as holonomy groups cannot in general be deformed continuously
from one to another. (In this construction, cerain ``pathological'' points
are deleted to ensure that $\bar{\cal C}$ has a manifold structure.) Since
$\bar{\Gamma}$ is the cotangent bundle over $\bar{\cal C}$, it also has
several disconnected components.

Let us now return to the $T$-algebra. Since they are Dirac observables, the
restrictions to the constraint surface of the $T^0[\gamma ]$ and $T^1[\gamma ]$
can be projected down unambiguously to functions $\bar{T}^0[\gamma ]$ and
$\bar{T}^1[\gamma ]$ on the reduced phase space $\bar\Gamma$. Since the
$T^0$ and $T^1$ form a complete set of (Gauss-)gauge invariant functions on
$\Gamma$, it follows, in particular, that $\bar{T}^0$ and $\bar{T}^1$ form
a complete set on $\bar{\Gamma}$. Finally, these functions on $\bar\Gamma$
have an important property that is {\it not} shared by $T^0$ and $T^1$ on
the full phase space $\Gamma$:
\item{$\bullet$} {Because the connections under consideration are now {\it
flat}, the values of $\bar{T}^0[\gamma ]$ --and hence, using the
relation between $T^0[\gamma]$ and $T^1[\gamma]$, also of $\bar{T}^1
[\gamma ]$-- are left unchanged if the closed loop $\gamma$ is replaced by a
homotopic loop. Thus the $\bar{T}$-observables are labelled by the {\it
homotopy classes} of closed loops rather than by individual closed loops.}
\medskip

This concludes the discussion of the Hamiltonian formulation of 2+1
dimensional general relativity. It is apparent that the framework is
quite different from the geometrodynamical one that results from the
Einstein-Hilbert action. Mathematically, the two are equivalent (apart
from the subtlety involving degenerate metrics). However, the emphasis is
very different. Here, connections are put to forefront and the description
closely resembles the phase space formulation of gauge theories. Indeed,
we need not even ``know'' that we are speaking of a dynamical theory
of metrics; metrics are just secondary, derived objects. The basic
concepts are parallel transport and holonomies, not distances and
light cones. Primary tools are loops, primary operations are breaking,
re-routing and gluing loops. This new way of looking at general relativity
is not particularly illuminating as far as the classical theory is concerned.
However, as we will see in the next three subsections, it is significantly
better suited for quantization.

\goodbreak
\subsection{3.3}{Timeless description: Frozen formalism}%\par

Let us now implement the algebraic quantization program of section 2.4 to
arrive at the first of the three quantum descriptions of 2+1 dimensional
general relativity.

Let us make the obvious choice of elementary variables: the canonical
coordinates $A_a^I$ and ${\tw E}^a_I$ on the phase space. The complex
vector space generated by (the  smeared version of) these variables
(together with constant functions) is closed under the Poisson bracket
and is obviously ``large enough''. The quantum algebra is straightforward
to construct. The elementary quantum operators are ${\hat A}_a^I$ and
${\hat E}^a_I$, satisfying the canonical commutation relations. (There
are no algebraic relations between the elementary classical variables
whence the CCRs are the only relations to be incorporated in the quantum
algebra.) There is an obvious choice also for the  representation space $V$
unconstrained states: Let $V$ be the space of complex valued functionals
$\Psi (A)$ of the connection $\A$ and represent $\hat{A}_a^I$ by multiplication
and $\hat{E}^a_I$ by $-i\hbar {\delta/\delta A_a^I}$.

For later use, let us note the action of the $\hat{T}$ operators on $V$.
Since ${T}^0[\gamma ]$ is a classical configuration observable, its quantum
analog ${\hat T}^0[\gamma ]$ is a multiplication operator on $V$. Similarly,
since ${T}^1[\gamma ]$ is linear in momentum, its quantum analog, ${\hat T}^1
[\gamma ]$, is a Lie derivative:
\bneq
{\hat T}^0[\gamma ]\cdot\Psi ({A}) := \bar{T}^0[\gamma]({A})
 \cdot\Psi ({A}), \quad {\rm and} \quad
 {\hat T}^1[\gamma ]\cdot\Psi ({A}) := -i\hbar\Lie{{X}[\gamma ]} \Psi
 ({A}),
\eneq{(3.3.1a)}
where $X_{[\gamma]}$ is, as before, the Hamiltonian vector field on ${\cal C}$
constructed from $T^0[\gamma] (A)$. These expressions follow from the action
of $\hat{A}_a^I$ and $\hat{E}^a_I$ on $V$ and the standard choice of
factor ordering one uses in quantum mechanics on manifolds. It is
straightforward to check that the commutators of these operators are
$-i\hbar$ times the Poisson brackets of their classical analogs. This relation
between the commutators and the Poisson brackets is quite general; it holds
in any physical system for observables which are independent of or linear in
momenta.

The next step is to impose the quantum constraints and extract the space
$V_{phy}$ of physical states. This turns out to be straightforward. Since the
constraints are either linear or independent of momenta, there are no
nontrivial factor ordering or regularization problems to solve. The quantum
Gauss constraint requires that $\Psi (A)$ be gauge invariant while the second
constraint requires that $\Psi (A)$ should have support {\it only} on flat
connections. Thus, the general solution, {\it a la} Dirac, of the two quantum
constraints is a function $\Psi (\bar{A})$ on the moduli space $\bar{\cal C}$
of flat connections on $\Sigma$. The space of these $\Psi(\bar{A})$ is the
required space $V_{phy}$.

We must now find the physical operators on $V_{phy}$. Fortunately, this task
is straightforward: all of the ${\hat T}$-operators introduced above commute
with the constraints; they are all physical operators. On the physical
subspace, $V_{phy}$, their action can be written as:
\bneq
{\hat T}^0[\gamma ]\cdot\Psi (\bar{A}) := \bar{T}^0[\gamma](\bar{A})
  \cdot\Psi (\bar{A}), \quad {\rm and} \quad
  {\hat T}^1[\gamma ]\cdot\Psi (\bar{A}) := -i\hbar\Lie{\bar{X}[\gamma ]} \Psi
  (\bar{A}),
\eneq{(3.3.1b)}
where $\bar{X}[\gamma ]$ is the vector field on $\bar{\cal C}$ induced by the
vector field $X[\gamma ]$ on ${\cal C}$, or,  alternatively, it is the
Hamiltonian vector field on $\bar{\cal C}$ generated by $\bar{T}^0[\gamma ]$
via the symplectic structure $\bar\Omega$.

Our next task is to introduce an inner-product on $V_{phy}$. For this, we
want to impose the ``reality conditions''. The classical observables
$T^A[\gamma ]$, with $A=1,2$ are all real. Therefore, we want to find an
inner product on
$V_{phy}$ which makes the quantum operators ${\hat T}^A[\gamma ]$ self-adjoint.
We shall first exhibit such an inner product and then discuss the issue of its
uniqueness. Recall that the configuration space ${\cal C}$ is equipped with a
natural symplectic structure. We can pull it back to the space of flat
connections, and since it is manifestly gauge invariant, project it to the
space of their gauge equivalence classes, $\bar{A}$. The result is a symplectic
structure $\bar\Omega$ on $\bar{\cal C}$. Denote by $d\bar{V}$ the associated
Liouville volume element and introduce on $V_{phy}$ the following inner
product:
\bneq
\langle\Psi\>,\>\Phi\rangle := \lint_{\bar{\cal C}} d\bar{V}\>
\>(\overline{\Psi (\bar{A})}) \Phi (\bar{A})\>\>.
\eneq{(3.3.2)}
Denote by ${\cal H}$ the resulting Hilbert space. The ${\hat T}^0[\gamma ]$,
being multiplication operators, are obviously self-adjoint on ${\cal H}$.
Since $\bar{X}[\gamma ]$ are Hamiltonian vector fields, their action preserves
the Liouville volume element; they are divergence-free with respect to
$d\bar{V}$. Hence it follows that ${\hat T}^1[\gamma ]$ are also self-adjoint
\footnote{9}{More precisely, symmetric. The distinction between symmetric
and self-adjoint operators is too refined for the level of rigor adopted in
these notes.}. Thus, the required reality conditions have been fulfilled and
we have a $\star$-representation of the algebra of physical operators. As a
curiosity, let us ask how the diffeomorphism group of the spatial 2-manifold
$\Sigma$ acts on these quantum states. It is clear that every (smooth)
diffeomorphism on $\Sigma$ gives rise to a diffeomorphism on the reduced
configuration space $\bar{\cal C}$. (A diffeomorphism which is generated by a
vector field maps flat connections to gauge equivalent flat connections whence
the induced action on $\bar{\cal C}$ is in fact trivial. It is only the
``large diffeomorphisms'' that have non-trivial action.) Furthermore, it
follows from the expression of the symplectic structure $\Omega$ on ${\cal C}$
that $\bar\Omega$ --and hence $d\bar{V}\>$-- is invariant under these induced
diffeomorphisms. Thus, the Hilbert space ${\cal H}$ provides a unitary
representation of the {\it full} diffeomorphism group of $\Sigma$.

It turns out that this $\star$-representation of the physical algebra is
{\it reducible}. To see this, recall, first, that the moduli space
$\bar{\cal C}$ has several disconnected pieces. It is clear that the subspace
${\cal H}_k$ of quantum states with support just on one of these pieces, say
$\bar{\cal C}_k$, is mapped to itself by the quantum ${\hat T}$-algebra. Thus,
the full representation is reducible. Furthermore, the subspace ${\cal H}_k$
itself is in general reducible because the observables $T^A[\gamma ]$ form a
complete set only {\it almost} everywhere on $\bar{\cal C}$. One can show that,
on each irreducible piece, the inner product of (3.3.2) is uniquely picked
out by the reality conditions, i.e., by the requirement that each ${\hat T}^A
[\gamma ]$ be represented by a self-adjoint operator. Thus, in the technically
as well as physically appropriate sense, the reality conditions determine
the inner product.

Finally, let us check that the ``universal'' properties of the classical
$T$ observables also carry over to the quantum theory. First, since
$\bar{T}^0[\gamma ]$ --and hence, also $\bar{X}[\gamma ]$-- depend not on the
individual loop $\gamma$ but rather its homotopy class, it is clear from
(3.3.1) that ${\hat T}^0[\gamma ]$ and ${\hat T}^1[\gamma ]$ also depend only
on the homotopy class of $\gamma$. Next, let us consider the algebraic
relations in (3.2.7). Since ${\hat T}^0$ are multiplication operators, it is
clear that the identities in (3.2.7a) go over to the quantum theory simply by
replacing each $T^0[\gamma ]$ by the operator ${\hat T}^0[\gamma ]$. What is
the situation with respect to the identities in (3.2.7b) on observables
$T^1[\gamma ]$? Note first that (3.2.7a) imply that the vector fields
$\bar{X}[\gamma ]$ satisfy the following conditions:
$\bar{X}[0] =0 ; \bar{X}[\alpha ]= \bar{X}[\alpha^{-1}];
\bar{X}[\alpha\#\beta ] = \bar{X}[\beta\#\alpha ];$ and,
$\bar{X}[\alpha\#\beta ] + \bar{X}[\alpha \#\beta^{-1}] = T^0[\alpha ]
\bar{X}[\beta ] + T^0[\beta ] \bar{X}[\alpha ]$. It now immediately follows
that (3.2.7b) also carries over to quantum theory (where, in the last
equation, we must keep ${\hat T}^0$ to the left of ${\hat T}^1$.)
Thus, we have a representation of the quantum $\hat{T}$-algebra. This
representation can also be obtained directly starting with the holonomy
algebra generated by $\hat{T}[\gamma ]$, then using Gel'fand spectral
theory outlined in section 2.5, and finally choosing (2.5.6) for the
generating functional $\Gamma [\alpha ]$ with $dV$, the Liouville volume
element.

This completes the quantum description. The Hilbert space is the space
of square-integrable functions on the moduli space of flat connections
(i.e., the reduced classical configuration space.) A complete set of
observables is given by $T^0[\gamma ]$ and $T^1[\gamma ]$; elements of
both sets are labelled just by homotopy classes of closed loops. They
derive their physical meaning through the operation of parallel transport
of internal vectors, which are the $SU(1,1)$ spinors. One can devise
Bohm-Aharanov type experiments with spin-${1\over 2}$ test particles to
distinguish one quantum state from another. We have access to space but
there is no time. The description is in terms of ``constants of motion'' ,
i.e., Dirac observables. Classically, each gauge equivalence class of
solutions to the field equations is completely determined by the values taken
by $T^0[\gamma ]$ and $T^1[\gamma ]$ on it. In the language of space-time
metrics, $T^0[\gamma ]$ and $T^1[\gamma]$ provide us with a complete
set of labels on the space of solutions modulo space-time diffeomorphisms.
Quantum mechanically, the $T^0[\gamma]$ constitute a complete set of commuting
observables and these are diagonal in our chosen representation.

The situation is analogous to the one in the momentum representation for
a free relativistic particle in Minkowski space: Our $\Psi (\bar{A})$ are
analogous to the wave functions $\varphi (k)$ on the future mass shell and
our $\hat{T}^A$-observables are analogous to the angular momentum operators
generating rotations and boosts (which also constitute a complete set of
Dirac observables). In both cases, there is no time and no dynamics. Yet, in
a well-defined sense, the mathematical description is complete. What is
lacking in both cases is a satisfactory measurement theory. And, within
the present framework, it seems impossible to provide one
in the absence of time.

\goodbreak
%\vfill\break
\subsection{3.4}{Discrete elements: Pre-geometry}%\par

I would now like to present a second quantum description which is in terms of
discrete elements ---the finished picture will refer neither to time nor to
space, both of which are to arise as secondary, derived objects. The key idea
here is to use a ``loop representation'' in which states arise as functions
on a discrete set which, due to its mathematical structure, can be finally
identified with the homotopy classes of closed loops on the manifold $\Sigma$
of genus $g$. The action of our fundamental operators $\hat{T}^0$ and
$\hat{T}^1$ will be specified through combinatorial operations which exploit
the structure available on the discrete domain space of quantum states.

For a general genus $g$, this description can be arrived at using the
algebraic approach of section 2.4 in which $T^0$ and $T^1$ are chosen as
the elementary variables. However, in the general case, there many
technical subtleties --not all of which have been fully analysed--
which complicate the discussion. These can be traced back to the fact that
the effective configuration space $\bar{\cal C}$ has several disconnected
components with various topologies and, furthermore, on some of them
there are sets of measure zero where $\bar{T}^0$ fail to be complete, i.e.
the gradients of $\bar{T}^0$ fail to span the cotangent space. Since the
purpose of this section is only to illustrate how discrete structures and
combinatorics arise, I will forego the generality and make choices which are
technically as simple as possible. Also, for pedagogical reasons, I will
first arrive at the new description starting from the timeless framework of
the previous section and, at the end, state results in a way which makes
no reference to connections or indeed the 2-manifold $\Sigma$ on which
they are defined.

Let  me then restrict myself to the case in which the spatial 2-manifold,
$\Sigma$, is a 2-torus. We will begin by discussing the structure of the
moduli space $\bar{\cal C}$ of flat connections on $\Sigma$. Recall, first,
that the gauge invariant information in the connection is coded in the
holonomies it defines around closed loops. Since the connections of interest
are flat, their holonomies around a closed loop depend only on the homotopy
class of that loop. Now, the homotopy group on a 2-torus has 2
generators. Denote them by $\alpha_1$ and $\alpha_2$. Then, by the defining
relation of the homotopy group, we have $\alpha_1 \alpha_2 =\alpha_2\alpha_1$;
the group is Abelian. Hence, the homotopy class of {\it any} closed loop
$\alpha$ is labelled just by two integers, $n_1, n_2$, which tell us how many
times the loop winds around the two generators; $\alpha =\alpha_1^{n_1}
\alpha_2^{n_2}$. Fix any base point $p$ on $\Sigma$. Any flat connection on
$\Sigma$ can now be characterized just by the pair of holonomies
$U[{\alpha_1}]$ and $U[{\alpha_2}]$ around the two generators, modulo the
action of the gauge group at $p$. Since the homotopy group is Abelian, the
holonomies must also commute, $U[\alpha_1]\cdot U[\alpha_2] = U[\alpha_2]
\cdot U[\alpha_1]$, whence they are $SO(2,1)$ rotations around the {\it same}
axis. Under the action of the gauge group at $p$, the axis itself rotates
preserving only its time-like, null or space-like character. Therefore, the
gauge equivalent classes $\bar{A}$ of flat connections fall into three
disconnected sectors. A simple calculation reveals that the sector with
time-like axis has topology $S^1\times S^1$, the one with null axis has
topology $S^1$ while the one with space-like axis, $\real^2/Z^2$.
For technical simplicity, I will now restrict myself to the time-like sector
(eventhough from a geometrodynamical viewpoint, it is the space-like sector
that is more interesting.)

We can now make the discussion of the previous subsection more specific by
making the states, the inner product and the operators explicit. Let us
coordinatize the time-like component $\bar{\cal C}_t$ of the reduced
configuration space by $a_I$, where $I=1,2$, with $a_I\in [-1,1]$. Each
element $\bar A$ of $\bar{\cal C}_t$ is thus labelled by two numbers, $a_I$.
The volume element
$d\bar V$ on $\bar{\cal C}$ now turns out to be precisely $da_1\wedge da_2$.
Thus, the Hilbert space ${\cal H}_t$ is just the space of square-integrable
functions $\Psi(a_1,a_2)$ on $\bar{\cal C}_t$. Finally, it is straightforward
to work out the explicit expressions of the basic ${\hat T}$ operators. The
operators associated with the generators $\alpha_I$ of the homotopy group of
$\Sigma$ are given by:
\beginalign{
{\hat T}^0[\alpha_J ]\cdot\Psi (a_1,a_2) =& \cos(a_J\pi) \- \-
   \Psi(a_1, a_2)\cr
   {\hat T}^1[\alpha_J ]\cdot\Psi (a_1,a_2) =& 2\pi i\hbar \sin(a_J\pi)\>
 \epsilon_{IJ}\>\> {\partial\Psi (a_1, a_2)\over\partial a_I}\>,\cr}
\endalign{(3.4.1)}
where $\epsilon_{IJ}$ is the anti-symmetric symbol. Thus, ${\hat
T}^0[\alpha_J]$
commute among themselves and so do ${\hat T}^1 [\alpha_J]$. Similarly, the two
operators associated with any one generator commute with each other. The non
commuting (and hence, conjugate) pairs are ${\hat T}^0[\alpha_1]$,
${\hat T}^1[\alpha_2]$, and, ${\hat T}^0[\alpha_2]$, ${\hat T}^1[\alpha_1]$.

We are now ready to introduce the loop representation. The idea is to
follow the Rovelli-Smolin (1990) prescription to perform a transform from
wave functions of connections to those of loops via:
\bneq
\Psi [\gamma ] = \lint_{{\cal C}/{\cal G}} d\mu (A)\- \>
[T^0[\gamma](A)]\>\-  \Psi (A).
\eneq{(3.4.2)}
where ${\cal C}/{\cal G}$ is the space of connections (modulo gauge
transformations) and $d\mu (A)$ a measure on this space. (See, section 2.5).
In the case now under considerations, we can use for ${\cal C}/{\cal G}$ the
reduced configuration space $\bar{\cal C}$ and use as our measure the
Liouville volume element $d\bar{V}$. With these choices, $\Psi (A)$ becomes
the function $\Psi(a_1, a_2)$ of two variables and the loop $\gamma$ can be
labelled by two integers $(n_1, n_2)$ since, as far as $T^0[\gamma]$ is
concerned, only its homotopy class of $\gamma$ matters. Consequently,
the Rovelli-Smolin transform reduces simply to a Fourier transform:
\bneq
\Psi [n_1, n_2 ] = \lint_{-1}^{1} da_1\lint_{-1}^1 da_2\>\cos(a_1n_1+a_2n_2)
  \pi\> \Psi (a_1, a_2),
\eneq{(3.4.3)}
where we have used the fact that the trace of the holonomy of the connection
labelled by $(a_1,a_2)$ around a loop $\gamma$ labelled by
$(n_1,n_2)$ is given by: $\bar{T}^0[n_1, n_2](a_I) = \cos(n_1a_1 \pi +
n_2a_2\pi )$. Thus, the quantum states are now functions of two integers
$(n_1, n_2)$ labelling the homotopy classes of loops
\footnote{10}{Furthermore, they are invariant under a reflection
$(n_1,n_2) \to (-n_1, -n_2)$. For a general genus $g$, the quantum states
are functions on the space of homotopy classes subject to similar conditions
which follow directly from identities (3.2.7a) satisfied by traces of
holonomies.}.
The inner product can be easily expressed: $\langle\Psi ,\Psi\rangle \equiv
\sum \ovr{\Psi(n_1,n_2)}\>\Psi (n_1,n_2) < \infty$.  This provides us with the
Hilbert space $H$ in the loop representation. The basic operators can also
be expressed directly in the loop representation. Using (3.4.1) and (3.4.3)
we have:
\beginalign{
2{\hat T}^0[\alpha_1]\cdot\Psi (n_1,n_2) =&\> \Psi (n_1+1, n_2) +
\Psi (n_1 -1, n_2)\cr
2{\hat T}^1[\alpha_1]\cdot\Psi (n_1,n_2) =&\> i\hbar n_2[\Psi (n_1+1, n_2)
- \Psi(n_1 - 1 , n_2)], \cr}
\endalign{(3.4.4)}
and similarly for operators associated with the generator $\alpha_2$.

Let us now examine the final picture. The quantum states are functions on
a discrete set, that of two integers. The inner product uses the obvious
measure on this discrete set: characteristic functions of points provide an
orthonormal basis. The basic operators of the theory have a simple,
combinatorial action: the arguments of the wave functions are shifted by
$\pm 1$. There is no space, no time. Indeed, space can be regarded as a
genuinely ``derived'' concept. Given the mathematical expressions above,
one may suddenly realize that the two integers in the argument of the
wave function can be ``interpreted'' as the labels for homotopy classes of
closed loops on a 2-torus. One may then notice the expressions of the
inner product and operators and realize that we are dealing with a theory
of flat connections on a torus.

In a more general context, one would recognize the discrete set featuring
in the argument of the wave function has just the right mathematical
properties for us to identify it with the homotopy classes of a 2-manifold
of genus $g$ and the combinatorial expressions of the basic operators would
lead us the space of flat connections on this 2-manifold. We can then work
backwards and realize that what we have is a quantum theory of 2+1-dimensional
general relativity on $\Sigma\times\real$ where $\Sigma$ is a genus $g$
2-manifold. Out of discrete structures would thus arise the continuum
interpretations
\footnote{11}{As indicated in footnote 10, there are technical
caveats to all these statements. My aim here is to communicate the overall
qualitative picture since it is the general viewpoint rather than detailed
framework that serves as a guideline in 3+1 dimensional gravity.}.

\goodbreak
\subsection{3.5}{Extracting time: Dynamics unfolded}%\par

As in the previous subsection, we will assume that the spatial topology is
that of a 2-torus. However, there we arrived at a ``pre-geometry''
description by starting with the frozen formalism and {\it removing} direct
reference to space. We will now proceed in the opposite direction and {\it
supplement} the frozen description with the notion of time. Thus, now, the wave
functions will depend on {\it three} parameters rather than two, but will be
subject to one quantum constraint which can be regarded as as a Schr\"odinger
equation if one of the three arguments of the wave functions is interpreted
as time and the other two are considered as representing the true dynamical
degrees of freedom. Thus, dynamics will be born once an internal time is
isolated.

Let me first outline the conceptual structure of the argument. Recall that
the content of the classical constraints (3.2.4) is as follows. The first set,
$\D_a\tw{E}^a_I= 0$,  is just the Gauss law which tells us that the internal
$SO(2,1)$ rotations should be interpreted as gauge. The second,
$\tw{\eta}^{ab}F_{ab}^I = 0$ is the ``covariant'' version of the combined
scalar and vector constraints. Thus, we expect the information about dynamics
to be contained in the ``zeroth'' component of this constraint. There are,
however, two technical problems that need to be solved to extract this
information. First, in the expression of this constraint, it is the internal,
rather than the space-time index which is free whence the canonical
transformations generated by this constraint corresponds to ``translations''
in the internal space, which leave space-time points fixed. We need to carry
their action to space-time. Second, to split
the constraint into its scalar and vector parts, we need to carry out its 2+1
decomposition without violating the Gauss gauge invariance.  Fortunately,
both these steps can be carried out in one stroke: Replace
$\tw{\eta}^{ab}F_{ab}^I = 0$ by the pair:
\bneq
\tw{E}^a_I F_{ab}^I = 0;\quad  {\rm and} \quad \epsilon^{IJK}\tw{E}^a_I
\tw{E}^b_J\- F_{abK} = 0,
\eneq{(3.5.1)}
where $\epsilon^{IJK}$ is the 3-form on the Lie algebra of $SO(2,1)$
defiend by the natural Killing form. (One might be concerned at first about
degenerate triads in passing from (3.2.4) to (3.5.1). However, the gauge
transformations generated by (3.2.4) are such that within each gauge orbit
there is a non-degenerate $\tw{E}^a_I$, whence, in the final picture, the
degenerate triads can be in effect ignored.) The canonical transformations
generated by the first of these equations on the phase space $\Gamma$
correspond to spatial diffeomorphisms on $\Sigma$ while those generated by
the second correspond to time evolution. Thus, modulo internal rotations
generated by the Gauss law, these two constraints are equivalent to the
standard diffeomorphism and the Hamiltonian constraints in the triad version
of geometrodynamics. (We will see in section 4 that the corresponding
constraints of
4-dimensional general relativity can be cast in a similar form.) Now, the
scalar equation holds at {\it each point} of $\Sigma$; it corresponds to an
infinite number of constraint equations. The idea is to interpret the ``zero
mode'' of this equation in quantum theory as the Schr\"odinger equation. Thus,
all other equations will be used to reduce the number of free parameters in
the argument of the wave function and thus eliminated while the ``zero mode''
of the scalar constraint will be retained in the final description as the
quantum evolution equation.

The detailed implementation of this idea is considerably simplified if one
uses an ingenious technique introduced by Manojlovi\'c and Mikovi\'c (1992).
To explain this technique, let me first consider a general Hamiltonian
system with phase space $\Gamma$ and a set of first class constraints
$C_{\bf i} = 0$. Suppose $\Gamma$ admits a symplectic sub-manifold $\Gamma'$
whose intersection with the constraint surface of $\Gamma$ gives a set of
first class constraints $C'_{\bf i'} =0$. It is then possible to ask for
the relation between the two reduced phase spaces. Suppose they are naturally
isomorphic. Then, the two classical systems are equivalent and we may use
$(\Gamma' , C_{\bf{i'}})$ as a starting point for quantization. Let me rephrase
this result. Let us suppose that the defining equations of the sub-manifold
$\Gamma'$ in $\Gamma$ are $C'_{\bf\alpha'} = 0$. Then, the set of constraints
$(C'_{\bf{\alpha'}}, C'_{\bf{i'}})$ on $\Gamma$ is equivalent to the original
set $C_{\bf{i}}$.

To apply this idea to 2+1 gravity, we will again follow Manojlovi\'c and
Mikovi\'c (1992). Let us introduce on the spatial 2-torus, two periodic
coordinates $\varphi^{\und{a}}$ (with $\und{a} =1,2$), denote the coordinate
dyads by $X^a_{\und{a}}$ and co-dyads by $\chi_a^{\und{a}}$. Using this
background structure, let us construct the ``zero modes'' of the canonical
variables:
\bneq
\underline{A}_{\und{a}}^I := {1\over V}\-\lint_\Sigma d^2\varphi\-
A_a^I X^a_{\und{a}}, \quad {\rm and} \quad \underline{E}^{\und{a}}_I :=
{1\over V}\-\lint_\Sigma d^2\varphi\- (q_o)^{-{1\over 2}}
\tw{E}^a_I X^a_{\und{a}},
\eneq{(3.5.2)}
where $q_o$ is the determinant of the (background) metric defined by the
dyad $X^a_{\und a}$ and $V$ the volume of $\Sigma$ it determines.
Now, let us consider the sub-manifold $\Gamma'$ of $\Gamma$ defined
by
\bneq
{A}_a^I X^a_{\und{a}} - \underline{A}_{\und{a}}^I = 0 ,\quad {\rm and} \quad
\tw{E}^a_I \chi_a^{\und{a}} - \sqrt{q_o}\-\underline{E}^{\und{a}}_I = 0.
\eneq{(3.5.3)}
$\Gamma'$ is a six dimensional, symplectic sub-manifold of $\Gamma$;
${\und A}_{\und a}^I$ and $\und{E}^{\und a}_I$ serve as the natural
canonical coordinates. The constraint surface of $\Gamma$ intersects
$\Gamma'$ and the induces the following constraints on $\Gamma'$:
\bneq
\epsilon^{IJK} \underline{E}^{\und{a}}_J \underline{A}_{\und{a}K} = 0, \quad
{\rm and} \quad
\epsilon_{IJK} \underline{A}_{\und{a}}^J \underline{A}_{\und{b}}^K = 0, \quad
\eneq{(3.5.4)}
where we have used the fact that $\underline{A}_{\und{a}}^I$ and
$\underline{E}^{\und{a}}_I$ are constant on $\Sigma$. It is easy to check
that, although (3.5.4) seems to provide us with six constraints, only four
of them are independent whence the reduced configuration space is two
dimensional. It is in fact naturally isomorphic with the moduli space of
flat connections on $\Sigma$ ---the reduced configuration space that resulted
from (3.2.4). Thus, together, (3.5.3) and (3.5.4) provide us with a system of
constraints on $\Gamma$ which is equivalent to (3.2.4).

To go to quantum theory, let us work in the connection representation and
use quantum constraints to select physical states. However, since we want
to extract time, we should first decompose the second equation in (3.5.4)
into a vector and a scalar part as in (3.5.1). We obtain:
\bneq
\epsilon_{IJK} \underline{E}^{\und{a}I}\underline{A}_{\und{a}}^J
\underline{A}_{\und{b}}^K = 0, \quad {\rm and} \quad
\underline{E}^{\und{a}}_{[I} \underline{E}^{\und{b}}_{J]}
\underline{A}_{\und{a}}^I \underline{A}_{\und{b}}^J = 0.
\eneq{(3.5.5)}
Note that the first of these equations is already contained in the Gauss
constraint of (3.5.4). Thus, effectively we only have the three Gauss
constraints and a scalar constraint whence the system has two true degrees
of freedom. In section 3.4, we just stated this result; we now have an
explicit method to see how the counting works out.

We are now ready to impose the quantum constraints. To begin with, because of
(3.5.3), the wave functions $\Psi(A)$ depend only on the six components
of the ``zero modes'' $\underline{A}_{\und{a}}^I$ of the connection. Next,
the quantum Gauss constraint implies, as usual, that the wave function be
gauge invariant. More explicitly, if we denote $\underline{A}_{\und{a}}^I$
as the (internal) vector $\vec{A}_{\und{a}}$, the Gauss constraint
requires that the wave function $\Psi (\underline{A})$ depend only on the
three gauge invariant parameters ${\vec A_{\und{a}}}\cdot {\vec A_{\und{b}}}$,
where the ``dot'' denotes the inner-product of the two internal vector with
respect to the Killing form on the Lie algebra of $SO(2,1)$. Thus, we have
eliminated all but the scalar constraint in (3.5.5) to conclude:
$\Psi(A) \equiv \Psi(\vec{A}_{\und{a}}\cdot\vec{A}_{\und{b}})$. The physical
states are thus functions $\Psi(\vec{A}_{\und{a}}\cdot\vec{A}_{\und{b}})$ of
three variables, ${\vec A}_1\cdot {\vec A}_1$, ${\vec A}_1\cdot {\vec A}_2$
and ${\vec A}_2\cdot {\vec A}_2$, subject to just the scalar constraint.

Let us explore the structure of this last quantum constraint. It is
quadratic in momentum ---i.e., of the form $G^{\alpha\beta}p_\alpha p_\beta
= 0$. Furthermore, it is easy to check that the ``supermetric'' $G^{\alpha
\beta}$ is flat, with signature -++.  Thus, the quantum scalar constraint is
just the Klein-Gordon equation in a 3-dimensional Minkowski space. The
Cartesian coordinates $b_{\und{I}}$ of this Minowskian metric are easily
expressible in terms of the three parameters ${\vec A_{\und{a}}}\cdot
{\vec A_{\und{b}}}$. In terms of these, the constraint is simply:
\bneq
\left(-\partial_0^2 +\partial_1^2 +\partial_2^2 \right)
\Psi(b_{\und{I}}) = 0.
\eneq{(3.5.6)}

Our next task is to introduce an inner product on this space. This can be again
achieved by expressing the $T$-variables as operators on the space of
solutions $\Psi(b_{\und{I}})$ to (3.5.6) and invoking the reality conditions.
As one might expect, this leads one precisely to the positive and negative
frequency decomposition and the standard Klein-Gordon inner product on the
positive frequency solutions. (As is usual in the algebraic quantization
program, it is the condition that we have an {\it irreducible} representation
of the observable algebra that forces us to restrict ourselves just to
positive frequency fields.) The end result is that, on physical states, the
quantum scalar constraint reduces to the Schr\"odinger equation:
\bneq
i\partial_0 \Psi (b_{\und{I}}) = + \sqrt{-\partial_1^2
-\partial_2^2}\circ\Psi(b_{\und{I}}).
\eneq{(3.5.6)}
Thus $b_{\und{1}}$ and $b_{\und{2}}$ can be thought of as the true degrees of
freedom and $b_{\und{0}}$, as the internal time. We have succeeded in carrying
out Wheeler's program. On the one hand,  the quantum scalar constraint
is just a constraint that a wave function must satisfy to be admissible as a
physical state. On the other hand, {\it if we choose to interpret}
$b_{\und{0}}$ {\it as time}, the same equation can be re-interpreted as
the dynamical or evolution equation: it tells us how the dependence of the
wave functional on the true degrees of freedom changes as one moves, in
the domain space of the wave function, from one constant $b_{\und{0}}$ slice
to another.

Since the quantum scalar constraint reduces to the wave equation in a
3-dimensional Minkowski space, semi-classical states are easy to construct.
Classical solutions correspond to null geodesics in this Minkowski
space. Hence, the semi-classical states can be obtained by forming wave
packets, using a superposition of ``plane waves,''  which are sharply
peaked at specific null wave vectors. In these states, the space-time
geometry will be approximately classical.
Finally, we can use this preferred class of foliations of the ``superspace''
spanned by the three $b_{\und{I}}$ to specify a complete set of mutually
exclusive alternatives to provide the framework with a measurement theory
along the lines discussed in Jim Hartle's lectures.
\medskip

These three faces of 2+1 quantum gravity suggest how apparently distinct
goals and approaches can lead to mathematically equivalent pictures in the
final analysis. This is the vision that underlies the discussion of the
3+1-theory in the next two sections.

\goodbreak
\vfill\break

%\input [ijtpd.lh]qgmac
%\singlespace\overfullrule=0pt
\def\E{$\tw E^a{}_i$}  \def\e{$e^a{}_I$}
\def\A{$A_a{}^i$}      \def \fA{${}^4\!A_a{}^{IJ}$}
\section{4.}{${}\quad$ 3+1 General relativity as connection-dynamics}
\medskip
{\narrower\narrower\smallskip\noindent
{\sl When Henry Moore visited the University of Chicago ... I had
the occasion to ask him how one should view sculpture: from
afar or from nearby. Moore's response was that the greatest
sculptures can be viewed --indeed, should be viewed-- from all
distances since new aspects of beauty will be revealed in every
scale. ... In the same way, the general theory of relativity
reveals strangeness in the proportion at any level in which
one may explore its consequences.}
\smallskip
S. Chandrasekhar (Truth and Beauty)\smallskip}

\subsection{4.1}{Introduction}

General relativity is traditionally considered to be the theory of  dynamics
of geometry, i.e., of space-time metrics. This view lies at the heart of the
classical treatments of a diverse range of topics ranging from cosmology to
black holes, gravity waves to singularity theorems. Yet, it appears that this
view of general relativity may not be the one that is best suited for
microscopic physics at the Planck scale. Indeed, we saw in the last section
that, in 3 dimensions, the problem of quantization takes on a much more
manageable form if general relativity is cast as a theory of dynamics of
{\it connections} rather than of geometries. We will now see that the same
is true also in 4 dimensions%
\footnote{12}{Note incidentally that, chronologically, it is this 3+1 framework
that came first and served as a motivation to consider 2+1 gravity in the
language of connections. The same is true of the T-variables in section 4.3.}.
Furthermore, it will turn out that, from this new perspective, the theory
has {\it new} striking features. It's form is quite different, but once again
it has an elegance, a beauty and a strangeness in proportion.

The idea of formulating general relativity in terms of connections is
of course quite old. Indeed, since all other basic interactions are
mediated by gauge bosons, described classically by connections, it is
natural to attempt to formulate gravity along similar lines.
In previous attempts, however, one was led to {\it new} theories of
gravity, based on a Yang-Mills type action. Since these actions are
typically quadratic in curvature, one is naturally led to field equations
which are of order four (or higher) in the metric. In the present approach,
on the other hand, one wants to retain general relativity; it is {\it
Einstein's} field equations that are to be re-interpreted as governing
dynamics of a connection.

The obvious question is: which connections should one use in such a
reformulation? Because of the success of the procedure used in 3
dimensions, one might imagine using the Palatini form of the action based
on a $SO(3,1)$ connection and a tetrad but regard the connection --rather
than the metric or the tetrad-- as the ``fundamental variable''. Such a
viewpoint was adopted by a number of authors --including Einstein and
Schr\"odinger-- and was developed in detail by Kijowski and his
collaborators. However, as far as I am aware, when one tries to cast the
resulting theory in a Hamiltonian form, one is led to a Arnowitt-Deser-Misner
type framework in which the 3-metric and the extrinsic curvature --rather than
a connection-- play a fundamental role. Thus, the situation is quite different
from the one we came across in the 2+1 theory. The difference arises because
in 4 dimensions the Palatini action contains two co-tetrads, rather than just
one:
\bneq
 S_P(e, {}^4\!\omega) := {1\over G}\int d{}^4\! x\>\- \tw\eta^{abcd}\-
e_{aI} e_{bJ}\-\> {}^\star({}^4\!R_{cd}{}^{IJ}),
\eneq{(4.1.1)}
where $G$ is Newton's constant, $I,J,...$ denote the internal indices
labeling the co-tetrads, $a,b,..$ denote space-time indices and where
${}^\star({}^4\!R_{ab}{}^{IJ})$ is the dual of the
curvature tensor of the $SO(3,1)$
connection ${}^4\!\omega_a^{IJ}$. (Note that the internal indices can be
raised and lowered freely using the fixed, kinematical metric $\eta_{IJ}$ on
the internal space.) Hence, when one performs the Legendre transform, the
momentum $\tw{\Pi}^a_{IJ}$ conjugate to the connection $A_a^{IJ}$ is the dual
$\tw{\eta}^{abc} \epsilon^{IJ}{}_{KL}e_b^K e_c^L$ of a product of two
co-triads rather than of a single co-triad as in the 2+1 dimensional case.
(Here, $\tw\eta^{abc}$ is the metric independent Levi-Civita density on the
3-manifold used in the 3+1 decomposition.) The theory then has an additional
constraint --saying that the momentum is ``decomposable'' in this manner--
which spoils the first class nature of the constraint algebra. Following the
Dirac procedure, one can solve for the second class constraint and obtain new
canonical variables. It is in this elimination that one loses the connection
1-form altogether and is led to geometrodynamics. (For details, see chapters
3 and 4 in Ashtekar (1991).)

These complications disappear, however, if one requires the connection
to take values only in the {\it self dual} (or, alternatively anti-self
dual) part of $SO(3,1)$. Furthermore, the resulting connection dynamics
is technically significantly simpler than geometrodynamics. It is this
simplicity that underlies many of the recent developments in
non-perturbative quantization. Thus, the answer to the question we posed
above is: it is the use of {\it chiral connections} that simplifies the
theory. The purpose of section 4 is to elaborate on this observation.

Let me first explain what I mean by self duality here. Let us denote the
the self dual connections by ${}^4\!A_a^{IJ}$. If one begins with a Lorentz
connection ${}^4\!\omega_a^{IJ}$, the self dual connection ${}^4\!A_a^{IJ}$
can be obtained via:
\bneq
2G\-\, {}^4\!A_a^{IJ} =  {}^4\!\omega_a^{IJ} - {i\over 2}
\epsilon^{IJ}{}_{KL}\-\, {}^4\!\omega_a^{KL},
\eneq{(4.1.2)}
where $G$ is Newton's constant. (This factor has been introduced for later
convenience and plays no role in this discussion of the mathematical meaning
of self duality.) However, we will regard the self dual connections
themselves as fundamental; they are subject just to the following algebraic
condition on internal indices:
\bneq
\textstyle{1\over 2}
\epsilon^{IJ}{}_{KL}\-\, {}^4\!A_a^{KL} = i\,\ \- {}^4\!A_a^{IJ}.
\eneq{(4.1.3)}
Thus, unlike in the analysis of self dual Yang Mills fields, the notion
of self duality here refers to the {\it internal} rather than space-time
indices: to define the duality operation, we use the kinematical internal
metric $\eta_{IJ}$ (and its alternating tensor) rather than the dynamical
space-time metric (to be introduced later).

The new action is obtained simply by substituting the real $SO(3,1)$
connection ${}^4\!\omega_a^{IJ}$ by the self dual connection $A_a^{IJ}$ in
the Palatini action (modulo overall constants):
\bneq
  S(e, {}^4\!A) := \int d{}^4\! x\>\- \tw\eta^{abcd}\- \, e_{aI}\-\,
  e_{bJ}\,\-\> {}^\star({}^4\!F_{ab}{}^{IJ}),
\eneq{(4.1.4)}
where,
\bneq
{}^4\!F_{abI}{}^J := 2 \d_{[a} {}^4\!A_{b]I}{}^J + G\- \,\ {}^4\!A_{aI}
{}^M\- \, {}^4\!A_{bM}{}^J - G\- \,\ {}^4\!A_{bI} {}^M\- \,\
{}^4\!A_{aM}{}^J
\eneq{(4.1.5)}
is the field strength of the connection ${}^4\!A_{aI}{}^J$.  Thus, $G$
plays the role of the coupling constant. Note incidentally that because of
the factors of $G$, ${}^4A_{aI}{}^J$ and ${}^4\!F_{abI}{}^J$ do not have
the usual dimensions of connections and field strength. This fact will be
important later.

By setting the variation of the action with respect to \fA\- to zero we
obtain the result that \fA\- is the self dual part of the (torsion-free)
connection ${}^4\Gamma_a{}^{IJ}$ compatible with the tetrad \e\- . Thus,
\fA\- is completely determined by \e\- . Setting the variation
with respect to \e\- to zero and substituting for the connection from the
first equation of motion, we obtain the result that the space-time metric
$g^{ab} =e^a{}_I e^b{}_J\eta^{IJ}$ satisfies the vacuum Einstein's
equation. Thus, as far as the classical equations of motion are concerned,
the self dual action (4.1.4) is completely equivalent to the Palatini
action (4.1.1).

This result seems surprising at first. Indeed, since \fA\- is the self dual
part of ${}^4\!\omega_a^{IJ}$, it follows that the curvature ${}^4F_{ab}
{}^{IJ}$ is the self dual part of the curvature ${}^4\!R_{ab}^{IJ}$. Thus,
the self dual action is obtained simply by adding to the Palatini action
an extra (imaginary) term. This term is {\it not} a pure divergence. How can
it then happen that the equations of motion remain unaltered? This comes
about as follows. First, the compatibility of the connections and the
tetrads forces the ``internal'' self duality to be the same as the
space-time self duality, whence the curvature ${}^4\!F_{abI}{}^J$ can be
identified with the self dual part, on space-time indices, of the Riemann
tensor of the space-time metric. Hence, the imaginary part of the field
equation says that the trace of the dual of the Riemann tensor must vanish.
This, however, is precisely the (first) Bianchi identity! Thus, the imaginary
part of the field equation just reproduces an equation which holds in any case;
classically, the two theories are equivalent. However, the extra term does
change the definition of the canonical momenta in the Legendre transform
--i.e., gives rise to a canonical transform on the Palatini phase space--
and this change enables one to regard general relativity as a theory governing
the dynamics of  3-connections rather than of 3-geometries.

In section 4.2, we discuss this Hamiltonian framework. In section 4.3, we
introduce the 3+1 analogs of the $T$-variables which played an important
role in the 2+1 theory and analyse their properties. These ideas will
constitute the basis of quantization in section 5. As before,
the emphasis is on presenting the global picture. I will therefore skip
derivations and technical caveats. The details can be found in Ashtekar
(1991), Romano (1991), Rovelli (1991) and references contained
therein.

\goodbreak
\subsection{4.2}{Hamiltonian framework}
For simplicity, I shall restrict myself to source-free general relativity.
The framework is, however, quite robust: all its basic features remain
unaltered by the inclusion of a cosmological constant and coupling of
gravity to Klein-Gordon fields, (classical or Grassmann-valued) Dirac fields
and Yang-Mills fields with any internal gauge group.

Since in the Lorentzian signature self dual fields are necessarily complex,
it is convenient to begin with complex general relativity --i.e. by
considering complex Ricci-flat metrics $g_{ab}$ on a real 4-manifold $M$--
and take the ``real section'' of the resulting phase-space at the end.
Let $e_a^I$ then be a complex co-tetrad on $M$  and \fA\- a self dual
$SO(3,1)$ connection, and let the action be given by (4.1.4). Let us assume
that $M$ has the topology $\Sigma\times\real$ and carry out the Legendre
transform. This is a fairly straightforward procedure. The resulting canonical
variables are then complex fields on a (``spatial'') 3-manifold $\Sigma$.
To begin with, the configuration variable turns out to be a 1-form $A_a^{IJ}$
on $\Sigma$ which takes values in the self dual part of the (complexified)
$SO(3,1)$ Lie-algebra and its canonical momentum $\tw{\Pi}_a^{IJ}$
is a self dual vector density which takes values also in the self dual
part of the $SO(3,1)$ Lie algebra. The key improvement over the Palatini
framework is that {\it there are no additional constraints on the algebraic
form of the momentum}. Hence, all constraints are now first class and the
analysis retains its simplicity. For technical convenience, one can set up,
once and for all, an isomorphism between the self dual sub-algebra of the Lie
algebra of $SO(3,1)$ and the Lie algebra of SO(3). When this is done, we can
take our configuration variable to be a complex, $SO(3)$-valued connection
\A\- and its canonical momentum, a complex spatial triad \E\- with density
weight one, where `$a$' is the (co)vector index and `$i$' is the triad or the
$SO(3)$ internal index.

The (only non-vanishing) fundamental Poisson brackets are:
\bneq
\{\tw E^a{}_i(x),\>A_b{}^j(y)\}=-i\delta^a{}_b \delta_i{}^j\delta^3(x,y).
\eneq{(4.2.1)}
The geometrical interpretation of these canonical variables is as follows.
As we saw above, in any solution to the field equations, \fA\- turns out to
be the self dual apart of the spin connection defined by the tetrad, whence
\A\- has the interpretation of being a potential for the self dual part of
the Weyl curvature. \E\- can be thought of as a ``square-root'' of the
3-metric (times its determinant) on $\Sigma$. More precisely, the relation
of these variables to the familiar geometrodynamical variables, the 3-metric
$q_{ab}$ and the extrinsic curvature $K_{ab}$ on $\Sigma$, is as follows:
\bneq
GA_a{}^i = \Gamma_a{}^i - i K_a{}^i \quad {\rm and} \quad
\tw E^a{}_i \tw E^{bi} = (q) q^{ab}
\eneq{(4.2.2)}
where, as before, $G$ is Newton's constant, $\Gamma_a{}^i$ is the spin
connection
determined by the triad, $K_a{}^i$ is obtained by transforming the space
index `$b$' of $K_{ab}$ into an internal index by the triad $E^a_i :=
(1/\sqrt{q})\tw{E}^a_i$, and $q$ is the determinant of $q_{ab}$. Note,
however, that, as far as the mathematical
structure is concerned, we can also think of \A\- as a (complex)
$SO(3)$-Yang-Mills connection and \E\- as its conjugate electric field.
Thus, the phase space has a dual interpretation. It is this fact that enables
one to import into general relativity and quantum gravity ideas from Yang-Mills
theory and quantum chromodynamics and may, ultimately, lead to a unified
mathematical framework underlying the quantum description of all fundamental
interactions. In what follows, we shall alternate between the interpretation
of \E\- as a triad and as the electric field canonically conjugate to the
connection \A .

Since the configuration variable \A\- has nine components per space point and
since the gravitational field has only two degrees of freedom, we expect
seven first class constraints. This expectation is indeed correct. The
constraints are given by:
$$\eqalignno{
&{\cal G}_i(A,\tw{E}) :=\D_a \tw E^a{}_i=0,&(4.2.3)\cr
&{\cal V}_a (A,\tw{E}) := \tw E^b{}_i\- F_{ab}{}^i\equiv \tr E\times B
 =0,&(4.2.4)\cr
&{\cal S} (A,\tw{E}) := \epsilon^{ijk}\tw E^a{}_i\-\tw E^b{}_j\-\,F_{abk}
\equiv \tr E\times E\cdot B =0,&(4.2.5)\cr}$$
%\eneq
where, $F_{ab}{}^i:=2\d_{[a} A_{b]}{}^i+G\epsilon^{ijk} A_{a{}j} A_{b{}k}$
is the field strength constructed from \A , $B$ stands for the magnetic
field $\tw\eta^{abc}F_{bc}^i$, constructed from $F_{ab}^i$, and $\tr$ refers
to the standard trace operation in the fundamental representation of
$SO(3)$. Note that
all these equations are simple polynomials in the basic variables; the worst
term occurs in the last constraint and is only quadratic in each of \E\- and
\A . The three equations are called, respectively, the Gauss constraint, the
vector constraint and the scalar constraint. The first, Gauss law, arises
because we are now dealing with triads rather than metrics. It simply tells
us that the internal $SO(3)$ triad rotations are ``pure gauge''. Modulo
these internal rotations, the vector constraint generates spatial
diffeomorphisms on $\Sigma$ while the scalar constraint is responsible for
diffeomorphisms in the ``time-like directions''. Thus, the overall situation
is the same as in triad geometrodynamics.

{}From geometrical considerations we know that the ``kinematical gauge group''
of the theory is the semi-direct product of the group of local triad rotations
with that of spatial diffeomorphisms on $\Sigma$. This group has a natural
action on the canonical variables \A\- and \E\- and thus admits a natural
lift to the phase-space. This is precisely the group formed by the canonical
transformations generated by the Gauss and the vector constraints. Thus, six
of the seven constraints admit a simple geometrical interpretation. What
about the scalar constraint? Note that, being quadratic in momenta, it is
of the form $G^{\alpha\beta} p_\alpha p_\beta=0$ where, the connection
supermetric $\epsilon^{ijk}F_{ab k}$ plays the role of $G^{\alpha\beta}$.
Consequently, the motions generated by the scalar constraint in the phase
space correspond precisely to the {\it null geodesics of the ``connection
supermetric''.} As in geometrodynamics, the space-time interpretation of
these canonical transformations is that they correspond to ``multi-fingered''
time-evolution. Thus, we now have an attractive representation of the Einstein
evolution as a null geodesic motion in the (connection) configuration space
\footnote{13}{At first sight, it may appear that this interpretation requires
$G^{\alpha\beta}$ to be non-degenerate since it is only in this case that one
can compute the connection compatible with $G^{\alpha\beta}$ unambiguously
and speak of null geodesics. However, in the degenerate case, there exists
a natural generalization of this notion of null geodesics which is routinely
used in Hamiltonian mechanics.}.

If $\Sigma$ is spatially compact, the Hamiltonian is given just by a linear
combination of constraints. In the asymptotically flat situation, on the
other hand, constraints generate only those diffeomorphisms which are
asymptotically identity. To obtain the generators of space and time
translations, one has to add suitable boundary terms. In a 3+1 framework,
these translations are coded in a lapse-shift pair. The lapse --which tends
to a constant value at infinity-- tells us how much of a time translation we
are making while the shift --which approaches a constant vector field at
infinity-- tells us the amount of space-translation being made. Given a lapse%
\footnote{14}{In this framework, the lapse naturally arises as a scalar density
$\ut{N}$ of weight $-1$. It is $\ut{N}$ that is the basic, metric independent
field. The ``geometric'' lapse function $N$ is metric dependent and given by
$N=\sqrt{q}\ut{N}$. Note also that, unlike in geometrodynamics, Newton's
constant never appears explicitly in the expressions of constraints,
Hamiltonians, or equations of motion; it features only through the expression
for $F_{ab}{}^i$ in terms of the connection.}$\ut{N}$ and a shift $N^a$, the
Hamiltonian is given by:
 $$\eqalignno{
 H(A,\tw E) = i \lint_{\Sigma} d^3 x \>(N^a& F_{ab}{}^i\tw E^b{}_i
 -\textstyle{i\over 2}\ut{N} \epsilon^{ijk}F_{ab k} \tw E^a{}_i
 \tw E^b{}_j)&\cr
 & -\loint_{\d\Sigma} d^2S_a\>(\ut{N} \epsilon^{ijk} A_{bk} \tw E^a{}_i \tw
 E^b{}_j + 2 i N^{[a} \tw E^{b]}{}_i A_b{}^i).&(4.2.6)\cr}$$
%\eneq
The dynamical equations are easily obtained since the Hamiltonian is also
a low order polynomial in the canonical variables. We have
\beginalign{
\dot{A}_a^i &= -i\epsilon^{ijk}\ut{N}\tw{E}^b_jF_{ab}{}_{k} - N^bF^i_{ab}\cr
\dot{E}^a_i &= i\epsilon_i^{jk}\D_b(\ut{N}\tw{E}^a_j \tw{E}^b_k) -
 2\D_b(N^{[a}\tw{E}^{b]i})\cr}
\endalign{(4.2.7)}
Again, relative to their analogs in geometrodynamics, these equations are
significantly simpler.

So far, we have discussed {\it complex} general relativity. To recover the
Lorentzian theory, we must now impose reality conditions, i.e., restrict
ourselves to the real, Lorentzian section of the phase-space. Let me explain
this point by means of an example. Consider a simple harmonic oscillator.
One may, if one so wishes, begin by considering a complex phase-space spanned
by two complex co-ordinates $q$ and $p$ and introduce a new complex
co-ordinate $z= q - ip$. ($q$ and $p$ are analogous to the triad \E\- and
the extrinsic curvature $K_a{}^i$, while $z$ is analogous to \A .) One can
use $q$ and $z$ as the canonically conjugate pair, express the Hamiltonian
in terms of them and discuss dynamics. Finally, the real phase-space of the
simple harmonic oscillator may be recovered by restricting attention to those
points at which $q$ is real and $ip = q-z$ is pure imaginary (or,
alternatively, $\dot{q}$ is also real.) In the present phase-space
formulation of general relativity, the situation is analogous. In terms of
the familiar geometrodynamic variables, the reality conditions are simply
that the 3-metric be real and the extrinsic curvature --the time derivative
of the 3-metric-- be real. If these conditions are satisfied initially, they
continue to hold under time-evolution. In terms of the present canonical
variables, these become: {\it i}) the (densitized) 3-metric $\tw E^a{}_i
\tw E^{bi}$ be real, and, {\it ii}) its Poisson bracket with the Hamiltonian
$H$ be real, i.e.,
$$\eqalignno{
 (\tw E^a{}_i \tw E^{bi})^\star &= \tw E^a{}_i \tw E^{bi}&(4.2.8)\cr
 [\epsilon^{ijk}\tw E^{(a}{}_i \D_c(\tw E^{b)}{}_k\tw E^c{}_j )]^\star
 &= - \epsilon^{ijk}\tw E^{(a}{}_i \D_c(\tw E^{b)}{}_k\tw E^c{}_j),
 &(4.2.9)\cr}$$
%\eneq
where $\star$ denotes complex-conjugation. (Note, incidentally, that
in Euclidean relativity, these conditions can be further simplified since
self dual connections are now real: The reality conditions require only
that we restrict ourselves to real triads and real connections.) As far
as the classical theory is concerned, we could have restricted to the
``real slice'' of the phase-space right from the beginning. In quantum theory,
on the other hand, it is simpler to first consider the complex theory, solve
the constraint equations and then impose the reality conditions as suitable
Hermitian-adjointness relations. Thus, the quantum reality conditions would
be restrictions on the choice of the inner-product on physical states.

Could we have arrived at the phase-space description of {\it real} general
relativity in terms of (\A\- , \E\-) without having to first complexify the
theory? The answer is in the affirmative. This is in fact how the new
canonical variables were first introduced (Ashtekar (1987).) The idea is
to begin with the standard Palatini action for real tetrads and real
Lorentz-connections, perform the Legendre transform and obtain the phase-space
of real relativity {\it a la } Arnowitt, Deser and Misner. The basic canonical
variables in this description can be taken to be the density weighted triads
\E\- and their canonical conjugate momenta $\pi_a^i$. The interpretation of
$\pi_a^i$ is as follows: In any solution to the field equations, i.e.,
``on shell,'' $K_{ab}:= \pi_{(a}^i E_{b)i}$ turns out to be the extrinsic
curvature. Up to this point, all fields in question are real. On this real
phase space, one can make a (complex) canonical transformation to pass to the
new variables: $(\tw{E}^a_i, \pi_a^i)\to (\tw{E}^a_i , GA_a^i := \Gamma_a^i -
i \pi_a^i \equiv (\delta F/\delta\tw{E}^a_i) - i \pi_a^i)$, where the
generating function $F(\tw{E})$ is given by: $F(\tw{E}) = \lint_{\Sigma} d^3x
\tw{E}^a_i \Gamma_a^i$, and where $\Gamma_a^i$ are the spin coefficients
determined by the triad \E\-. Thus, \A\- is now just a complex coordinate on
the traditional, real phase space. This procedure is completely analogous to
the one which lets us pass from the canonical coordinates $(q,p)$ on the phase
space of the harmonic oscillator to another set of canonical coordinates
$(q, z = dF/dq - ip)$, with $F(q) = \textstyle{1\over2}q^2$, and makes the
analogy mentioned above transparent. Finally, the second of the reality
conditions, (4.2.9), can now be re-expressed as the requirement that $GA_a^i
- \Gamma_a^i$ be purely imaginary, which follows immediately from the
expression of \A\- in terms of the real canonical variables $(\tw{E}^a_i ,
K_a^i)$.

\medskip
I will conclude this subsection with a few remarks.
\item{i)} In broad terms the Hamiltonian framework developed above
is quite similar to the one discussed in section 3 (see especially
section 3.5) for the 2+1 theory:
in both cases, the configuration variable is a connection, the momentum
can be regarded as a ``square root'' of the metric, and the constraints,
the Hamiltonian and the equations of motion have identical form. However,
there are a number of key differences. The first is that the connection
in the 3+1 theory is the self dual $SO(3,1)$ spin-connection while that in
the 2+1 theory is the real $SO(2,1)$ spin-connection. More importantly, while
the scalar and the vector constraints together imply in the 2+1-theory
that the connection is flat, a simple counting argument shows that there is
no such implication in the 3+1 theory. This is the reason behind the key
difference between the two theories: Unlike in the 2+1 case, the 3+1 theory
has {\it local} degrees of freedom and hence gravitons.
\item{ii)} A key feature of this framework is that all equations of the
theory --the constraints, the Hamiltonian and hence the evolution equations
and the reality conditions-- are simple polynomials in the basic variables
\E\- and \A . This is in striking contrast to the ADM framework where the
constraints and the evolution equations are non-polynomial in the basic
canonical variables. An interesting --and potentially, powerful--
consequence of this simplicity is the availability of a nice algorithm to
obtain the ``generic'' solution to the vector and the scalar constraint
(Capovilla et al 1989). Choose any connection \A\- such that its magnetic
field $\tw B^a{}_i := \tw\eta^{abc} F_{bci}$, regarded as a matrix, is
non-degenerate. A ``generic'' connection \A\- will satisfy this condition;
it is not too restrictive an assumption.  Now, we can expand out \E\- as
\E = $M_i{}^j \tw B^a{}_j$ for some matrix $M_i{}^j$. The pair (\A\- , \E\ )
then satisfies the vector and the scalar constraints {\it if and only if}
$M_i{}^j$ is of the form $M_i{}^j = [\phi^2 -\textstyle{1\over 2} \tr
\phi^2 ]_i{}^j$, where $\phi_i{}^j$ is an arbitrary trace-free, symmetric
field on $\Sigma$. Thus, as far as these four constraints are concerned,
the ``free data'' consists of \A\- and $\phi_i{}^j$.
\item{iii)} The phase-space of general relativity is now identical to
that of complex-valued Yang-Mills fields (with internal group $SO(3)$).
Furthermore, one of the constraint equations is precisely the Gauss law
that one encounters on the Yang-Mills phase-space. Thus, we have a natural
embedding of the constraint surface of Einstein's theory into that of
Yang-Mills theory: Every initial datum ($A_a^i,$ \E ) for Einstein's theory
is also an initial datum for Yang-Mills theory which happens to satisfy,
in addition to the Gauss law, a scalar and a vector constraint. From the
standpoint of Yang-Mills theory, the additional constraints are {\it the
simplest diffeomorphism and gauge invariant expressions one can write down
in absence of a background structure such as a metric} (Bengtsson 1989).
Note that the
degrees of freedom match: the Yang-Mills field has $2$ (helicity) $\times 3$
(internal) $= 6$ degrees and the imposition of four additional first-class
constraints leaves us with $6-4 =2$ degrees of freedom of Einstein's theory.
I want to emphasize, however, that in spite of this close relation of the
two initial value problems, the Hamiltonians (and thus the dynamics) of the
two theories are {\it very} different. Nonetheless, the similarity that
does exist can be exploited to obtain interesting results relating the two
theories (See, e.g. the review by Samuel (1991)).
\item{iv)} Since all equations are polynomial in \A\- and \E\- they continue
to be meaningful even when the triad (i.e. the ``electric field'') \E\-
becomes degenerate or even vanishes. As in the 2+1 theory, this feature
enables one to use in quantum theory a representation in which states are
functionals of \A and $\hat{E}^a_i$ is represented simply by a functional
derivative with respect to \A , thereby shifting the emphasis from triads
to connections. In fact, Capovilla, Dell and Jacobson (1989) have introduced
a Lagrangian framework which reproduces the Hamiltonian description discussed
above but which {\it never even introduces a space-time metric}! This
formulation of ``general relativity without metric'' lends strong support
to the viewpoint that the traditional emphasis on metric-dynamics, however
convenient in classical physics, is not indispensable.

This completes the discussion of the Hamiltonian description of general
relativity which casts it as a theory of connections. We have transformed
triads from configuration to momentum variables and found that self dual
connections serve as especially convenient configuration variables. In
effect, relative to the Arnowitt-Deser-Misner geometrodynamical
description, we are looking at the theory ``upside down'' or ``inside
out''. And this unconventional way of looking reveals that the theory has
a number of unexpected and, potentially, profound features: it is much closer
to gauge theories (particularly the topological ones) than was previously
imagined; its constraints are the simplest background independent expressions
one can write down on the phase space of a gauge theory; its dynamics has a
simple geometrical interpretation on the space of connections; etc. It opens
new doors particularly for the task of quantizing the theory. We are led to
shift emphasis from metrics and distances to connections and holonomies and
this, in turn suggests fresh approaches to unifying the mathematical framework
underlying the four basic interactions (see, in particular, Peldan (1992)).

\goodbreak
\subsection{4.3}{Rovelli-Smolin Loop variables}

Let us now introduce the analogs of the 2+1 $T$-variables and analyse
their properties. As one might expect, they will play an important role
in the quantum theory.

Consider continuous, piecewise analytic mappings $\alpha: S^1 \to \Sigma$
from a circle to the spatial 3-manifold $\Sigma$. Each of these mappings
gives us a parametrized, closed curve on $\Sigma$. We will denote these
curves by $\alpha(s), \beta(t),..$, let the parameters $s,t,...$ take
values modulo $1$ and call them {\it loops}. The inverse of $\alpha$ will be
defined to be the loop $\alpha^{-1}(s) = \alpha (1-s)$. The equivalence
class of such closed curves, where any two defer only by an orientation
preserving reparametrization, will be referred to as an {\it unparametrized
loop}. While we will generally use loops in the intermediate steps, the final
results will not be sensitive to the choice of parametrization. Such
technicalities
are necessary also in 2+1 dimension. We skipped them because we could
pass quickly to the homotopy classes of loops, which in turn was possible
because the constraints required that the connections be flat. In the
3+1 theory now under consideration, the connections are {\it not} forced
by the constraints to be flat; there are local degrees of freedom. Hence,
for a significant part of the discussion, we have to deal directly with
loops themselves and now the details given above become relevant. Finally,
in 3 --and only in 3-- dimensions, we have the interesting phenomenon of
knotting and linking of loops. As we saw in section 2.5, already in the
case of a Maxwell field, linking numbers play an important role in the
quantum description. We will see in the next section that in 3+1 gravity
(and, although we will not discuss them here, also in certain topological
field theories) both knot and link invariants play a key role.

As in 2+1 gravity, the configuration variables $T[\gamma ]$ are just the
traces of holono\-mi\-es. Let $\sigma^j$ denote the three $2\times 2$ Pauli
matrices and set $\tau^j = (1/2i)\sigma^j$. We will use these matrices to
pass to the 2-dimensional representation of (complexified) $SO(3)$. Thus,
given the connection 1-form $A_a^i$, we can construct a $2\times 2$ matrix
$A_a := A_a^j\tau_j$ in the Lie algebra of $SL(2,C)$ and use it in the
expression of the parallel-transport operation:
\bneq
U_\gamma(s,t) := {\cal P}\- \exp \big(G\-\- \lint_{\gamma(s)}^{\gamma(t)}\-\-
dS^a A_a \-\big),
\eneq{(4.3.1)}
where $\gamma(s)$ and $\gamma(t)$ are two points on $\gamma$ and, as before,
${\cal P}$ stands for ``path-ordered'' and where the factor of $G$ is
necessary because it is $GA_a$ that has the dimensions of inverse length.
$U_\gamma (s,t)$ is
reparametrization invariant; it depends only on the points $\gamma(s)$
and $\gamma(t)$ fixed on the loop $\gamma$.) If $t = s+1$, we
obtain a group element at $\gamma(s)$ which encodes the parallel transport
operation along the closed loop $\gamma$. We denote it simply by
$U_\gamma(s)$; this is the holonomy defined by $A_a$ around the closed loop
$\gamma$, evaluated at $\gamma(s)$. Our configuration variable is
simply the trace of this $SL(2,C)$ matrix:
\bneq
T[\gamma ](A) := {1\over 2}\-\tr\- U_\gamma (s).
\eneq{(4.3.2)}
Using  (4.3.1) and the cyclic property of the trace operation, it is easy
to verify that the trace can be evaluated at {\it any} point along the loop;
its value depends only on the loop $\gamma$ and not on the point $\gamma(s)$
used in the evaluation. As noted in section 3, these configuration variables
are automatically gauge invariant. Thus, if ${\cal C}$ denotes the space
of all (suitably regular) connections $A_a$ on $\Sigma$ and ${\cal C}/
{\cal G}$ its quotient by the {\it local} $SL(2,C)$ transformations,
$T[\gamma]$ can be naturally projected down to ${\cal C}/{\cal G}$.
Furthermore, they form a complete set  --in the sense that their gradients
span the cotangent space to ${\cal C}/{\cal G}$ -- almost  everywhere
(Goldberg et al 1992). As remarked in section 3, had the gauge group been
$SO(3)$ or $SU(2)$, these variables would be complete. It is the presence
of the null rotation subgroup of $SL(2,C)$ that makes them incomplete on
sets of measure zero. Finally, as in 2+1 theory, where ever the $T[\gamma
]$ are complete, they are in fact overcomplete; there are algebraic
identities between them which should be carried over to the quantum theory.
These are:
\beginalign{
T[0]& = 1, \qquad T[\gamma] = T[\gamma^1],\qquad T[\gamma\circ\beta]
= T[\beta\circ\gamma ]\cr
&{\rm and} \quad 2T[\gamma ]\cdot T[\beta] = T[\gamma \sharp\beta ] +
T[\gamma \sharp \beta^{-1}],\cr}
\endalign{(4.3.3)}
where, as before $\gamma \circ \beta$ is the composition of two loops at an
intersection point (so the third relation is empty if the loops dont
intersect) and $\gamma\sharp\beta $ is the ``eye-glass'' loop obtained by
joining the two loops by a line segment (whence the fourth relation {\it is}
relevant even if the loops do not intersect). The right hand side of the
last equation is independent of the choice of the line segment used to
join the loop, provided, of course the {\it same} segment is used in both
terms.

Next, we wish to introduce variables which are linear in momenta \E .
Since \E\- has a free internal index, it is gauge covariant rather than
gauge invariant. To obtain a gauge invariant momentum variable, we need
to ``tie'' this index with that of some other gauge covariant, momentum
independent object. The obvious choice is the holonomy $U_\gamma(s)$
around a closed loop $\gamma$. However, unlike its trace $T[\gamma ]$,
$U_\gamma(s)$ ``sits'' at a specific point $\gamma(s)$ of the loop $\gamma$.
Hence, the electric field must ``sit'' at the same point. Thus, the
momentum variable is associated with a closed loop $\gamma$ together with
a marked point $\gamma(s)$ on it. Given this pair, we set:
\bneq
T^a[\gamma; s] (A,\tw{E}) := \textstyle{1\over 2} \tr \big( U_\gamma (s)
\tw{E}^a (\gamma(s))\big),
\eneq{(4.3.4)}
where $\tw{E}^a = \tw{E}^a_i\tau^i$ and where, for notational simplicity,
I have dropped the tilde over $T^a$ even though it is a density of weight
one. By construction, these variables are also (Gauss) gauge invariant.
It is therefore convenient to reduce the phase space with respect to the
Gauss constraint --i.e., to first restrict ourselves to the (\A\- ,\E\- )
pairs which satisfy the Gauss constraint and then factor out by the
canonical transformations generated by this constraint. Since the constraint
is linear in momenta, the reduced phase space is again a cotangent bundle
and the reduced configuration space is just ${\cal C}/{\cal G}$. The analysis
of Goldberg, Lewandowski and Stornaiolo (1992) shows that the $T[\gamma]$
and $T^a[\gamma; s]$ provide us with an overcomplete set almost
everywhere on this reduced phase space. The ``marked ''point on $\gamma$ at
which $\tw{E}^a$ sits is called a {\it hand}. In various calculations,
--such as that of the Poisson bracket between $T^a[\gamma; s]$ and
$T[\beta]$-- non-trivial terms generally result only if another loop
--$\beta$ in our example-- intersects $\gamma$ {\it at the marked point}
$\gamma(s)$. In this case, one says {\it the hand on} $\gamma$ {\it has
grasped} $\beta$.

In contrast to
the 2+1 dimensional case, the momentum variable is now a function
on the phase-space which takes values in vector densities of weight one
rather than complex numbers and depends not only on the loop but also a point
on the loop. As we will see, the form in (4.3.4) is convenient in practice
when dealing with constraints. However, in a proper mathematical treatment
along the lines sketched in section 2.5, one must begin with a genuine
algebra of operators which in turn means that the elementary classical
variables should be genuine complex-valued functions on the phase space
yielding a true Lie algebra under the Poisson bracket operation. For these
purposes, it is convenient to smear $T^a[\gamma; s]$ appropriately. This
can be done as follows. Note first that, since the $T^a$ are vector densities
and since vector densities in three dimensions are dual to 2-forms, it is
now natural to smear them on 2-surfaces. Further, since we must get rid of
the marked point on the loop, smearing should contain in it an integration
over the loop; the loop itself should be contained in the 2-surface. These
considerations lead us to the following construction. Consider a continuous,
piecewise analytic mapping from $S^1\times (-\epsilon,\epsilon)$ to the
3-manifold. Its image in $\Sigma$ is a ribbon which comes with
a parametrization $(\sigma,\tau )$, with $\sigma \in [0,1]$ and $\tau
\in (-\epsilon, \epsilon)$. These parameters will be convenient in the
intermediate steps of our calculations. However, just as the variable
$T[\gamma ]$ and its properties are invariant under the reparametrizations
of the loop $\gamma$, the smeared momentum variable $T[S]$ will be
insensitive to the values of the parameters $\sigma, \tau$. The invariant
structure will be the $S^1\times \real$ topology of $S$, its foliation
by a family of closed curves and its orientation ($d\sigma\wedge d\tau$).
The equivalence class of ribbons $S$ where any two are considered as being
equivalent if they differ only by an orientation preserving reparametrization
will be called a {\it strip}. Given a strip $S$, we define:
\bneq
T[S](A,E) := \lint_S\- dS^{ab}\-\- \eta_{abc}\-\- T^c[\tau; \sigma] ,
\eneq{(4.3.5)}
where the closed loops needed in the definition of $T^c$ are provided by
$\tau = {\rm const}$ circles. These are now gauge invariant, complex-valued
functions on the phase space, which are linear in momenta. As in the 2+1
theory, the algebraic relations (4.3.3) in turn imply that the strip
variables are also algebraically constrained:
\bneq
T[S] = 0, \quad T[S] = T[S^{-1}] \quad {\rm and} \quad
T[S\circ S'] = T[S'\circ S],
\eneq{(4.3.6)}
where $S\circ S'$ is the strip obtained by composing the two strips at
the points of their intersection (the portions of either strips which
are not ruled by circles passing through the intersection points being
discarded.) As in the 2+1 theory, the momentum variables $T[S]$ can be
``derived'' from the $T[\gamma ]$ although the construction is somewhat
more involved. This derivation can again be used to obtain (4.3.6) from
(4.3.3). (See chapter 15 in Ashtekar (1991).)
There is, however, one difference. In the 2+1 theory, the momentum
variables $T^1$ are also associated with loops and the last equation in
(4.3.3) leads to an additional algebraic constraint on the $T^1$ variables.
In the 3+1 theory, due to the asymmetry between loops and strips, we do not
obtain such a relation.

The above construction of the loop-strip variables, as well as their
properties has a universality. For gauge fields in $n$ (spatial) dimensions,
the configuration variables are associated with closed loops while
the momentum variables are associated with $n$-1 dimensional strips.
In 2+1 dimensional theories, the numbers conspire and the momenta are again
associated with closed loops. Independent of the dimension $n$, the
loop-strip variables are closed under the Poisson bracket. Furthermore,
the operation of taking these brackets corresponds simply to breaking,
re-routing and joining the loops and the strips%
\footnote{15}{Thus, had we required that the loops and strips be smooth, the
$T[\gamma ]$ and $T[S]$ would {\it not} have been closed under the Poisson
bracket. This is why we allowed them to be non-differentiable at isolated
points. We required them to be piecewise analytic rather than, say,
$C^\infty$ to ensure that the sums on the right side of the Poisson brackets
range over a finite number of terms.}:
\beginalign{
\{ T[\gamma ], T[\beta ]\} & = 0; \qquad
\{ T[S], T[\gamma ]\} = \sum_i \Delta_i(S,\gamma )( T[\gamma\circ_i
\tau_i] - T[\gamma\circ_i\tau_i^{-1}]) \cr
&\{T[S], T[S'] \} = \sum_i \Delta_i(S, S')(T[S\circ_i S'] - T
[S\circ_i (S')^{-1}],\cr}
\endalign{(4.3.7)}
where the subscript $i$ labels the intersections between $\gamma$ and $S$
or, $S$ and $S'$; $\tau_i$ is the closed loop on $S$ passing through the
$i$-th intersection between $\gamma$ and $S$; $\circ_i$ denotes the
composition of the loops or strips at the $i$-th intersection; and,
$\Delta_i$ equals $\pm {1\over 2}$ depending on the orientation of the two
arguments at the intersection. Thus, the structure we encountered in the
Poisson brackets between $T$ variables in 2+1 dimensions is not
accidental.

There {\it is} however a key property of the 2+1 loop variables which is not
shared by the 3+1 ones: The $T[\gamma ]$ and the $T[S]$ are no longer Dirac
observables. While they continue to strongly commute with the
Gauss constraints, in the 3+1 theory, they fail to commute even weakly
with the vector and the scalar constraints. This failure can be traced
back to the fact that the gravitational field now has {\it local} degrees
of freedom. At a technical level, it is this failure that causes the
key difficulties in the quantization of the 3+1 theory and explains why, in
contrast to the 2+1 theory, the program there is still so far from
completion. From a physical viewpoint, however, this feature lies at the
heart of the richness of the 3+1 theory.

The $T[\gamma]$ and the $T[S]$ serve as the point of departure for
the quantization problem in a proper mathematical setting, e.g., in the
spirit of section 2.5. Thus, one may first construct a $C^\star$ algebra
and then analyse its representations systematically. However, given
a representation, one is still left with the task of introducing physically
interesting operators, such as the quantum analogs of constraints. At this
stage, the unsmeared form (4.3.4) of the momentum variable (as well as its
analogs which are quadratic or higher order in momenta) are often more
directly useful. We will conclude with a discussion of how this comes about.

Let us begin with the vector constraint on the classical phase space.
(Note incidentally that both the vector and the scalar constraints commute
with the Gauss constraint and therefore admit a well-defined projection
to $T^\star({\cal C}/{\cal G})$, the reduction of the phase space with
respect to the Gauss constraint. Since the $T$-variables are also admit
this projection, the entire analysis can be carried out on this reduced
space.) Fix a point $p$ on $\Sigma$ and let $x^i$ be local coordinates
centered at $p$. Let $\gamma_{12}^\delta$ be a loop in $\Sigma$ of
coordinate area $\delta$, centered at $p$, lying entirely in the $1-2$
coordinate plane. Then, using the definition of the path ordered exponential,
it follows that:
\beginalign{
U [\gamma^\delta{}_{12}] &= 1 + \delta F_{12}(p) + o(\delta),
\quad {\rm whence},\cr
T^2[\gamma^\delta_{12}] &= {\delta\over 2} \, \tr \- F_{12}(p)\tw{E}^2(p)
+o(\delta),\cr}
\endalign{(4.3.8)}
where $o(\delta)$ denotes terms which vanish in the limit faster than
$\delta$. Hence, the vector constraint can now be expressed as a sum of
three terms:
\bneq
{\cal V}_a (p) := \tr F_{ab}(p)\tw{E}^b(p) = \lim_{\delta\to 0}\-
{2\over\delta}\- \sum_a T^b[\gamma^\delta_{ab}; 0],
\eneq{(4.3.9)}
involving three loops centered at $p$ in the 3-planes defined by the
coordinates. To express the scalar constraint in a similar fashion, it
is convenient to introduce a loop variable $T^{a,a'}$ which is quadratic
in momenta:
\bneq
T^{aa'}[\gamma; s,s'](A,E) = \textstyle{1\over 2}\tr \big( U_\gamma (s',s)
\tw{E}^a(\gamma(s))U_\gamma(s,s')\tw{E}^{a'}(\gamma(s'))\big).
\eneq{(4.3.10)}
This is again gauge invariant. These variables can be used to express
the scalar constraint:
\bneq
{\cal S}(p) := \tr F_{ab}(p)\tw{E}^a(p)\tw{E}^b(p) = \lim_{\delta\to 0}\-
{1\over\delta}\- \sum_{a,a'} T^{[aa']}[\gamma^\delta{}_{aa'};
0, \delta ],
\eneq{(4.3.11)}
as well as the density weight 2, contravariant metric $\tw{\tw{q}}{}^{ab}
\equiv \tw{E}^a_i \tw{E}^{bi}$ :
\bneq
-4\- \, \tw{\tw{q}}{}^{aa'} (p) := \lim_{\delta\to 0}\-
T^{aa'}[\gamma ; 0,\delta ].
\eneq{(4.3.12)}

{}From the point of view of the classical theory, in spite of their manifest
gauge invariance, the loop variables are rather awkward to use: they
are non-local in their dependence on $\Sigma$ and, consequently, one has
to resort to limiting procedures to extract even the most basic functions
on the phase space such as the 3-metric and the constraints. Furthermore,
under the evolution generated by the Hamiltonians, each of the loop
variables is mapped to functions on the phase space which can be
expressed only as a limit of an infinite sum of loop variables; the
vector space generated by these variables is not stable under evolution
of general relativity or even the linearized approximation thereof.
Consequently, so far, these variables have played virtually no role at all
in classical general relativity. However, their weakness turns to strength
in the process of quantization. For, we expect expressions such as the
3-metric or the constraints which involve products of \A\- and \E\- at the
{\it same point} to give rise to divergent operators in the quantum theory
and that a suitable ``point splitting'' procedure would be necessary to
regulate such expressions. The loop variables provide a natural
regularization procedure. In the expression (4.3.12) for the metric, for
example, we have basically point-split the two triads in the expression
$\tw{E}^a_i\tw{E}^{bi}$ in a gauge invariant way. We will see that the
loop variables can be promoted directly to quantum operators: One
does not first promote \A\- and \E\- to quantum theory and then reconstruct
operators corresponding to the loop variables in terms of them. It is
natural to interpret these operators as the regulated versions of
constraints, metric, etc and then {\it define} these latter operators
by performing suitable limits. Thus, the loop variables come into their
own in the quantum theory.

\goodbreak
\vfill\break

%\input [ijtpd.lh]qgmac.tex
%\singlespace
\section{5.}{${}\quad $ The loop representation for 3+1 gravity}%\par
\medskip
{\narrower\narrower\smallskip\noindent
{\sl Aboriginals ... could not imagine territory as a block of land
but rather as an interlocking networks of ``lines'' or ``ways through.'' All
... words for ``country'' ... are the same as the words for ``line.'' }
\smallskip
Bruce Chatwin (The songlines)\smallskip}

\subsection{5.1}{Introduction}%\par

In this section, we want to combine the ideas introduced in the last
three sections to develop a non-perturbative approach to quantum general
relativity in four dimensions.

As in 2+1-gravity, we will use the algebraic quantization
method of section 2.4. Recall that this program requires two
key inputs: the choice of a class of elementary classical variables right
in the first step and the choice of the representation of the quantum
algebra in the fourth step. The geometric properties of the $T$-variables
introduced in the last section suggest that they are natural candidates to
be the elementary classical variables. Even after this choice is made,
however, there remains a considerable freedom in the selection of the
representation of the required quantum algebra. Indeed, in the 3-dimensional
theory, we found that at least two possibilities arise naturally: the
connection representation and the loop representation. The discussion of
the ``non-linear duality'' between connections and loops (of section 2.5)
suggests that, from the viewpoint of a general representation theory, it
is the connection representation that should be regarded as ``basic'' and
the loop representation as  being ``derived'' through the loop transform.
This is the strategy we adopted in 2+1-gravity. In
4-dimensional general relativity, however, this procedure faces certain
technical problems. These are being investigated but have not been fully
resolved. More importantly, so far, the theory has advanced much
further in the loop representation than in the connection picture.
Therefore, in the main body of this section, I will essentially follow
the steps used by Rovelli and Smolin (1990) in their pioneering work
and construct the loop representation ab-initio. In the last subsection,
I will briefly present the current status of the construction of a systematic
representation theory along the lines of section 2.5.

In section 5.2, we will construct the loop representation: the space $V$
of states as well as the action of the $\hat{T}$-operators on this space
will be specified. This is quantum kinematics. Unlike in the 2+1 theory,
the system now has an infinite number of degrees of freedom and this poses
non-trivial regularization problems. These are addressed in the following
two subsections. We begin in section 5.3 by showing how certain operators of
physical interest can be regulated in a way that respects the diffeomorphism
invariance of the theory. We then use these regulated expressions to exhibit
states --called weaves-- which approximate smooth classical geometries
on a macroscopic scale but have a discrete structure of a definite type at the
Planck scale. In section 5.4, we discuss quantum
dynamics, i.e., solutions to the quantum constraints. The first step, again,
is the regularization of the constraint operators. The Gauss constraint is
trivial in the loop representation and the vector constraint can be regulated
such that the resulting operator is manifestly free of dependence on the
background structure introduced in the process. Furthermore, it has the
expected action: it generates diffeomorphisms. Therefore, the {\it general}
solution to the Gauss and the vector constraints can be given explicitly:
These are the loop functions which project down to the space of (generalized)
knots and links (where intersections are permitted). Since the space of knots
and links is discrete, one expects key mathematical as well as conceptual
simplifications. Finally, we discuss the solutions to the quantum scalar
constraint ---the analog of the Wheeler-DeWitt equation of quantum
geometrodynamics. This constraint needs both regularization and
renormalization and, futhermore, the resulting operator carries background
dependence. However, the dependence has a specific form so that a
background {\it independent} kernel can be extracted. The result yields an
infinite dimensional space of solutions to all constraints. Some of these are
directly related to knot invariants. It is important to note that these
solutions are {\it not} just formal; detailed regularization has been carried
out and, futhermore, attention is paid so that the final results are
independent of background structures used in the intermediate steps.
We will conclude in section 5.5 with a brief discussion of the open problems
and of some of the work in progress.

Thus, we will see in this section that significant progress has been already
made along several different lines. These results have provided valuable
insights. The framework leads to discrete structures at two levels. First,
the quantum states representing a macroscopic, smooth geometry exhibit
discreteness at the Planck scale on 3-dimensional space. Second, the domain
space of quantum states --the space on which solutions to quantum constraints
live-- is also discrete. The two are related but quite distinct both
mathematically and conceptually. The first type of discreteness suggests that
the assumption that space-time is represented by a smooth continuum, normally
made in field theory, may be seriously flawed. More importantly, it tells
us how this error can be rectified ---what the continuum is to be replaced
by. The second type of discreteness should enable to one to formulate the
mathematics of quantum theory in a combinatorial framework, without any
reference to space and time. It opens up doors to new arenas, lets one
visualize Planck scale physics in an entirely different fashion and provides
tools to pose and answer new questions. At a technical level, many of the hard
problems of functional analysis are simplified because physical states
are functions on a discrete space. The first type of discreteness has
implications which are easier to grasp in terms of notions we are used to;
it is as though there is an ultra-violet cut-off built into the very fabric of
the non-perturbative treatment or that the correct expression of the density
of states at high energies is qualitatively different from the naive one, used
in the perturbation theory. However I believe that, in the end, the second
type of discreteness is likely to have a more profound impact on our
understanding of the Planck scale physics.

These developments are exciting. However, as will be clear from the
discussion, a comprehensive picture which unites the various concrete
results in a systematic fashion is not yet available. In particular, in the
loop representation, rather than constructing a proper representation theory,
one makes convenient choices as they are needed. Consequently, there are
neither uniqueness results nor a good control of what would have happened
if different choices had been made. In a sense this is no different from
what one does in interacting quantum field theory in Minkowski space, where
again there is no a priori justification for using the Fock representation.
(Moreover, in 4 dimensions, there are good mathematical reasons for {\it not}
using it!) However, in that case, over the years one has acquired a great deal
of theoretical experience and, more importantly, there exist a vast number of
experimental tests which support the strategy as a good working hypothesis.
In non-perturbative quantum gravity, on the other hand, one lacks both and
hence the issue of ``control'' and uniqueness become much more important.
(Incidentally, although this point is rarely raised in other non-perturbative
approaches to quantum gravity, such as quantum geometrodynamics or string
theory, it is equally valid there.) However, I should also emphasize that the
choices made in the construction of the loop representation {\it are} well
motivated. Furthermore, within the chosen representation, the analysis has now
reached a high level of precision and is on a sound  mathematical footing by
standards conventionally used in theoretical (as opposed to mathematical)
physics. Finally, although important technical problems remain, a general,
systematic approach (along the lines of section 2.5) which may justify
the choices we have made {\it is} now in sight.

A much more serious limitation is the {\it incompleteness} of the program. In
particular, the space of solutions to all constraints, although infinite
dimensional, is far from being exhaustive. Even on the space of available
solutions, we do not know the correct inner-product. The general quantization
program of section 2.4 does provide a strategy involving the so-called
``reality conditions''. However, in the obvious implementation of this idea,
one needs to know the complex conjugation relation on a large class of Dirac
observables and {\it very} few of these are known. Finally, there is the
issue of physical interpretation. Over the last two years, however, the
program has been completed in several truncated models and more recently,
certain approximation methods have been introduced in the full theory. These
have provided fresh insights into the open problems. Much of the
current work is in the area of constructing more and more sophisticated models
--e.g. one Killing field reduction of general relativity-- with the hope that
they will tell us how to tackle the problem in its complete generality.

\goodbreak
\subsection{5.2}{The loop representation}%\par

Let us now introduce the loop representation in a somewhat loose manner,
specifying only those technicalities which are needed in the next two
subsections.

The first step is the construction of the quantum algebra. Let us choose
as our elementary variables the loop-strip functions $T[\gamma ]$ and $T[S]$
introduced in section 4.3 (see (4.3.2) and (4.3.5) respectively.) The
$T[\gamma ]$ represent the configuration variables and the $T[S]$ represent
the momentum variables. They are complete --in fact, overcomplete-- almost
everywhere on the phase space and closed under the Poisson bracket. Let us
denote by ${\cal S}$ the complex vector space they generate. Because of
overcompleteness, there are algebraic relations (4.3.3) and (4.3.6) between
these $T$-variables. This is not surprising. For, the
effective configuration space ${\cal C}/{\cal G}$, the quotient of the
space ${\cal C}$ of connections modulo the local gauge group ${\cal G}$,
is a genuinely non-linear space with a complicated topology, whence any set of
functions on the effective phase space, $T^\star({\cal C}/{\cal G})$,
which is complete is {\it necessarily} overcomplete. It {\it is} possible
to reduce this overcompleteness significantly and it is often convenient to
do so, especially for concrete calculations in a lattice theory (Loll 1991).
However, the overcompleteness cannot go away entirely because of the
topological reasons mentioned above. Therefore, in the construction of the
general framework here, it will be easier to deal with the entire set of
loop-strip variables and incorporate the algebraic relations between them in
the very fabric of the quantum algebra. Thus, in any representation of this
algebra, the quantum operators will satisfy the corresponding relations,
ensuring, in particular, the correct classical limit.

Let us then begin the construction of the quantum algebra, by associating
with each loop $\gamma$ an operator $\hat{T}[\gamma ]$ and a strip $S$ another
operator $\hat{T}[S]$ and consider the algebra they generate. On this algebra,
we impose the hatted versions of the algebraic relations (4.3.3) and
(4.3.6) and require that the commutator between any two loop-strip
variables be given by $i\hbar$ times their Poisson bracket (4.3.7). This
is our quantum algebra ${\cal A}$.

Our next task is to find representations of this algebra. Let us begin by
specifying the vector space $V$ of states. Let $V$ be the space of functions
$\psi(\gamma)$ of loops satisfying the following two conditions:
\item{i)} {Continuity: If a sequence of loops $\gamma_i$ converges to a loop
$\gamma$ pointwise on $\Sigma$, then the complex numbers $\psi(\gamma_i)$
converge to $\psi(\gamma)$.}
\item{ii)} {If $\sum c_iT[\alpha_i](A) = 0$ for all $A\in {\cal C}/{\cal G}$
then $\sum c_i\psi[\alpha_i] =0$, where $c_i$ are constants.}

Note that the first condition can be stated directly on the space of loops
without any reference to the connection. The second condition, however,
knows about connections, whence also about the gauge group and the
effective configuration space ${\cal C}/{\cal G}$. It is an open question
whether this condition can also be recast purely in the language of loops,
without reference to connections, {\it in a concise fashion}. Finally,
it is very likely  that in a more precise treatment one would need to impose
additional ``regularity conditions'' on the loop functions: examples of
topological field theories as well as of linearized gravity (Ashtekar, et al
1991) and Maxwell theory (Ashtekar \& Rovelli, 1992)
make this amply clear. However, I will refrain from imposing any
conditions that will not be needed directly in the calculations that follow,
thereby maintaining a flexibility that may be useful subsequently.

Let us now specify the action of operators on these loop states. We will
set
\footnote{16}{Rovelli and Smolin (1990) considered wave functions
$\psi(\{\beta\}) := (\psi_0 , \psi_1(\beta), \psi_2(\beta_1\cup\beta_2),
...\psi_n(\beta_1\cup ....\cup\beta_n ) ... )$ of multi-loops and defined
the action of operators $\hat{T}[\gamma]$ as follows: $(\hat{T}[\gamma ]
\circ\psi) (\{\beta\}) := \psi (\gamma\cup\{\beta\})$. At first, it was
thought that the existence of such states would let one introduce
``raising'' and ``lowering'' operators and
endow the space of states with a natural Fock like structure, states with
support on single loops playing the role of the ``first excited'' or,
``one particle'' states. This interpretation turned out to be incorrect
because multi-loop states can be identified with linear combinations of
single-loop states in 3 as well as 4 dimensional gravity (and also in the
Fock space of spin-2 gravitons). Thus, at a fundamental, conceptual level,
we only need to consider functions of single loops. In calculations, however,
it is often convenient to consider the combination $\textstyle{1\over 2}(\psi
(\gamma\sharp\beta)+ \psi (\gamma\sharp\beta^{-1}))$ as a 2-loop state and
denote it by $\psi (\gamma\cup\beta )$. We will use this notation in sections
5.3 and 5.4.}:
\beginalign{
\big(\hat T [\gamma ]\circ \psi\big) (\alpha ) &= \textstyle{1\over 2}
\big(\psi(\alpha\sharp \gamma ) + \psi (\alpha\sharp \gamma^{-1})\big)\cr
\big(\hat T [S ]\circ \psi\big) (\alpha ) &= \hbar G \sum_i \Delta_i (S,
\alpha) \big( \psi (\alpha\circ\tau_i) - \psi(\alpha\circ\tau_i^{-1})\big)
,\cr}
\endalign{(5.2.1)}
where, as before $\alpha\sharp\gamma$ is the ``eye-glass loop'' obtained
by joining $\alpha$ and $\gamma$ by any line segment,
$\Delta_i$ is $\pm 1$ depending on the orientation of the loop
$\alpha$ and the strip $S$ at the $i$-th intersection and, as in section
4.3, $\tau_i$ labels the loop on the strip $S$ at the $i$-th intersection
between $S$ and the loop $\alpha$. (In particular, if the loop $\alpha$
in the argument of the wave function does not intersect the strip $S$,
then the value of $\hat{T}[S]\cdot \psi$ vanishes at the loop $\alpha$.)
One must verify that these operators are well-defined, i.e., they leave
the vector space $V$ defined above invariant. This is, however,
straightforward to check. Finally, it is convenient to use a bra-ket notation
to express the action of these operators. Setting $\psi(\gamma ) =\langle
\gamma\mid \psi \rangle$, the action can be expressed more directly as:
\beginalign{
\langle\alpha\mid\circ \hat T [\gamma ] &= \textstyle{1\over 2} (\langle
\alpha\sharp \gamma\mid + \langle\alpha\sharp \gamma^{-1}\mid )\cr
\langle \alpha\mid\circ \hat T [S ] &= \hbar G \sum_i \Delta_i (S, \alpha)
(\langle\alpha\circ\tau_i\mid - \langle\alpha\circ\tau_i^{-1}\mid).\cr}
\endalign{(5.2.2)}
The content of the two sets of equations is precisely the same%
\footnote{17}{Thus, strictly speaking, the argument $\alpha$ in the bra
$\langle \alpha\mid$ is not a single loop but an equivalence class of loops
$\{\alpha\}$ on which each $\psi(\alpha )$ in $V$ takes the same value.}.
In what follows, we will often use the notationally more compact form (5.2.2)
which avoids the use of the dummy variable $\psi$ in each term and enables
us to visualize various operators more directly in terms of what they do to
loops.

{}From the viewpoint of the general quantization program, one is free to
choose any $V$ one likes and represent the quantum algebra in any manner
one pleases so long as the result is a proper representation of the algebra
${\cal A}$; there is no a priori need to ``justify'' these choices. However,
one's results may depend sensitively on these choices ---and we will see
that, in a certain sense, ours do. Therefore, as indicated above, one would
like a degree of control and, ideally, a uniqueness theorem which says that
these are the only possibilities subject to certain well-motivated physical
and mathematical restrictions. At the current stage of the program, we are
quite far from such a theorem. However, we can motivate the choices
as follows. First, the choices we made mirror the structure we found in
the loop representation of the 2+1 theory. Second, in the connection
representation, it is natural to represent $A_a^i$ by a multiplication
operator and $\tilde{E}^a_i$ by a functional derivative. Now, if one performs
a heuristic transform to the loop representation {\it assuming} that the
measure is (quasi)\-invariant under the action of the motions on the
configuration space ${\cal C}/{\cal G}$ generated by the vector fields which
implicitly feature in the momentum variables $T[S]$, one is led to the
representation map (5.2.1). Thus, in a certain sense, the choices made above
are the ``simplest'' ones consistent with (4.3.3), (4.3.6) and the canonical
commutation relations.

This completes steps 1, 2 and 4 in the algebraic quantization program. Notice
that we skipped step 3 ---the introduction of $\star$ relations on the algebra
${\cal A}$ of quantum operators. The primary reason is that the reality
conditions (4.2.9) are quite complicated to capture in the language of
$T$-variables. The issue is being investigated currently and I will report
on its status in section 5.5. The purpose of the reality conditions is to
provide us an inner product on the vector space $V$. However, since this inner
product is introduced prior to the imposition of quantum constraints, it
does {\it not} have direct physical significance. Indeed, in most toy
models, where $\star$-relations can be introduced, it turns out that the
resulting inner product on $V$ is such that the solutions to quantum
constraints fail to be normalizable. However, the inner product {\it can be}
of considerable technical help especially when the system has an infinite
number of degrees of freedom. Fortunately, at our present level of rigor,
these specific technical issues do not play a significant role. Hence, the
absence of the $\star$ relation on the algebra will pose no obstruction until
we have solved the constraints and found the physical sector of the theory.

Finally, we saw in section 4.3 that, although  $T[\alpha]$ and $T[S]$ are
overcomplete almost every where on the phase space, it is easier to express%
\footnote{18}{The situation is rather similar to Schr\"odinger quantum
mechanics on manifolds. In that case, with each (suitably regular) function
$f$ on the configuration space, one associates a configuration variable
$Q[f]$ and with each vector field $\vec v$, a momentum variable $P[\vec v]$.
These are analogous to our $T[\alpha]$ and $T[S]$, respectively, and constitute
an overcomplete set. However, to obtain the operator analog of the kinetic
part $g^{ab}p_ap_b$ of the Hamiltonian, one does not first decompose this
classical function as a sum of products of the elementary variables and then
write the analogous quantum operator. It is easier to appeal to general
covariance requirements and promote the operator directly as the Laplacian
$-\hbar^2 g^{ab}D_a D_b$. We are adopting a similar procedure here.}%
geometrical variables such as the (density weighted, contravariant) metric
as well as the constraint functions in terms of
distributions $T^a[\alpha; s]$ and $T^{aa'}[\alpha; s,s']$. Let us therefore
see how these distributional variables can be promoted to (distribution
valued) operators. We have:
\bneq
\langle\beta\mid \circ \hat{T}^a[\gamma; s] := {\hbar G\over 2} \big(
\oint_\beta dt \delta^3(\beta(t), \gamma(s))\dot\beta^a(t)\big)\-\cdot\-
\big(\langle\beta\circ_s\gamma\mid -
\langle \beta\circ_{1-s}\gamma^{-1}\mid \big),
\eneq{(5.2.3)}
where the subscript to $\circ$ denotes the intersection at which the two
loops are re-routed. Similarly, the action of $\hat{T}^{aa'}$ is given by:
\beginalign{
\langle\beta\mid\!\circ\- \hat{T}^{aa'}[\gamma; s,s'] :=& ({\hbar G
\over 2})^2 \-\-
\left(\loint_\beta dt\- \delta^3(\beta(t), \gamma(s)) \dot\beta^a(t)\cdot
\loint_\beta dt' \- \delta^3(\beta(t'),\gamma(s')) \dot\beta^{a'}(t')\right)\cr
\times & \bigg(\langle \gamma_s^{s'} \beta_{t'}^t (\gamma_{s'}^s)^{-1}
(\beta_t^{t'})^{-1}\mid + \langle \gamma_s^{s'}(\beta_t^{t'})^{-1}
(\gamma_{s'}^s)^{-1} \beta_{t'}^t \mid \cr
& + \langle \gamma_s^{s'}\beta_{t'}^t \cup \gamma_{s'}^s\beta_t^{t'}\mid +
\langle\gamma_s^{s'}(\beta_t^{t'})^{-1} \cup \gamma_{s'}^s \beta_{t'}^t \mid
\bigg)\cr}
\endalign{(5.2.4)}
where we have made the reroutings in the four terms explicit by specifying
the compositions of various segments involved and where in the last two terms
we have used the notation of footnote 16 for disjoint union of loops.
Equation (5.2.3) gives just the action of the unsmeared version
of $\hat{T}[S]$, while the second operator is new. Note that the action is
non-trivial only if (at least) one ``hand'' on $\gamma$ grasps $\beta$, i.e.
if $\beta$ intersects $\gamma$ at a point at which a triad lives. Both
definitions  are motivated, as before, by a heuristic loop transform.
Furthermore, the commutators of these operators with one another as well as
with $\hat{T}[\alpha ]$ is just $i\hbar$ times the quantum analog of the
Poisson bracket, modulo terms of the oder $\hbar^2$ and higher (which arise
if one or both operators are quadratic in momenta). For details, see the
original paper by Rovelli and Smolin (1990).

To conclude this section, I will introduce (see, e.g., Chen (1973))
the notion of a certain directional derivative on the space of loop
functions which turns out to be extremely useful in the analysis of
quantum constraints. Note first that, given a loop $\gamma$, a parameter
value $s$, and two vectors $V^a$ and $W^a$ at the point $\gamma(s)$ of
$\Sigma$, we can construct an infinitesimally displaced loop as follows.
First extend $V^a$ and $W^a$ smoothly in a neighborhood of $\gamma(s)$ such
that $[V, W]^a = 0$ , then construct from their integral curves an
infinitesimal parallelogram, $\delta P$, by moving first along the integral
curve of $V^a$ an affine distance $\epsilon$, then along the integral curve
of $W^a$ an affine parameter distance $\epsilon$, then again along $-V^a$ and
back to $\gamma(s)$ along $-W^a$, and finally consider the composed loop
$\gamma\circ_s \delta P$. A loop function $\psi(\beta)$
is said to be {\it area differentiable in the direction of this displacement}
$(s, \delta P$) {\it at the loop} $\gamma$ if
\bneq
\lim_{\epsilon\to 0} {{\psi(\gamma\circ_s\delta P) - \psi (\gamma)}\over
\epsilon^2}\quad\quad {\rm exists},
\eneq{(5.2.5)}
depends {\it only} on the bi-vector $V^{[a}W^{b]}$ at $\gamma(s)$, and the
dependence is linear. Let us denote the limit via:
\bneq
\lim_{\epsilon\to 0} {{\psi(\gamma\circ_s\delta P) - \psi (\gamma)}\over
\epsilon^2} = \- \- [\sigma^{ab}(\gamma(s))\-\-\Delta_{ab}(s)\circ \psi ]
(\gamma ),
\eneq{(5.2.6)}
where $\sigma^{ab} = V^{[}a W^{b]}(\gamma(s))$. If the
directional derivative exists for all $s$ and for all choices of bi-vectors,
$\psi$ is said to be {\it area differentiable} and $\Delta_{ab}\circ
\psi$ is said to be area derivative of $\psi$. It is important to note
that, in spite of the term ``area'' in this nomenclature, the notion does
{\it not} refer to a background field such as a metric or an alternating
tensor. An example of a function which is area differentiable is given
by the trace of the holonomy $\tr H_A (\gamma)$ of a smooth connection $A$
around the loop $\gamma$: $[\Delta_{ab}(s)\cdot \tr H_A](\gamma) = \tr
(F_{ab}(\gamma(s))\cdot H_A(\gamma )(s))$. Of course, not all functions are
area differentiable. In particular, a diffeomorphism invariant function is
in general area differentiable in the direction $\sigma^{ab}(\gamma(s))$ only
if the 2-flat spanned by $\sigma^{ab}$ contains the tangent to the loop.

\goodbreak
\subsection{5.3}{Regularization and weaves}%\par

I will now discuss two striking results that have emerged --already at a
kinematical level-- from the loop representation. The first is that certain
operators representing geometrical observables can be regulated in a way
that respects the diffeomorphism invariance of the underlying theory. What
is more, these regulated operators are finite {\it without any
renormalization}. Using these operators, one can ask if the space $V$
admits loop states which approximate smooth geometry at large scales.
One normally takes for granted that the answer to such questions would be
obviously ``yes''.  However, in genuinely non-perturbative treatments,
this is by no means clear a priori; one may be working in a sector of a
theory which does not admit the correct or unambiguous classical limit.
For example, the sector may correspond to a confined phase which has no
classical analog or the limit may yield a wrong number even for the macroscopic
dimensions of space-time. The second main result of this subsection is
that not only is the answer to the question raised above in the affirmative
but, furthermore, {\it these states exhibit a discrete structure of a
definite type at the Planck scale}.

Let us begin with the issue of regularization. As noted in section 4, in
the present framework, the spatial metric (of density weight two) is a
``composite'' field given by $\tw{\tw{q}}^{}{ab}(x) = \tw{E}^{ai} (x)
\tw{E}^b_i(x)$. In the quantum theory, therefore, this operator must be
regulated. The obvious possibility is point splitting. One might set
${\tw{\tw q}}{}^{ab}(x) = \lim_{y\to x} \tw{E}^{ai}(x) \tw{E}^b_i(y)$.
However, the procedure violates gauge invariance since the internal indices
at two {\it different} points have been contracted. As we saw in section 4.3,
a gauge invariant prescription is to use the Rovelli-Smolin loop variable
$T^{ab} [\gamma ](x,y)$ defined in the classical theory by
\bneq
T^{aa'}[\gamma ] (y,y') := {1\over 2} \tr\big[({\cal P} \exp\- G\int_{y'}^y
A_b dl^b) \tw{E}^a(y')\- ({\cal P}\exp \- G\int_y^{y'}A_c dl^c)\-
\tw{E}^{a'}(y)\big],
\eneq{(5.3.1)}
where $y$ and $y'$ are two points on the loop $\gamma$, and note that
in the limit $\gamma$ shrinks to zero, $T^{aa'}[\gamma](y,y')$ tends to
$-4\tw{\tw{q}}{}^{aa'}$. Now, we just saw that, in quantum theory, one can
define the action of the operator $\hat{T}^{aa'}[\gamma](y,y')$
directly on the loop states via (5.2.4). Its action is rather simple: if a
loop $\beta$ does not intersect $\gamma$ at $y$ or $y'$, the operator simply
annihilates the bra $\langle\beta\mid$ while if an intersection does occur,
it breaks and re-routes the loop $\beta$, each routing being assigned a
specific weight. One may therefore try to define a quantum operator
$\hat{q}^{aa'}$ as a limit of $\hat{T}^{aa'}[\gamma ]$ as $\gamma$ shrinks
to zero.

The resulting operator does exist after suitable regularization {\it and}
renormalization. However, because of the density weights involved, the
operator necessarily carries memory of the background metric used in
regularization. Before going into details of the regularization scheme
let me sketch a rough argument to see the origin of this problem intuitively.
The operator in question is analogous to the product $\delta^3(x)\cdot
\delta^3 (x)$ of distributions at the same point. To regulate it, we may
introduce a background metric. The final result is a distribution of the form
$\tw{N}(x)\delta^3(x)$ where, because $\delta^3(x)$ is a density of weight one,
the renormalization parameter $\tw{N}$ is now a density of weight one,
proportional to the determinant of the background metric. Since the final
answer carries a memory of the particular metric used to regulate the operator,
we have violated diffeomorphism invariance. Although there is no definitive
proof, there do exist arguments which suggest that any {\it local} operator
carrying the information about geometry will face the same problem.

There do exist, however, {\it non-local} operators which can be regulated
in a way that respects diffeomorphism invariance.

As the first example, consider the function $q(\omega)$ --representing the
smeared 3-metric-- on the classical phase space, defined by
\bneq
Q(\omega):= \int d^3x\> (\tw{\tw{q}}{}^{ab}\omega_a\omega_b)^{1\over 2}\- ,
\eneq{(5.3.2)}
where $\omega_a$ is any smooth 1-form of compact support. (Note that the
integral is well-defined without the need of a background volume element
because $\tw{\tw{q}}^{ab}$ is a density of weight two. Also, in spite of
the notation, $Q(\omega)$ is not obtained by smearing a {\it distribution}
with a test field; there is a square-root involved.) To define the
corresponding operator, we can proceed as follows. Let us choose on $\Sigma$
test fields $f_\epsilon (x,y)$ which are densities of weight one in $x$, which
satisfy:
\bneq
\lim_{\epsilon\to 0} \lint_\Sigma \- d^3x\-\- f_\epsilon(x,y)\- g(x)
= g(y)
\eneq{(5.3.3.)}
for all smooth functions of compact support $g(x)$. If $\Sigma$ is
topologically $\real^3$, for example, we can construct these test
fields as follows:
\bneq
f_\epsilon (x,y) = {\sqrt{h(x)}\over {\pi^{3\over 2}\epsilon^3}}
\-\- \exp -{\mid \vec x -\vec y\mid^2\over 2\epsilon^2},
\eneq{(5.3.4)}
where $\vec x$ are the cartesian coordinates labeling the point $x$ and
$h(x)$ is a ``background'' scalar density of weight 2. Next, let us
define
\bneq
\tw{\tw{q}}{}^{aa'}_\epsilon(x) = -{1\over 4}\lint_\Sigma d^3y \lint_\Sigma
d^3y' f_\epsilon(x,y) f_\epsilon(x,y') T^{aa'}(y,y').
\eneq{(5.3.5)}
As $\epsilon$ tends to zero, the right side tends to $\tw{\tw {q}}{}^{ab}$
because the test fields force both the points $y$ and $y'$ to approach $x$,
and hence the loop passing through $y, y'$, used in the definition of
$T^{aa'}(y,y')$, to zero. It is now tempting to try to define a local
metric operator $\hat{q}^{aa'}$ corresponding to $\tw{\tw{q}}{}^{aa'}$ by
replacing $T^{aa'}(y,y')$ in (5.3.5) by its quantum analog and then taking
the limit. One finds that the limit does exist provided we first renormalize
$\hat{q}_\epsilon^{aa'}$ by an appropriate power of $\epsilon$. However, the
answer depends on the background structure (such as the density $h(x)$) used
to construct the test fields $f_\epsilon (x,y)$. If, however, one tries to
construct the quantum analog of the {\it non-local} classical variable
$Q(\omega)$, this problem disappears. To see this, let us first express
$Q(\omega )$ using (5.3.5) as:
\bneq
Q(\omega) = \lim_{\epsilon\to 0} \lint_\Sigma \- d^3x \-\-
(\tw{\tw{q_\epsilon}}{}^{aa'}\omega_a\omega_{a'})^{1\over 2}.
\eneq{(5.3.6)}
The required quantum operator $\hat{Q}(\omega)$ on the loop states can now be
obtained by replacing $T^{aa'}(y,y')$ by the operator $\hat{T}^{aa'}(y,y')$
defined in (5.2.3). A careful calculation shows that: i) the resulting
operator {\it is} well-defined on loop states; ii) no renormalization is
necessary, i.e., the limit is automatically {\it finite}; and, iii) the final
answer carries no imprint of the background structure (such as the density
$h(x)$ or, more generally, the specific choice of the test fields
$f_\epsilon (x,y)$) used in regularization. To write out its explicit
expression, let me restrict myself to smooth loops $\gamma$ without any
self-intersection. Then, the action is given simply by:
\bneq
\langle\gamma\mid\circ\hat{Q}(\omega ) = \l_P^2 \oint_\gamma ds
|\dot{\gamma}^a \omega_a|\>\cdot \langle \gamma\mid ,
\eneq{(5.3.7)}
where $l_P= \sqrt{G\hbar}$ is the Planck length, $s$, a parameter along
the loop and $\dot{\gamma}^a$ the tangent vector to the loop. In this
calculation, the operation of taking the square-root is straightforward
because the relevant operators are diagonal in the loop representation.
This is analogous to the fact that, in the position representation of
non-relativistic quantum mechanics, we can set $<x|\circ \sqrt{\exp (\hat{X})}
= <x|\cdot \exp x/2$ without recourse to the detailed spectral decomposition
of $x$. The $G$ in $l_P$ of (5.3.7) comes from the fact that $GA_a^i$ has
the usual dimensions of a connection while $\hbar$ comes from the fact that
$\hat{E}^a_i$ is $\hbar$ times a functional derivative. The final result is
that, on non-intersecting loops, the operator acts simply by
multiplication: the loop representation is well-suited to find states in
which the 3-geometry --rather than its time evolution-- is sharp.

The second class of operators corresponds to the area of 2-surfaces.
Note first that, given a smooth 2-surface S in $\Sigma$, its area
${\cal A}_S$ is a function on the classical phase space. We first express
it using the classical loop variables. Let us divide the surface $S$ into
a large number $N$ of area elements $S_I, I=1,2...N$, and set
${\cal A}_I^{\rm appr}$ to be
\bneq
{\cal A}_I^{\rm appr} = -{1\over 4}\left[ \lint_{S_I} d^2S^{bc}(x)\-\-
\eta_{abc}\lint_{{\cal S}_I} d^2S^{ b'c'} (x')\-\- \eta_{a'b'c'}\,
   T^{aa'}(x,x') \right]^{1\over 2},
\eneq{(5.3.8)}
where $\eta_{abc}$ is, as usual, the (metric independent) Levi-Civita density
of weight $-1$. Since $T^{aa'}$ approximates $-4({\rm det}\- q) q^{ab}$ for
smooth metrics, ${\cal A}_I^{\rm appr}$ approximates the area function (on the
phase space) defined by the surface elements $S_I$, the approximation
becoming better as $S_I$ --and hence loops with hands at $x$ and $x'$ used
in the definition of $T^{aa'}$-- shrink. Therefore, the total area
${\cal A}_S$ associated with $S$ is given by
\bneq
   {\cal A}_S = \lim_{N \rightarrow\infty}\, \, \sum_{I=1}^{N} \-
   {\cal A}_I^{\rm appr}.
\eneq{(5.3.9)}
To obtain the quantum operator $\hat{\cal A}_S$, we simply replace $T^{aa'}$
in (5.3.8) by the quantum loop operator $\hat{T}^{aa'}$. This somewhat
indirect procedure is necessary because, as indicated above, there is no
well-defined operator-valued distribution that represents the metric or its
area element {\it at a point}. Again, the operator $\hat{\cal A}_S$ turns out
to be finite. Its action, evaluated  on a nonintersecting loop $\gamma$ for
simplicity, is given by:
\bneq
  \langle \gamma | \circ \hat{\cal A}_S  =
   {l_p^2\over 2} \, \> I(S,\gamma )\- \cdot \langle\gamma | ,
\eneq{(5.3.10)}
where $I(S,\gamma)$ is simply the {\it unoriented} intersection number
between the 2-surface $S$ and the loop $\alpha$. (One obtains the
{\it un}oriented intersection number here and the absolute sign in the
integrand of (5.3.7) because of the square-root operation involved in the
definition of these operators.) Thus, in essence, a loop $\gamma$
contributes half a Planck unit of area to any surface it intersects.

The fact that the area operator also acts simply by multiplication on
non-intersecting loops lends further support to the idea that the loop
representation is well-suited to ``diagonalize'' operators corresponding
to 3-geometry. Indeed, we can immediately construct a large set of
simultaneous eigenbras of the smeared metric and the area operators.
There is one, $\langle\gamma|$, associated to every nonintersecting loop
$\gamma$. Note that the corresponding eigenvalues of area are {\it quantized}
in integral multiples of $l_P^2/2$.  There are also eigenstates associated
with intersecting loops which, however, I will not go into to since the
discussion quickly becomes rather involved technically.

With these operators on hand, we can now turn to the construction of
weaves. Recall that the goal here is to introduce loop states which
approximate a given 3-metric, say, $h_{ab}$ on $\Sigma$ on scales large
\footnote{19}{Note, incidentally, that the large scale limit is equivalent to
the semi-classical limit since in source-free, non-perturbative quantum
general relativity, $\hbar$ and $G$ always occur in the combination
$\hbar G = l_p^2$.}
compared to $l_p$. The basic idea is to weave the classical metric out of
quantum loops by spacing them so that on an average only one line crosses
every surface element whose area, {\it as measured by the given} $h_{ab}$
is one Planck unit. Such loop states will be called {\it weaves}. Note that
these states are not uniquely picked out since our requirement is rather weak.
Indeed, given a weave approximating a given classical metric, one can obtain
others, approximating the same classical metric. Let us begin with a concrete
example of such a state which will approximate a {\it flat} metric $h_{ab}$.
To construct this state, we proceed as follows. Using this metric, let us
introduce a random distribution of points on $\Sigma = R^3$ with density $n$
(so that in any given volume $V$ there are $nV(1+ {\cal O}(1/\sqrt{nV}))$
points). Center a circle of radius $a = (1/n)^{1\over 3}$ at each of these
points, with a random orientation. We assume that $a<< L$, so that there is a
large number of (non-intersecting but, generically, {\it linked}) loops in a
macroscopic volume $L^3$. Denote the collection of these circles by $\Delta$.
As noted in footnote 18, due to $SL(2,\comp )$ trace identities, multi-loops
are equivalent to single loops, whence there is a well-defined bra $\langle
\Delta|$. I would like to claim that this is a weave state with the required
properties. Let us first consider the observable $\hat{Q}[\omega]$. To see if
$\langle\Delta |$ reproduces the geometry determined by the classical metric
$h_{ab}$ on a scale $L>>l_p$, let us introduce a 1-form $\omega_a$ which is
{\it slowly varying on the scale} $L$ and compare the value $Q[\omega](h)$
of the classical $Q[\omega]$ evaluated at the metric $h_{ab}$, with the action
of the quantum operator $\hat{Q}[\omega]$ on $\langle\Delta|$. A detailed
calculation yields:
\bneq
\langle\Delta|\circ \hat{Q}[\omega] =  \left[{\pi\over 2} \- \
({l_p\over a})^2 \, Q[w](h) + {\cal O}({a\over L})\right]\-  \cdot
\langle\Delta|.
\eneq{(5.3.11)}
Thus, $\langle\Delta|$ is an eigenstate of $\hat{Q}[\omega ]$ and the
corresponding eigenvalue is closely related to $Q[\omega](h)$. However, even
to the leading order, the two are unequal {\it unless} the average distance,
$a$, between the centers of loops {\it equals} $\sqrt{\pi/2}\, l_p$. More
precisely,  (5.3.11) can be interpreted as follows. Let us write the leading
coefficient on the right side of this equation as $(1/4)(2\pi a/l_p)(nl_p^3)$.
Since this has to be unity for the weave to reproduce the classical value (to
leading order), we see that $\Delta$ should  contain, on an average, one
fourth Planck length of curve per Planck volume, where lengths and volumes are
measured using $h_{ab}$.

The situation is the same for the area operators $\hat{\cal A}_S$. Let $S$ be
a 2-surface whose extrinsic curvature varies slowly on a scale $L >>l_P$. One
can evaluate the action of the area operator on $\langle \Delta |$ and compare
the eigenvalue obtained with the value of the area assigned to $S$ by the
given flat metric $h_{ab}$. Again, the eigenvalue can be re-expressed as a
sum of two terms, the leading term which has the desired form, except for an
overall coefficient which depends on the mean separation $a$ of loops
constituting $\Delta$, and a correction term which is of the order of
${\cal O}({a\over L})$. We require that the coefficient be so adjusted that
the leading term agrees with the classical result. This occurs, again,
precisely when $a = \sqrt{\pi/2}\- l_p$. It is interesting to note that the
details of the calculations which enable one to express the eigenvalues in
terms of the mean separation are rather different for the two observables. In
spite of this, the final constraint on the mean separation is {\it precisely}
the same.

Let us explore the meaning and implications of these results.
\item{1)} {As was emphasized by Jim Hartle in his lectures, to obtain
classical behavior from quantum theory, one needs two things: i) appropriate
coarse graining, and, ii) special states. In our procedure, the slowly varying
test fields $\omega_a$ enable us to perform the appropriate coarse graining
while weaves --with the precisely tuned mean separation $a$-- are the special
states. There is, however, something rather startling: The restriction on the
mean separation $a$ --i.e., on the {\it short distance} behavior of the
multi-loop $\Delta$-- came from the requirement that $\langle\Delta|$ should
approximate the classical metric $h_{ab}$ on {\it large scales} $L$!}
\item{2)} {In the limit $a\to\infty$, the eigenvalues of the two
operators on $\langle\Delta|$ go to zero. This is not too surprising. Roughly,
in a state represented by any loop $\gamma$, one expects the quantum geometry
to be excited just at the points of the loops. If the loops are {\it very}
far away from each other as measured by the fiducial $h_{ab}$, there would
be macroscopic regions devoid of excitations where the quantum geometry
would seem to correspond to a zero metric.}
\item{3)} {The result of the opposite limit,  however, {\it is} surprising.
One might have naively expected that the best approximation to the classical
metric would occur in the continuum limit in which the separation $a$ between
loops goes to zero. However, the explicit calculation outlined above shows
that this is not the case: as $a$ tends to zero, the leading terms in
the eigenvalues of $\hat{Q}[\omega]$ and ${\cal A}_S$ actually diverge
\footnote{20} {One's first impulse from lattice gauge theories may be to
say that the limit is divergent simply because we are not rescaling, i.e.,
renormalizing the operator appropriately. Note, however, that, in contrast to
the calculations one performs in lattice theories, here, we {\it already}
have a well defined operator in the continuum. We are only probing the
properties of its eigenvectors and eigenvalues, whence there is nothing to
renormalize. Even in non-relativistic quantum mechanics the spectrum of
respectable operators are typically unbounded whence the Hilbert space
admits sequences of eigenstates with the property that the corresponding
sequence of eigenvalues diverges in the limit.}!
It is, however, easy to see the underlying reason. Intuitively, the factors
of the Planck length in (5.3.7) and (5.3.10) force each loop in the weave to
contribute a Planck unit to the eigenvalue of the two geometrical observables.
In the continuum limit, the number of loops in any fixed volume
(relative to the fiducial $h_{ab}$) grows unboundedly and the eigenvalue
diverges.}
\item{4)} {It is important to note the structure of the argument. In
non-perturbative quantum gravity, there is no background space-time. Hence,
terms such as  ``slowly varying'' or  ``microscopic'' or ``macroscopic'' have,
a priori, no physical meaning. One must do some extra work, introduce some
extra structure to make them meaningful. The required structure
should come from the very questions one wants to ask. Here, the questions had
to do with approximating a classical geometry. Therefore, we could {\it begin}
with classical metric $h_{ab}$. We used it repeatedly in the construction: to
introduce the length scale $L$, to speak of ``slowly varying'' fields
$\omega_a$ and surfaces $S$, and, to construct the weave itself. The final
result is then a consistency argument: If we construct the weave according to
the given prescription, then we find that it approximates $h_{ab}$ on
macroscopic scales $L$ provided we choose the mean separation $a$ to be
$\sqrt{\pi/2} l_p$, where all lengths are measured relative to the same
$h_{ab}$.}
\item{5)} {Note that there is a considerable non-uniqueness in the
construction. First of all, as we noted already, a given 3-geometry can lead
to distinct weave states; our construction only serves to make the existence
of such states explicit. For example, there is no reason to fix the radius
$r$ of the individual loops to be $a$. For the calculation to work, we only
need to ensure that the loops are large enough so that they are generically
linked and small enough so that the values of the slowly varying fields on
each loop can be regarded as constants plus error terms which we can afford to
keep in the final expression. Thus, it is easy to obtain a 2-parameter family
of weave states, parametrized by $r$ and $a$. The condition that the leading
order terms reproduce the classical values determined by $h_{ab}$ then gives
a relation between $r$, $a$ and $l_P$ which again implies discreteness.
Clearly, one can further enlarge this freedom considerably: There are a lot
of eigenbras of the the smeared-metric and the area operators whose
eigenvalues approximate the classical values determined by $h_{ab}$ up to
terms of the order ${\cal O}({l_p \over L})$ since this approximation ignores
Planck scale quantum fluctuations. Thus, this non-uniqueness is not very
surprising. There is, however, a possibility of a more subtle non-uniqueness
in the opposite Direction: A given weave may approximate two {\it different}
3-metrics! This could happen precisely because the notion of ``slowly
varying'' is tied to the metric $h_{ab}$ we are trying to approximate.
Suppose, two metrics $h_{ab}$ and $h'_{ab}$ which are {\it not} slowly varying
with respect to one another, lead to quite distinct classes of slowly varying
test fields $\omega_a$ and $\omega'_a$. Then, it could happen that there is
a single weave $\langle\Delta|$ which has the property that, to leading order,
the eigenvalues of $\hat{Q}[\omega ]$ are equal to $Q[\omega ](h)$ while
those of $\hat{Q}[\omega']$ are equal to $Q[\omega ](h')$. While it is likely
that such an ambiguity exists, whether it in fact does is not quite clear.
In a broader context, I suspect that such ambiguities will arise in the
classical interpretations of quantum states in {\it any} framework simply
because of the freedom in the choice of coarse graining that is needed to
pass to this limit. In most current discussions, ``obvious'' choices are made
and ambiguities may very well be simply overlooked.}

\item{6)} {Finally, I would like to emphasize that, at a conceptual level,
the important point is that the eigenvalues of $\hat{Q}[\omega ]$ and
${\cal A}[S]$ can be discrete%
\footnote{21}{I have used the conservative phrase ``can be'' because
at the present level of rigor we cannot be sure that the eigenvalues that
have naturally emerged above constitute the entire spectrum. One would have
to tie oneself to a specific inner product on $V$ to nail down this issue.};
quantized in multiples of half Planck units. Given a specific eigenstate, one
can examine the micro-structure of the geometry it defines. The precise
characteristics of the discreteness in that structure will vary from state
to state. Since many of these weaves may define the same macroscopic geometry,
it is also clear that we cannot, at least at this stage, associate a {\it
specific} discrete structure to a given classical geometry. All detailed
claims refer only to eigenstates of the geometric operators.}

Let me conclude the discussion on weaves with two remarks. First, it is not
difficult to extend the above construction to obtain weave states for curved
metrics $q_{ab}$ which are slowly varying with respect to a flat metric
$h_{ab}$. Given such a metric, one can find a slowly varying tensor field
$t_a{}^b$, such that the metric $q_{ab}$ can be expressed as $t_a{}^c
t_b{}^d h_{cd}$. Then, given a weave of the type $\langle\Delta|$ considered
above approximating $h_{ab}$, we can ``deform'' each circle in the multi-loop
$\Delta$ using $t_a{}^b$ to obtain a new weave $\langle \Delta |_{t}$ which
approximates $q_{ab}$ in the same sense as $\langle\Delta|$ approximates
$h_{ab}$. The second remark is that since the weaves are eigenbras of the
operators that capture the 3-geometry, none of them is a candidate for
representing the vacuum state. In the linearized Theory, the vacuum is a
coherent state. It is neither an eigenstate of the linearized metric operator,
nor of the linearized extrinsic curvature (or, connection). Rather, it is
peaked, with {\it minimum} uncertainty spreads for both operators, at their
zero values. The candidates for vacuum state in the full theory would have a
similar characteristic. Some insight into this issue has come from a recent
analysis of the relation between the exact and the linearized theory in the
loop representation (Iwasaki \& Rovelli (1992)).

Since these results are both unexpected and interesting, it is important to
probe their origin. We see no analogous results in familiar theories. For
example, the eigenvalues of the fluxes of electric or magnetic fields are
not quantized in QED nor do the linearized analogs of our geometric operators
admit discrete eigenvalues in spin-2 gravity. Why then did we find
qualitatively different results? The technical answer is simply that the
familiar results refer to the {\it Fock representation} for photons and
gravitons while we are using a completely different representation here. Thus,
the results {\it are} tied to our specific choice of the representation. Why
do we not use Fock or Fock-like states? It is not because we insist on working
with loops rather than space-time fields such as connections. Indeed, one {\it
can} translate the Fock representation of gravitons and photons to the loop
picture. (See, e.g., Ashtekar et al (1991) and Ashtekar \& Rovelli, (1992).)
And then, as in the Fock space, the discrete structures of the type we found
in this section simply disappear. However, to construct these loop
representations, one must use a flat background metric and essentially every
step in the construction violates diffeomorphism invariance. Indeed, there is
simply no way to construct ``familiar, Fock-like'' representations without
spoiling the diffeomorphism invariance. Thus, the results we found are,
in a sense, a direct consequence of our desire to carry out a genuinely
non-perturbative quantization without introducing any background structure.
As remarked in section 5.1, however, we do not have a uniqueness theorem
singling out the representation we are using. One cannot rule out the
possibility of the existence of other, sufficiently rich representations
which also maintain diffeomorphism invariance but in which geometric
operators considered above have only continuous spectra. The one which we
are using is just the ``simplest'' background independent representation
and it leads to interesting results.

My overall viewpoint is that one should simultaneously proceed along two
lines: i) one should take these results as an indication that we are on the
right track and push this particular representation as far as possible; and,
ii) one should try to better understand this representation by, e.g.,
perturbing it in small steps, with the hope of arriving at a uniqueness result.

\goodbreak
\subsection{5.4}{Quantum dynamics}%\par

So far we have dealt with quantum kinematics. The next step in the algebraic
program is the imposition of constraints. As we saw in section 4, because of
the absence of a background metric in classical general relativity, dynamics
is in effect governed by constraints. Furthermore, we saw in section 3 that,
in the 2+1-theory, quantum dynamics --in the sense of time evolution-- can
be recovered from the quantum constraints. In this subsection I will address
the mathematical problem of solving the quantum constraints in the 3+1-theory.
We will see that, in striking contrast with geometrodynamics, there is
available an infinite dimensional space of solutions to all constraints.
Technically, this is possible essentially because the constraints now have
a form that is significantly simpler than that of geometrodynamics.
Furthermore, thanks to the interplay between connections and loops, the
solutions have a natural interpretation in terms of knot theory.

The classical framework gives us three sets of constraints. The first of
these, (4.2.3), constitutes the Gauss law which ensures gauge invariance of
physical states. Fortunately, in the loop representation of any theory of
connections, everything is manifestly gauge invariant whence the Gauss
constraint is redundant. (The loop transform (2.5.2), for example, is from
functions on the {\it moduli} space ${\cal C}/{\cal G}$ of connections to
functions of loops.)

The second set, (4.2.4), constitutes the vector constraint. As noted in
section 4.3, it can be expressed in terms of the $T^a$-variables:
\beginalign{
{\cal V}(\vec N)\equiv & \lint_\Sigma d^3x \- N^a(x){\cal V}_a (x)\cr
& = \lim_{\delta\to 0}\- {1\over{\delta}} \lint_\Sigma d^3 x N^a(x)\- \>
\sum_b T^b [\gamma^\delta_{ab}], \cr}
\endalign{(5.4.1)}
where, as before, $\gamma^\delta_{ab}$ is a loop at $x$ in the $a,b$ plane of
(coordinate) area $\delta$, (the position of the ``hand'' i.e. of the
insertion of electric field-- on the loop being allowed to be arbitrary.)
We can take this expression over to quantum theory. It is already regulated
and the limit turns out to be well-defined {\it without} renormalization.
Thus, unlike the smeared metric and the area operator, the regulated vector
constraint {\it is} a good operator valued distribution. This result may seem
surprising at first but it is a rather straightforward consequence of the
fact that $T^a$ is only linear in momentum. Using the definition of the
operator $\hat{T}^a$ and of the area derivative (see section 4.3), one finds:
\beginalign{
[\hat{\cal V}(N)\circ\psi](\beta ) &= [\loint_\beta ds\- \dot\beta^b(s) N^a
(\beta(s)) \Delta_{ab}(s)\circ \psi](\beta)\cr
&= [\loint_\beta ds\- N^a(\beta(s))\- {\delta\over{\delta\beta^a(s)}}
\circ\psi](\beta),\cr}
\endalign{(5.4.2)}
where in the second step we have used the fact that the direction in which
the areas derivative is being taken includes the tangent to the loop. I should
emphasize that, unlike in section 5.3, here and in subsequent calculations
in this subsection, {\it the loop in the arguments of the quantum states is
allowed to have an arbitrary number of self-intersections} and, in particular,
there may be multi-intersections through the same point. Thus, the effect of
the regulated vector constraint on the loop states is precisely the expected
one; the loop is displaced infinitesimally along the vector field $N^a$.
Therefore, the state $\psi$ is annihilated by the vector constraint if and
only if it is diffeomorphism invariant. Now, two loops (possibly with
self-intersections) which can be mapped into each other by a diffeomorphism
belong to the same (generalized) knot class. Thus, we conclude that the {\it
general} solution to the Gauss and vector constraints is a loop function
which can be projected down unambiguously to the space of (generalized) knots.
Unlike the superspace of geometrodynamics --the space of diffeomorphism
equivalence classes of 3-metrics-- this space is {\it discrete} and a good
deal is known about its structure. Indeed, in the last four years, there has
been an explosion of activity relating knot theory to other branches of
mathematics ranging from $C^\star$ algebras to topological field theories.
One hopes that this powerful machinery will be useful in understanding the
meaning and the structure of these solutions better.

Let us now consider the last constraint, ${\cal S}(A,E)$ of (4.2.5). We noted
in section 4.3 that this constraint can be expressed in terms of $T^{ab}$.
However since (the constraint as well as) $T^{ab}$ is {\it quadratic} in
momenta, one would expect that the corresponding regulated operator would
not be finite without renormalization and, even after renormalization, would
{\it not} define a background independent operator valued distribution. After
all, this is the reason why we were led in section 5.3 to consider {\it
non-local} variables to carry information about the 3-geometry. The general
expectations are indeed correct. Furthermore, in the present case, we do
{\it not} wish to construct non-local expressions since the constraints have
to be imposed locally; only a linear smearing with a lapse field is allowed.
Thus, we seem to be stuck with background dependence in the expression of the
operator. This may seem like a disaster. Fortunately, however, we are not
interested in the entire spectrum of the operator but only its kernel. And,
as we shall now see, the kernel {\it can} be extracted in a background
independent manner.

As in previous cases, we begin by first rewriting the scalar constraint in
terms of the classical $T$-variables. Using the same notation as before,  we
have:
\bneq
{\cal S}(\ut{N}) \equiv \lint_\Sigma d^3x\- \ut{N}(x) {\cal S}(x) =
\lim_{\delta \to 0} {\cal S}^\delta (\ut{N}),
\eneq{(5.4.3a)}
where,
\bneq
{\cal S}_\delta (\ut{N}) = \lint_\Sigma d^3x\ut{N}(x) ({1\over \delta}) \>\-
\sum_{a\not= b}T^{[ab]} [\gamma^\delta_{ab}; 0,\delta].
\eneq{(5.4.3b)}
We now take this expression over to quantum theory. A proper treatment of
intersecting loops is considerably more complicated now since $T^{ab}$ has
{\it two} hands. To take the limit properly, one has to introduce
(background dependent) regulators $f_\epsilon(x,y)$ used in section 4.3,
and furthermore, renormalize the expression by multiplying it by
$\epsilon$ before taking the limit $\epsilon\to 0$. This procedure has been
carried out in detail by Br\"ugmann and Pullin (1992). The final, renormalized
operator $\hat{\cal S}$ has the following action. First, if a loop $\beta$
has no self intersections, $\hat{S}(\ut{N})\circ\psi$ vanishes at $\beta$. If
there {\it are} intersections, $\beta (s_i) = \beta(t_i)$, with $s_i \not=
t_i$, we obtain:
\bneq
[\hat{S}(\ut{N})\circ\psi](\beta) = \sum_i \big[N \hat{\cal F}_i
 \- \>\dot\beta^a(s_i)\dot\beta^b(t_i)\- \Delta^1_{ab}\circ\psi\big]
(\beta_{t_i}^{s_i}\cup\beta_{s_i}^{t_i} ),
\eneq{(5.4.4)}
where $\hat{\cal F}_i$ is a background dependent factor involving the
properties of the loop at the $i$th intersection, $\beta_{t}^{s}$ is the loop
obtained by going along $\beta$ from $\beta (t)$ to $\beta (s)$, $\cup$ is
the disjoint union of two loops (see footnote 16) and where the loop
derivative $\Delta_{ab}^1$ acts just on the first loop in the argument of
$\beta$. Thus, the operator has a form
\bneq
\hat{S}(\ut{N}) = \sum_i \hat{\cal F}_i \-  \hat{\cal S}_i
\eneq{(5.4.5)}
of a sum of products of background dependent and background independent
operators. Therefore, if we choose a wave function which is annihilated
by {\it each} $\hat{\cal S}_i$, we would have a state which is annihilated
by constraints for {\it any} choice of background.
\footnote{22}{This is a good illustration of the difference between a
quantum field theory in Minkow\-ski space and a diffeomorphism invariant
theory. On the one hand, diffeomorphism invariance makes it difficult to
regulate the constraints since we do not have a canonical background metric
at our disposal. On the other hand, dynamics is now governed by constraints
and the problem is only that of finding the {\it kernel} of the constraint
operator rather than the entire spectrum of a Hamiltonian. It is not that
one problem is easier than another. Rather, the problems are {\it different}.
This is why, as pointed out in section 1, the experience we have gained from
Minkowskian field theories is only of limited use in dealing with the
mathematical problems of non-perturbative quantum gravity.}
Generically, to be annihilated by the regulated constraint for any choice of
background, the wave function has to be in the kernel of all $\hat{S}_i$
as well. The work of Rovelli \& Smolin (1990), Blencowe (1990), Br\"ugmann,
Gambini  \& Pullin (1992a,b) has provided us with an infinite dimensional space
of such states.

First, given any state obtained by choosing values on non-intersecting loops
in any smooth manner and then letting the defining regularity conditions
constrain its values on intersecting loops, satisfies the scalar constraint
(Rovelli \& Smolin (1990).) Thus, in particular, from the weave $\langle
\Delta|$ we can construct a ket $\Psi_\Delta (\gamma )$satisfying the
Hamiltonian constraint. However, these kets do not satisfy the vector
constraint. To get a simultaneous solution to both constraints, we can
proceed as follows. First, given a weave, we can consider the knot class
to which it belongs, ask that on non-intersecting loops, $\psi$ be the
characteristic function on the knot class and then extend the value of $\psi$
to intersecting loops in a way consistent with the regularity conditions.
This provides us a solution to {\it all} constraints of quantum gravity.
Heuristically, one may interpret this solution as representing the
3-{\it geometry} to which the 3-metric represented by the original weave
$\langle\Delta|$ belongs. Another class of such simultaneous solutions
can be obtained from holonomies. Let $A_a^i$ be a flat connection and set
$\psi_A(\beta) = \tr {\cal P}\exp \oint_\beta A.dl$. Then, $\psi_A(\gamma )$
is a solution to {\it all} constraints (Blencowe, 1990). Such solutions
capture an aspect (the first homology) of the topology of the 3-manifold.

The 3-metrics underlying the Rovelli-Smolin solutions to the Hamiltonian
constraint are, microscopically, distributional with support only on the
location of the loops. Furthermore, the excitations of the metric operator
are, in a certain sense, restricted to ``point along'' the loops whence
the metrics are also (algebraically) degenerate. However, as we saw in
section 4.3, some of these solutions can represent smooth, non-degenerate
geometries {\it macroscopically}, when coarse-grained appropriately. The
3-geometries underlying the Blencowe solutions are harder to visualize
because they are not eigenstates of geometrical operators. However in the
simplest examples, the algebraic degeneracy persists.

More interesting solutions, related to knot invariants --the second
coefficient of the Alexander-Conway polynomial-- have been obtained more
recently by Br\"ugmann, Gambini and Pullin (1992a,b). Apart from
intriguing connections with knot theory that they have opened up, these
solutions have also considerably improved our overall understanding of
the structure of the space of solutions to the scalar constraint. For,
unlike the earlier ones, these solutions have support on loops with triple
intersections whence they represent 3-geometries which are distributional
but (algebraically) {\it non-degenerate} at points where they have
support, i.e., even microscopically. There is a close relations between
these solutions, Jones polynomials and Chern-Simons theory whose
ramifications are still to be fully understood. This is an area where one
can expect significant progress in the coming years.

\goodbreak
\subsection{5.5}{Outlook}

The developments reported in the last two subsections represent definite
progress. Let us contrast the situation with, say, quantum geometrodynamics.
This framework has shaped the general thinking in the field and provided a
general picture of what one should expect from non-perturbative quantum
gravity. These lessons have been extremely valuable. However, quantum
geometrodynamics has not provided detailed insight into the micro-structure
of space-time. The pictures of space-time foam (Wheeler, 1963) that came out
of this framework are largely qualitative; there are essentially no hints,
for example, as to what would replace differential geometry in the Planck
regime. On the dynamical side, one is yet to succeed in regulating the
scalar constraint --the Wheeler-DeWitt equation-- whence, systematic work to
look for solutions in a well-defined mathematical framework has not even begun.
Furthermore, even at a heuristic level, {\it no} solution is known to the
Wheeler-DeWitt equation in full, non-truncated quantum geometrodynamics. In
the connection dynamics approach, as the last two subsections indicate, one
has been able to go further.

In spite of this progress, one should emphasize that there is still a long
way to go. Three key issues remain unresolved.

First, we do not know which - if any - of these solutions will have finite
norm with respect to the correct inner produce on the space of physical states
\footnote{23}{The requirement that the norm be finite can in fact carry the
crucial physical information. For example, if one solves the eigenvalue
equation for the harmonic oscillator in the position representation, one finds
that there exist eigenstates $\Psi(x)$ for {\it any} value of energy. It is
the requirement that the norm be finite that enforces both positivity and
quantization of energy. Note however that whether all solutions to the
eigenvalue equation are normalizable depends on the choice of representation.
For example, if states are taken to be holomorphic functions of $z=q-ip$,
every eigenstate $\Psi(z)$ turns out to be normalizable whence the conclusion
that energy is positive and quantized can be arrived at simply by solving the
eigenvalue equation. Of course, a priori it is not clear whether the loop
representation is analogous to the $x$ representation of the $z$ representation
in this respect.}.
The problem is that the reality conditions are hard to express in terms of
the $T$-variables and hence have so far not been translated to the loop
representation.

Let us assume for the moment that at least some of these solutions will have
finite norm. The second conceptual issue is then that of interpretation.
This issue appears to be quite difficult to address.  To illustrate the
problem, let me use an analogy. A Rovelli-Smolin type solution associated
with the knot class of a {\it single} loop would appear to represent ``an
elementary excitation'' of the gravitational field and is hard to interpret
in classical terms. We need to superpose a large number of such excitations
to obtain a weave, which then can be interpreted classically. This is
analogous to the fact that a single photon state has no obvious interpretation
from the standpoint of classical Maxwell theory and, to obtain a state that a
classical physicist can interpret using only his conceptual framework, we must
have a large number of photons. To interpret a single photon state, one has to
be ``at home'' with genuinely quantum concepts. Unfortunately, in
non-perturbative gravity, we have very little intuition for the analogous
genuinely quantum world.  And this makes the problem of interpretation
difficult for most solutions.  The problem would disappear only when we begin
to feel comfortable with observables --to be associated, say, with knot
invariants-- which distinguish between the physical states.

The final issue is in the realm of mathematical physics.  Results such as the
discreteness of the spectra of geometric operators depend critically upon the
choices we made a various steps in the construction of the loop representation.
Perhaps the most important of these is that we choose to require that $\hat
T[\gamma]$ and $\hat T[S]$ operators be well-defined in quantum theory even
though classically they are smeared only along 1-dimensional loops and
2-dimensional strips.  This means that the quantum theory will not in general
admit operator-valued distributions corresponding to $A^i_a(x)$ and $\tilde
E^a_i(x)$; the operators $\hat T[\alpha]$ and $\hat T[S]$ are the primary ones.
In the Fock space of photons, for example, the analogous operators are {\it
not} well-defined without additional smearing.  The reason we were led here to
consider, e.g., unsmeared loop operators is that there {\it is} no satisfactory
way of smearing loops without using a background field such as a 3-metric. The
viewpoint is that in ``topological'' theories of connections, such as general
relativity, one should not have to use background structures at a fundamental
level whence natural objects like traces of holonomies should be themselves
promoted to operators.  Some evidence for this view comes from 3-dimensional
gravity and other topological field theories.  However, all these systems have
only a finite number of degrees of freedom and the evidence is therefore weak.
One would like to have a better understanding and a better control of all the
assumptions that have been made in the construction of the loop representation.
Ideally, one would like to prove an uniqueness theorem in which the assumptions
can be motivated by physical considerations.

Some of the work now in progress addresses these issues.

First, certain approximation methods are being developed to get a better handle
on the first two issues.  The idea here is the following.  We saw in Section
5.3 that there exist weave states which approximate the flat 3-metric.  On the
other hand, there also exists a loop representation for spin-2 gravitons in
flat space, obtained by linearizing the Hamiltonian framework of section 4.
Now, since the connections are {\it self dual} (rather than negative
frequency), it turns out that one is forced to thicken out the loops in the
quantum theory (Ashtekar et al 1991).  The idea now is to recast this
description of spin-2 gravitons {\it as perturbations of the weave state}.
First steps in this analysis have been already completed (Iwasaki \& Rovelli
(1992), Zegwaard (1992)) and there exists a map from the states of the exact
theory to those of spin-2 gravitons with interesting properties. Work is in
progress to isolate states of the full theory which, under this map, are sent
to the vacuum state and the n-graviton states of linearized gravity. This will
provide the much needed intuition for (some of) the loop states of the full
theory and is also likely to suggest strategies for selecting the
inner-product. This work is also beginning to shed some light on how standard
quantum field theory in Minkowski space where only the smeared operators make
sense can emerge from a theory which, to begin with, has no smearing at all.
It is the background weave state that provides the structure for smearing and
one is essentially forced into smearing by the requirement that the operators
of the linearized theory should approximate the operators of the full theory in
an appropriate sense. If this analysis can be carried out to completion, the
unease about using unsmeared operators in the exact theory, mentioned above,
will disappear.

Another promising direction, within the exact theory, has been opened up by the
recent work of Gambini and collaborators.  The idea is to first characterize
(holonomically equivalent) loops by a series of distributions and then, using
these ``loop coordinates,'' construct a more general space in which the space
of loops is properly embedded (Gambini \& Leal, 1991).  (These loop coordinates
are the non-Abelian analogs of the ``form factors of loops'' introduced in the
context of the Maxwell theory (Ashtekar \& Rovelli, 1992.)  It then turns out
that this ``extended space'' is naturally endowed with the structure of an
infinite dimensional Lie group (Di Bortolo, Gambini \& Griego, 1992). The
structure of this group is being analyzed.  Its existence leads to an
extension of the loop representation just of the type needed to test how
robust the various results obtained so far are. In the near future, therefore,
one should have a much better feeling for the ``uniqueness'' issue that I
have raised above. In addition, these techniques have opened up new avenues to
explore generalized knot invariants from a differential geometric viewpoint
and to analyze their role in quantum gravity. They may therefore provide a
powerful tool to interpret various loop states and the combinatorial
operations thereon. The mathematical structures that have been unravelled so
far are so rich that it seem reasonable to expect that they will bring us a
wealth of new insights.

Work is also in progress to make the work on loop transform well defined for
gravity.  Ashtekar \& Isham (1992) used the Gel'fand spectral theory to exploit
the non-linear duality between loops and connections along the lines of section
2.5. However, that work is complete only in the case of real gauge fields with
a compact group, say, $SU(2)$. In the gravitational case, the connections are
complex-valued and this creates obstacles in completing the program. However,
since the connection is analogous to the complex coordinate $z=q-ip$ on the
phase space of an oscillator, the wave functions in the connection
representation are {\it holomorphic} functions rather than general,
complex-valued ones. By exploiting this fact, Ashtekar and Lewandowski have
recently proposed a strategy to extend the representation theory to general
relativity.  If this program can be completed, the loop transform would be
well-defined and one would therefore have a good mathematical control over the
available freedom in the construction of a general loop representation. One
would then have the technology to analyse the sense in which the loop
representation used here is unique.

Finally, recently Baez (1992) has used the mathematical structure available
on the space of tangles to provide an inner product on loop states in the
asymptotically flat context. In our terminology, he also uses reality
conditions.  However these are suggested by mathematical properties of certain
operators and it is not clear how these properties are related to the
``physical'' reality conditions which come from the properties of functions
on the gravitational phase space. Nonetheless, this is a very interesting
development and the general approach holds a great deal of promise.

For reasons mentioned in the beginning, throughout this section I have
restricted myself to the loop representation. However, like the metric
representation of geometrodynamics, this representation is not well suited to
analyse the issue of time (see, e.g., chapter 12 in Ashtekar (1991)). This
problem --along with some others-- has been partially analysed in the
connection representation. I will mention these results in the next section.

\vfill\break

%\input [ijtpd.lh]qgmac.tex
%\singlespace
\section{6.}{${}\quad$ Discussion}%\par

I will first summarize the main results of sections 3-5, then discuss the o
pen problems and the current thinking on how they might be
overcome and finally present an evaluation of the program from various
perspectives.

\goodbreak
\subsection{6.1}{Summary}%\par

In section 3, we considered 3-dimensional general relativity and found
that the algebraic quantization program can be completed in 3 different
ways: there is the combinatorial description in terms of ``pre-geometry''
that came from the loop representation; there is the timeless description
obtained in the ``frozen'' version of the connection representation; and,
there is a dynamical description where the scalar constraint is recast as
an evolution equation, also in the connection representation. These frameworks
serve to illustrate the type of mathematical results we are seeking in
4-dimensions. In all three cases, the starting point is a formulation of
general relativity as a dynamical theory of connections. This shift of
emphasis from metrics to connections underlies the entire discussion. It let
us regard quantum states as wave functions of connections; it provided
a natural strategy to introduce the inner product using the symplectic
structure on the new domain space; it enabled us
us to single out time as one of the connection components; it allowed us to
construct the $T$-variables which then served as the basic Dirac
observables; and, it naturally led us to the loop representation via the loop
transform.

In section 4, therefore, we began by recasting 4-dimensional general
relativity as a theory of connections. We found that we could again use
the Yang-Mills phase space and express Einstein constraints and Hamiltonians
in the language of connections and the conjugate electric fields. Indeed,
there even exists a pure connection formulation in which the space-time
metric never appears as a primary field either in the Lagrangian or the
Hamiltonian descriptions (Capovilla, et al 1989). If, however, we interpret
the electric fields as triads and the connections as potentials for the
self dual part of the Weyl tensor, we recover geometrodynamics. We may choose
to forego this interpretation and regard the theory simply as a diffeomorphism
invariant, dynamical theory of connections. This perspective immediately
puts at our disposal the powerful machinery of gauge theories. Furthermore,
there are now significant technical
simplifications relative to geometrodynamics: constraints and
Hamiltonians are low-order polynomials in the new canonical variables. Thus,
the new 4-dimensional Hamiltonian framework is qualitatively similar to the
3-dimensional one. However, there is also a {\it key} difference: the
connections are no longer constrained to be flat whence the theory now has an
infinite number of {\it local} degrees of freedom rather than just a finite
number of {\it topological} ones. This makes physics vastly richer but the
task of quantization correspondingly more difficult.

In section 5, we presented the current status of the quantum theory. For
brevity, we concentrated on the loop representation. As far as the
implementation of the algebraic quantization program is concerned, it is
in this framework that the non-truncated theory has advanced the most.
(In the terminology of section 3, the loop representation yields
the ``pregeometry'' picture. As we will see below, however, the program has
gone further in the connection representation in several truncated models
where, the the other two pictures from section 3 arise naturally.) We found
three significant results.

First, one {\it can} invent techniques so that operators of the
theory can be regulated in a way that respects the diffeomorphism invariance.
This result by itself is quite surprising from the perspective of Minkowskian
quantum field theories since the standard techniques such as normal ordering
or point-splitting produce regulated operators whose structure refers to the
background  space-time metric. However, this is not all; the regulated
operators --such as the area of a 2-surface-- carrying geometric information
turn out to have discrete eigenvalues, quantized in Planck units.

The second result is that the non-perturbative theory admits states --weaves--
which approximate smooth geometries at large scales but exhibit a discrete
structure of a specific type at the Planck scale. The existence of these
states suggests that a basic assumption of the perturbation theory
--that quantum states can be represented as fluctuations off a smooth
background geometry-- may be seriously flawed in the Planck regime.
Non-renormalizability of quantum general relativity implies that one should
expect new physics at small scale and these non-perturbative states indicate
the type of new structures that one can expect to find. That some discreteness
of this sort should occur in the Planck regime has been anticipated for quite
some time now. However, as John Wheeler puts it, such ideas of ``space-time
foam'' (which he introduced) were based on simple estimates and it is only
with the weave states that they have acquired a precise mathematical meaning.

The third result
concerns non-perturbative solutions to quantum constraints. I first
presented the {\it general solution} to the quantum Gauss and vector
constraints. These are functions on the space of generalized knots (where
the generalization consists of allowing for self-intersections). Thus,
as in the 3-dimensional case, the domain space of physical wave functions
is discrete in the loop description ---the homotopy classes are replaced by
the generalized knot classes. There is a good reason to believe that this
second and mathematically deeper level of discreteness will significantly
simplify the task of introducing inner products and render the subject to
combinatorial techniques. Consequently, it appears that, as in the
3-dimensional case, the loop representation will lead us to a ``pre-geometry''
description in which neither space nor time play a fundamental role ---both
will be derived concepts. Finally, I presented an infinite dimensional space
of solutions to the scalar constraint. The calculations that led us to these
solutions are instructive in that they bring out the differences between the
mathematical problems of diffeomorphism invariant quantum theories and those
of quantum field theories on a given space-time background. In particular, we
found that while the regularization of the scalar constraint produces a
background dependent operator, we could still get by since we are only
interested in its kernel; while most properties such as the spectrum of the
operator are background dependent, the kernel
turned out to be background independent. Thus, while the constraints generate
dynamics in a suitable sense, the mathematical role they play is quite
different from the role played by the Hamiltonian in more familiar theories.

Overall, the present approach has managed to go significantly beyond quantum
geometrodynamics. This seems quite surprising. After all, to go to connection
dynamics, one only performed a canonical transformation. How can things be so
different then? I am not sure of the complete answer for it depends largely
on whether one continues to make progress. If one does, then the following
scenario would seem appropriate.

One would conclude
that this upside down and inside out way of looking at general relativity is
better suited to the problem of quantization precisely because it brings to
forefront those concepts which are appropriate to the Planck regime. That is,
the view would be that while distances, light cones and geodesics may be the
most fruitful concepts to extract physical information from general relativity
at the macroscopic level, they just aren't the ``correct'' concepts in the
Planck regime. The ``correct'' concepts may be spin-connections, holonomies of
spinors around closed loops, knots and links. We have learn to {\it think} in
terms of these concepts, to formulate questions and analyse them in a genuinely
quantum mechanical fashion. We already expressed some of the geometrical
operators as well as quantum constraints in this new language. We should be
able to grasp fully the real microscopic ``meaning'' of the constraints in
this new language. In particular, we should think of the scalar constraint
{\it primarily} in terms of intersections, grasps and re-routings of loops;
its role as the generator of time evolution, for example, may be appropriate
only in a semi-classical regime in which there {\it is} such thing as time.
We should develop a physical intuition for the basic mathematical operations
directly, without going through metrics and space-time descriptions. After all,
this has happened before. A classical physicist would want to compute the
trajectories of electrons in the Bohr atom. He might be disappointed that
the uncertainty principle makes a precise calculation impossible but would
still insist on working out smeared out trajectories. From a quantum mechanical
point of view, these calculations are not fundamental; we have to shift our
perspective and worry about, e.g., eigenvalues and eigenvectors of the
Hamiltonian first. Similarly, we should not insist that space-time geometry
--or more generally, any specific feature of the standard differential
geometry-- must play a
role at a fundamental level in quantum theory. Indeed, it is more a rule
than an exception that quantum theory forces us to change our perspective.
We changed the perspective when we learned to think of the electromagnetic
field in terms of photons; when we put aside interference and diffraction
of fields and learned to think directly in terms of Feynman diagrams;
when we learned to put the scattering theory for classical Yang-Mills fields
on the back-burner and to think instead in terms of confinement and/or
symmetry breaking. It may well be that metrics and light cones come in to
their own only in the semi-classical regimes. Then, differential geometry
reigns. In the Planck world, it may be superseded by distributional, possibly
degenerate geometries; by the theory of knots and links; by the algebra of
combinatorial operations.

\goodbreak
\subsection{6.2}{Open issues and directions for future}%\par

In spite of these successes, as we saw in section 5.5, the program is quite
incomplete even at a mathematical level. First, we need to
find the ``correct'' inner product on the space of physical states, i.e., of
solutions to quantum constraints. Since we know the general solution to the
Gauss and vector constraints, it is tempting to first introduce an inner
product on these ``pre-physical states'' and then focus on the physical
subspace on which the scalar constraint is satisfied as well. The strategy
looks attractive especially because the domain space of pre-physical states
is discrete and the problem of finding measures is significantly simpler. I
believe that, at the present state of our understanding, this is a useful
strategy to pursue. (Furthermore, there exists in the literature a model
system due to Husain and Kucha\v r (1990) which is well suited to test
strategy; it only has the analogs of Gauss and vector constraints,
whence the loop representation provides us with a complete set of physical
states for this model. The first step in the above strategy is then
to find the correct inner product for the Husain-Kucha\v r model.)
However, the task of selecting one measure over another even on this space is
quite non-trivial. One might imagine using the reality conditions for this
purpose. However, neither do we know how to express the reality conditions
in terms of loop-variables (in the Husain-Kucha\v r model, we do!) nor do
we have anything beyond a handful classical observables which Poisson-commute
with Gauss and vector constraints. We do have ``enough'' linear
operators on the space of pre-physical states. However, since we do not
know their classical analogs, we do not know what the quantum
$\star$-relations between these observables should be. The first step then
would be to understand the physical meaning of these operators.

The quantization program has been, however, carried to completion in several
truncated models and these investigations provide considerable hope.
These include%
\footnote{24}{Partial results have been obtained in several other models.
In particular, solutions to constraints have been obtained in the Bianchi
IX model (Kodama, 1988) and have been transformed back to the metric
representation to obtain a first solution to the Wheeler-DeWitt equation
in the model (Moncrief \& Ryan, 1991). Also, a new proposal has been
put forward for singling out a preferred state in quantum cosmology to
serve as ``the wave function of the universe'' (Kodama, 1990).}
Bianchi I and II models (Gonzalez \& Tate, 1992); spherically
symmetric minisuperspace (Thiemann \& Kastrup 1992); the midisuperspace
of two hypersurface orthogonal, commuting Killing fields (Thiemann, 1992),
and, the weak field truncation of general relativity in which one keeps
terms in the constraints up to {\it quadratic} order (Ashtekar, 1991;
chapter 12). In
all these cases, sufficient simplifications occur to enable one to impose
the reality conditions and these then single out the inner-product
uniquely. In cases --such as the Bianchi models and the weak field
truncation-- where there are other reasons based on symmetry groups to
constrain what the correct answer should be, the inner product singled
out by the reality conditions can be shown to be the correct one. The
spherically symmetric model is of special interest because in this case
the reality conditions involve genuinely non-linear relations between the
classical phase space variables. The weak field truncation is interesting
for two reasons. First, this is the only one among the models listed above
in which one has worked directly in the loop representation to arrive at
the correct inner product. (All other models have been worked out {\it only}
in the connection representation.) Second, one can also work in the connection
representation and then the scalar constraint {\it can} be re-interpreted
as a Schr\"odinger equation by isolating time among the components of the
connection. (In this treatment, matter fields can be incorporated and the
the same gravitational connection component serves as the time variable for
all fields.) Thus, at least in the weak field limit of the quantum theory,
one can single out time and recast the ``timeless'' loop description in
the more familiar terms. At a conceptual level, this is an important result
since it tells us that the familiar quantum field theoretic description in
Minkowski space {\it does} emerge as a well-defined truncation of the abstract,
non-perturbative framework of canonical quantum gravity. The main reason why
I did not include this discussion in any detail here is that is has already
been in the literature for some time now.

Currently, work is in progress (by Ashtekar and Varadarajan) on the
midisuperspace with one Killing
field. It turns out that, in presence of a Killing symmetry, the vacuum
Einstein theory in 4-dimensions is equivalent to the Einstein theory
coupled to a doublet of scalar fields in 3-dimensions (where the doublet
constitutes a sigma model). Using this fact, the reality conditions can be
recast in a form that is fully manageable as in 3-dimensional gravity.
Therefore, one can begin to implement the program. The hope is that this
simplification will lead us to the inner product. (Note incidentally that,
to find the inner product, it is {\it not} essential to first find the general
solution to the constraints explicitly; what we need is a handle on the
structure of the space of solutions.) Furthermore, there are recent results
from perturbation theory which suggest that the theory is not only
renormalizable but {\it finite} (Bonacino et al, 1992).
This indication is supported by non-perturbative considerations in the
classical theory: in the spatially open case, the Hamiltonian turns out to
be not only positive but {\it bounded above}! Therefore, one can hope to
solve the theory exactly and compare the results with those of the
perturbation theory.

Perhaps the most important limitation of the ``loopy'' framework for the
full theory is that it is yet to produce testable, physical predictions.
True, we have a new picture of the micro-structure of space-time in the Planck
regime. True, we have all these interesting relations between mathematics of
knot theory and states of quantum gravity. But where are testable predictions?
This is a problem faced by all approaches to quantum gravity and the difficulty
lies in the fact that we do not have experimental tools to probe things like
the nature of geometry at the Planck scale. But it does seem to me that it
should be possible to make a number of theoretical checks. I would like to
propose three in an increasing order of speculative character. My primary
motivation is only to illustrate ways in which viability of these
non-perturbative quantization ideas can be tested.

The first check comes from the expectation that we should be able to replace
Minkow\-ski space in the familiar quantum theories
by weaves and obtain physically acceptable results. To begin with, this would
provide a viability test for the idea that the microstructure of space-time
is in fact discrete and that the discreteness is of a specific type. At a
rough level, it should not be too difficult to obtain such a result because
the difference between the weave states and the continuum picture is
significant only at the Planck scale and, since renormalizable field theories
are short distance insensitive, their predictions at the laboratory scale of
observation should not be sensitive to the Planck scale structure. But the
analysis is by no means straightforward and the calculation will provide us
with considerable new insight in to the viability of the weave ideas. Of
particular interest is the question of infinities: do they simply go away
if we use weaves?

The second check concerns Hawking radiation. I indicated
above that if we carry out a truncation of the theory, keeping only terms
up to second order in deviation from flat space, we can identify one of the
connection components as time and re-interpret the scalar constraint as a
Schr\"odinger equation for the evolution of quantum states. What happens if
we carry out a similar truncation about a black hole background? Do the
modifications in the boundary conditions at the horizon naturally lead to
the Hawking radiation? If so, one would have a systematic derivation of
the Hawking effect from full quantum gravity, where the nature of
approximation involved is clear so that one can probe the predictions of
the theory beyond this approximation. This would be a view of the effect
``from above;'' the usual view is ``from below'' since one begins
with the classical theory and climbs just one step up to quantum field
theory in curved space-time.

The final proposal addresses the issue of the
cosmological constant. Ashtekar and Renteln pointed out in '86 that if one
begins with any connection $A_a^i$ , constructs its magnetic field
$\tw{B}^a_i$ and simply sets $\tw{E}^a_i := (1/\Lambda) \tw{B}^a_i$, then
the pair $(A_a^i, \tw{E}^a_i)$ satisfies all the constraint equations with
a cosmological constant $\Lambda$, and, furthermore, this ans\"atz provides
a complete characterization of solutions with non-zero $\Lambda$ in which the
Weyl tensor is self dual. Thus, if one imposes the (Lorentzian) reality
conditions, this ans\"atz picks out uniquely the (anti-)de Sitter space-time.
(For details, see Samuel (1991).) Now, the corresponding ans\"atz in quantum
theory produces for the state $\Psi(A)$ precisely the exponential of the
Chern-Simons action, with $(1/\Lambda)$ playing the role of the coupling
constant. Several workers in the field noted independently that this functional
satisfies all the quantum constraints formally. More recently, this statement
was made precise by Br\"ugmann, Gambini and Pullin through the
implementation of a point splitting regularization. Kodama (1990) has shown
that, as one might intuitively expect, this state is associated with the
(anti-)de Sitter space-time in a WKB approximation. He has then gone on to
argue that it should be considered as a ground state of the theory in the
$\Lambda\not= 0$ sector. There is also some independent support for
this idea coming from the mathematical theory of representations of the
holonomy algebra. Let us accept this viewpoint. Then, we are led to the
conclusion that, if $\Lambda \not= 0$, the vacuum is no longer CP-invariant,
whence there is gravitational CP-violation (see, e.g., chapter 13 in Ashtekar
(1991).) Now, if we believe that there is no such violation  --being
gravitational, it would have to be universal, not restricted just to the weak
interaction sector-- we are led to the conclusion that
the cosmological constant must be zero. Thus, the suggestion is that the
two facts --vanishing of the cosmological constant and the CP symmetry of
the gravitational interaction-- are really two facets of the same feature
of non-perturbative quantum gravity. This reduction of the two mysteries to
one may be considered as a ``solution'' of the cosmological constant problem.

\goodbreak
\subsection{6.3}{An evaluation}%\par

In section 1, I outlined the difficulties of quantum gravity from three
perspectives: that of a high energy theorist, of a mathematical physicist
and a general relativist. How has the connection-loop program fared from
these three angles?

The emphasis on the absence of a background structures should please the
general relativist. The approach is, nonetheless, based on a 3+1 decomposition
and therefore fails to be manifestly covariant. However, one can regard the
use of the Hamiltonian framework only as an intermediate step. As in the
pre-geometry version of the
3-dimensional theory, at a fundamental mathematical level, there is neither
space nor time. Manifest covariance is violated but only by the procedure by
which we descend from the quantum theory to classical relativity. At a
fundamental level in quantum theory, we do not even need a manifold structure
and in the final step we do recover classical general relativity with all
its covariance. Furthermore, even the space-time split which arose in the
intermediate level --when we introduced a preferred ``internal'' time-- seems
to be essential, at this stage, to satisfy the demands of the quantum
measurement theory (see Jim Hartle's lectures). The full covariance is gained
in the classical limit when these quantum measurement issues can be ignored.
The mathematical
developments related to knot theory and the way in which time arises in the
weak field truncation suggests that the overall situation would be
similar in the 4-dimensional theory.

The mathematical physicist can see that the program is only a step or two
beyond the heuristic level. But she would probably appreciate that it has
enough ingredients --the algebraic quantization program, the representation
theory of holonomy algebras, the non-linear duality between connections and
loops, etc.-- of the sort that one comes across in mathematical physics so that
it should be possible to put the various constructions on a much more
satisfactory footing. In particular, one can hope to prove theorems which
will justify, or at least spell out, the assumptions that are implicit in the
choices one has made in the construction of the loop representation. While
this may seem ``reasonable,'' a little voice would probably say: hold on!
How can it be that one hopes to construct a quantum theory of {\it gravity}
when not a {\it single} realistic, interacting theory has been solved so far
in four dimensions? Let me answer this in two parts. First, as was emphasized
at various points (particularly in section 5.4) the mathematical problems one
comes across in non-perturbative gravity are quite different from those one
has come across in Minkowskian field theories. Second, the problems one faces
in Minkowskian field theories generally (although not exclusively) come from
the bad ultra-violet behavior. From a physical standpoint, however, at these
higher and higher energies, the basic approximation that gravity can be
ignored and space-time can be represented by a continuum, flat geometry
seems hopelessly bad. Thus, it may be that the mathematical difficulties one
encounters have a physical origin. Only when we handle the short distance
structure of space-time correctly may these difficulties go away. One might
say: wait a minute; if this was so, why were we successful in constructing
quantum field theories in low dimensions? Won't your objections apply there
as well? My present belief is that they are in fact {\it not} applicable
because, in less than 4 dimensions, the gravitational field does not have its
own degrees of freedom. There are no gravitational modes that can be
excited by the quantum fluctuations of the matter fields whence the smooth
continuum picture is probably a good approximation.

The high energy theorist may find the tools used here somewhat unusual
because at a fundamental level, there are no Fock spaces, no particle like
excitations and no
S-matrices. The methods may also seem somewhat unfamiliar for there are
no effective actions and no path integrals. But he would be pleased with
the fact that one does focus on issues that are of concern also in the field
theoretic approaches: the short distance behavior of the theory, the new
physics at the Planck scale whose existence is suggested by perturbative
non-renormalizability, and, the finiteness of eigenvalues of geometrical
operators. He would also be relieved that one is not just formally
manipulating products of operators but worries about regularization and
renormalization. There is, however, a key difference of philosophy between
our approach and the ones that have been used in high energy theory for well
over a dozen years. We have dealt with general relativity by itself and
ignored other fields; these are to be brought in later. It is {\it not} that
matter fields are irrelevant or uninteresting. That would be too naive a stand.
Furthermore, significant work {\it has been} done in this direction in the
present canonical framework, particularly on supersymmetric coupling of
gravity to matter (see, e.g., Nicoli \& Matschull, 1992). This represents an
important technical development in the subject and is bound to play a
substantial role in the coming years. However, the belief is that it is pure
gravity that plays a key role in determining the quantum structure of
space-time; other fields are dynamically important but, as far as the
microstructure of geometry and the corresponding kinematic structure of the
theory are concerned, they will not qualitatively alter what we have
learned from pure gravity. In the high energy approaches, on the other hand,
one assumes, at least implicitly, that a consistent theory would not result
until a whole tower of particle states is included; that quantum gravity by
itself would not make sense. While I do not find the standard reasoning
behind this view compelling, I certainly have no argument to rule out this
possibility.

The two ways of looking at the problem, I think, are complementary and it
is important that everyone does {\it not} follow the same approach, work
with the same set of prejudices. In a field like quantum gravity where the
experimental data is so scarse, diversity of ideas is all the more important.
The advice that Richard Feynman gave to younger researchers at CERN on his way
back from Stockholm in 1965 seems especially apt in the context of quantum
gravity:
\smallskip
{\narrower\narrower\smallskip\noindent
{\sl It's very important that we do not all follow the same fashion.
 ... It's necessary to increase the amount of variety ... and the only way to
do it is to implore you few guys to take a risk with your lives that you will
be never heard of again, and go off in the wild blue yonder and see if you
can figure it out.}\smallskip}
\bigskip

\goodbreak
\vfill\break

%\input [ijtpd.lh]qgmac.tex
%\singlespace

\section{}{Acknowledgments}%\par

The innumerable discussions I have had with Bernd Br\"ugmann, Rodolfo Gambini,
Chris Isham, Jorge Pullin, Carlo Rovelli, Lee Smolin and Ranjeet Tate have
added a great deal of clarity to these notes and the comments I received on
the manuscript from the members of the Syracuse relativity group have led to
a number of improvements. I am also grateful to the organizers of the school,
Bernard Julia and  J. Zinn-Justin, for their advice, help, and patience and to
participants of the school for many stimulating discussions. This work was
supported in part by the NSF grants PHY90-16733 and INT88-15209 and by
research funds provided by Syracuse University.

\section{}{References}%\par

\item{}{Achucarro, A. \& Townsend, P. (1986) Phys. Lett. {\bf B180}, 85-89.}
\item{}{Agishtein, A. \& Migdal, A. (1992) Mod. Phys. Lett. {\bf 7},
1039-1061.}
\item{}{ Arnowitt, R., Deser, S. \& Misner, C. (1962) In: L. Witten, ed
{\it Introduction to current research}, Wiley, New York; pp 227-265.}
\item{}{Ashtekar, A. (1987) Phys. Rev. {\bf D36}, 1587-1602.}
\item{}{Ashtekar, A. (1981) {\it Non-perturbative canonical gravity}, World
Scientific, Singapore.}
\item{}{Ashtekar, A. \& Stillerman, M. (1986) J. Math. Phys. {\bf 27},
1319-1330.}
\item{}{Ashtekar, A., Husain, V., Rovelli, C. \& Smolin, L. (1989) Class.
Quan. Grav. {\bf 6}, L185-L193.}
\item{}{Ashtekar, A.,  Rovelli, C. \& Smolin, L. (1991) Phys. Rev. {\bf D44},
1740-1755.}
\item{}{Ashtekar, A. \& Isham, C. (1992) Class. Quan. Grav. {\bf 9},
1069-1100.}
\item{}{Ashtekar, A. \& Rovelli, R. (1992) Class. Quan. Grav. {\bf 9},
1121-1150.}
\item{}{Ashtekar, A., Rovelli, C. \& Smolin L. (1992) Phys. Rev. Lett. {\bf 69}
237-240.}
\item{}{Ashtekar, A. \& Tate, R. (1992) Syracuse University Pre-print.}
\item{}{Baez, J. (1992) Class. Quan. Grav. (to appear)}
\item{}{Bengtsson, I. (1989) Phys. Lett. {\bf B220}, 51-53.}
\item{}{Blencowe, M. (1990) Nucl. Phys. {\bf B341}, 213-251.}
\item{}{Bonacina, G., Gamba, A. \& Martellini, M. (1992) Phys. Rev.
{\bf D45}, 3577-3583.}
\item{}{Br\"ugmann, B., Gambini, R. \& Pullin, J. (1992a) Phys. Rev. Lett.
{\bf 68}, 431-434.}
\item{}{Br\"ugmann, B., Gambini, R. \& Pullin, J. (1992b) In: S. Cotta \&
A. Rocha eds, {\it XXth International conference on differential geometric
methods in physics}, World Scientific, Singapore.}
\item{}{Br\"ugmann, B. \& Pullin, J. (1992) Nucl. Phys. {\bf B} (to appear)}
\item{}{Capovilla, R., Dell, J. \& Jacobson, T. (1989) Phys. Rev. Lett.
{\bf 63}, 2325-2328.}
\item{}{Chen, K. (1973) Ann. Math. {\bf 97}, 217-246.}
\item{}{Di Bortolo, R. Gambini \& Griego, J. (1992) Comm. Math. Phys.
(to appear)}
\item{}{Dirac, P. (1964) {\it Lectures on Quantum Mechanics}, Yeshiva
University Press, New-York.}
\item{}{Gambini R. \& Leal, A. (1992) Comm. Math. Phys. (to appear)}
\item{}{Goldberg, J., Lewandowski, J. \& Stornaiolo, C. (1992) Comm. Math.
Phys. {\bf 148}, 337-402.}
\item{}{Gonzalez G. \& Tate, R. (1992) Syracuse University Pre-print.}
\item{}{Husain, V. \& Kucha\v r K. (1990) Phys. Rev. {\bf D 42}, 4070-4077.}
\item{}{Isham, C.\& Kakas A. (1984a) Class. Quan. Grav. {\bf 1}, 621-632.}
\item{}{Isham, C.\& Kakas A. (1984b) Class. Quan. Grav. {\bf 1}, 633-650.}
\item{}{Iwasaki, J. \& Rovelli, C. (1993) Int. J. Mod. Phys. {\bf D}
(to appear)}
\item{}{Kucha\v r, K. (1981) In: C. Isham, R. Penrose \& D. Sciama eds,
{\it Quantum gravity 2}, Clarendon press, Oxford; pp329-374.}
\item{}{Kodama, H. (1988) Prog. Theo. Phys. {\bf 80}, 1024-1040.}
\item{}{Kodama, H. (1990) Phys. Rev. {\bf D42}, 2548-2565.}
\item{}{Kodama, H. (1992) Int. J. Mod. Phys. {\bf D} (to appear)}
\item{}{Loll, R. (1992) Nucl. Phys. {\bf B368}, 121-142.}
\item{}{Manojlovi\'c N. \& Mikoci\'c A. Nucl. Phys. {\bf B} (to appear)}
\item{}{Moncrief, V. \& Ryan, M. (1991) Phys. Rev. {\bf D44}, 2375-2379.}
\item{}{Nicoli, H. \& Matschull, H. (1992) DESY Pre-print.}
\item{}{Peldan, P. (1992) Phys. Rev. {\bf D46}, 2279-2282; Nucl. Phys.
{\bf B} (to appear)}
\item{}{Penrose, R. \& Rindler, W. (1986) {\it Spinors and Space-time;
Vol 1,} Cambridge University Press, Cambridge.}
\item{}{Randall, A. (1992a) Syracuse University Pre-print.}
\item{}{Randall, A. (1992b) Class. Quan. Grav. (to appear)}
\item{}{Romano, J. (1991) {\it Geometrodynamics vs connection dynamics},
Syracuse University Thesis;  Gen. Rel. Grav. (to appear)}
\item{}{Rovelli, C. (1991) Class. Quan. Grav. {\bf 8}, 1613-1675.}
\item{}{Rovelli, C. \& Smolin, L. (1990) Nucl. Phys. {\bf B331}, 80-152.}
\item{}{Samuel, J. (1991) In: A. Janis \& J. Porter eds, {\it Recent advances
in general relativity }, Birkh\"auser, Boston; pp72-84.}
\item{}{Smolin, L. (1992) In: {\it XXIInd International seminar on theoretical
physics}, World-Scientific, Singapore.}
\item{}{Streater, R. \& Wightman, A. (1964) {\it PCT, spin and statistics,
and all that}, Benjamin, New York.}
\item{}{Thiemann, T. (1992) Private Communication.}
\item{}{Thiemann, T. \& Kastrup, H. (1992) Nucl. Phys. {\bf B} (to appear).}
\item{}{Wald, R. (1984) {\it General Relativity,} University of Chicago
Press, Chicago.
\item{}{Wheeler, J. (1964) In: C. DeWitt and B. DeWitt eds, {\it Relativity,
groups and topology}, Gordon \& Breach, New York; pp316-520.}
\item{}{Wightman, A. (1992) Quantum field theory: some history and some
current problems; the annual joint mathematics-physics colloquium, Syracuse
university.}
\item{}{Witten, E. (1988) Nucl. Phys. {\bf B311}, 46-78.}
\item{}{Zegwaard, J. (1992) Nucl. Phys. {\bf B378}, 288-308.}

\end
\bye